\documentclass[12pt]{article}

\pdfoutput=1 
\usepackage{graphicx} 
\usepackage{color}                                                                                                                                                                                                                                                                                                                                                                                                                                                                                                                                                                                                                                                                                                                                                                                                                                                                                                                                                                                                                        
\usepackage{cancel,tabularx,moreverb,fancybox,amsmath,float,bm,braket,slashbox,txfonts,amssymb,bm,accents,ascmac}
\usepackage[top=30truemm,bottom=30truemm,left=25truemm,right=25truemm]{geometry}
\usepackage{latexsym}
\usepackage{here}
\usepackage{cite}
\usepackage{hyperref}                                                                                                                                                                                                                                                                                                                                                                                                                                                                                                                                                                                                                                                                                                                                                                                                                                                                                                                                                                                                                                                                                                                                                                                                                                                                                                                                                                                                                                                                                                                                                                                                                                                                                                                                                                                                                                                                                                                                                                                                                                                                                                                                                                                                                                                                                                                                                                                                                                                                                                                                                                                                                                                                                                                                                                                                                                                                                                                                                                                                                                                                                                                                                                                                                                                                                                                                                                                                                                                                                                                                                                                                                                                                                                                                                                                                                                                                                                                                                                                                                                                                                                                                                                                                                                          

\newcommand{\ex}[1]{\mathrm{e}^{#1}}

\newcommand{\pa}[1]{\left(#1 \right)}

\newcommand{\bb}[1]{\mathbb{#1}}

\newcommand{\ca}[1]{\mathcal{#1}}

\newcommand{\abs}[1]{\left|#1\right|}

\newcommand{\pd}[1]{\frac{\partial}{\partial #1}}

\newcommand{\ti}[1]{\tilde{#1}}

\newcommand{\fr}{\frac}
\newcommand{\s}[1]{\sqrt{#1}}

\def\be{\begin{equation}}
\def\ee{\end{equation}}
\def\ba{\begin{eqnarray}}
\def\ea{\end{eqnarray}}

 \def\w{{\omega}}

 \def\a{{\alpha}}
 \def\ba{{\bar{\alpha}}}

 \def\G{{\Gamma}}

\def\dd{{\mathrm{d}}}

  \makeatletter
    
    \@addtoreset{equation}{section}
  \makeatother

\begin{document}

\begin{titlepage}
\thispagestyle{empty}

\begin{flushright}
RIKEN-iTHEMS-Report-21,
\\

\end{flushright}

\bigskip

\begin{center}
\noindent{{\large \textbf{
Analytic Bootstrap in 2D Boundary Conformal Field Theory:\\
Towards Braneworld Holography
}}}\\
\vspace{2cm}
Yuya Kusuki ${}^{1,2}$
\vspace{1cm}

${}^{1}${\small \sl 
Walter Burke Institute for Theoretical Physics \\
California Institute of Technology, Pasadena, CA 91125, USA
}

${}^{2}${\small \sl RIKEN Interdisciplinary Theoretical and Mathematical Sciences (iTHEMS), \\Saitama 351-0198, Japan}

\vskip 2em
\end{center}

\begin{abstract}
Recently, boundary conformal field theories (BCFTs) have attracted much attention in the context of quantum gravity.
This is because a BCFT can be dual to gravity coupled to a heat bath CFT, known as the island model.
On this background, it would be interesting to explore the duality between the boundary and the braneworld.
However, this seems to be a challenging problem.
The reason is because although there has been much study of rational BCFTs,
there has been comparatively little study of irrational BCFTs, and irrational BCFTs are expected to be the boundary duals of the braneworlds.
For this reason, we explore properties of boundary ingredients: the boundary primary spectrum, the boundary-boundary-boundary OPE coefficients
and the bulk-boundary OPE coefficients.
For this purpose, the conformal bootstrap is extremely useful.
This is the first step in providing an understanding of BCFTs in the context of braneworld holography by using the conformal bootstrap. The techniques developed in this paper may be useful for further investigation of irrational BCFTs.

 \end{abstract}

\end{titlepage}

\restoregeometry

\tableofcontents

\section{Introduction \& Summary}

\subsection{Introduction}

\begin{figure}[t]
 \begin{center}
  \includegraphics[width=16.0cm,clip]{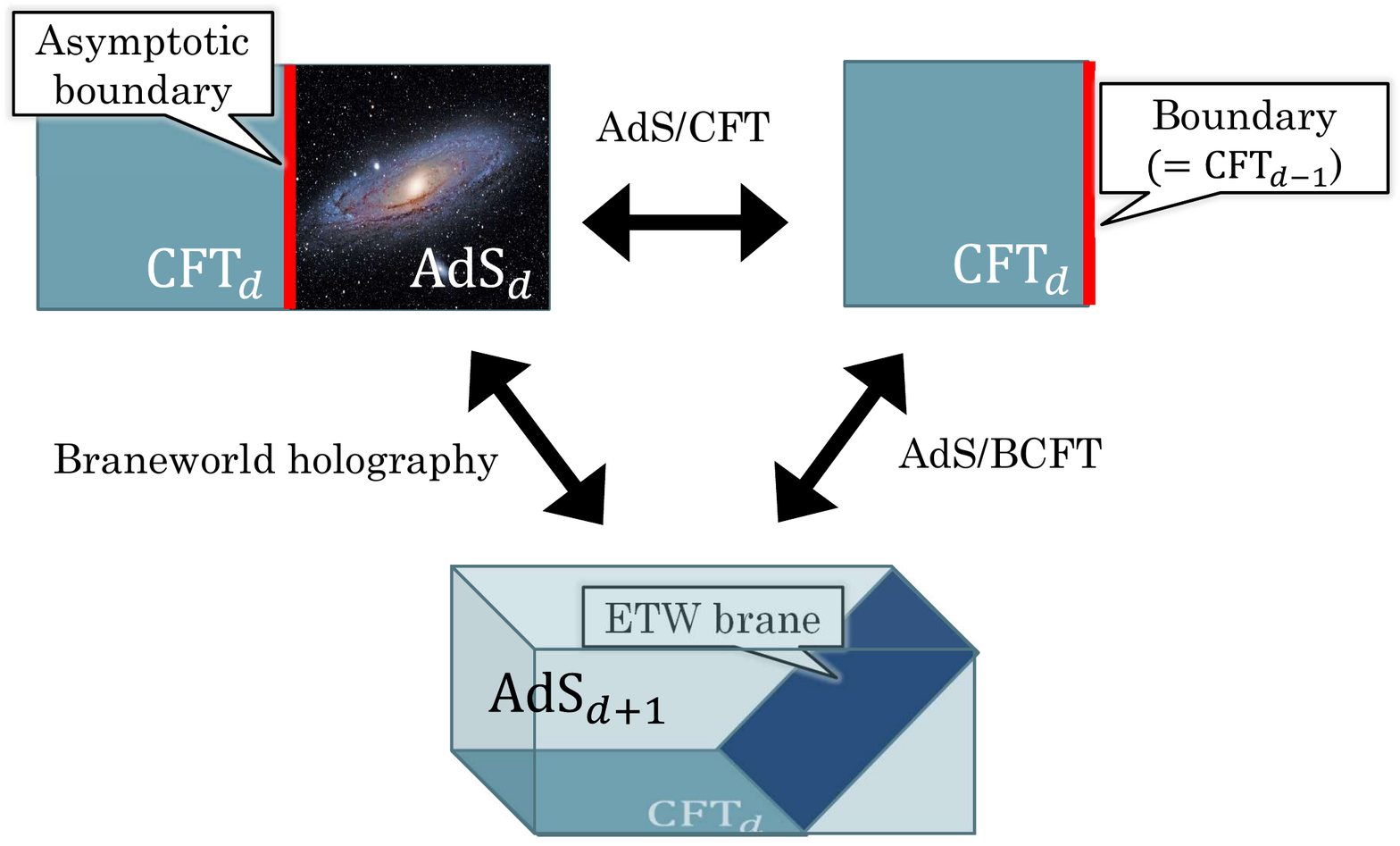}
 \end{center}
 \caption{Triality through the AdS/CFT correspondence. The CFT drawn in the left is dual to the BCFT drawn in the right through the AdS/CFT correspondence. Moreover, the BCFT is dual to the quantum gravity shown at the bottom through the AdS/BCFT correspondence.
There is also a relation between the left picture and the bottom picture through the braneworld holography
\cite{Karch2001,Randall1999,Randall1999a}.
}
 \label{fig:triality}
\end{figure}

Recently, there has been tremendous progress on the information paradox problem.
The idea is to consider a setup where a CFT${}_d$ on AdS${}_d$ and a CFT${}_d$ on a non-gravitational bath are glued along the (asymptotic) boundary \cite{Penington2020, Almheiri2019, Almheiri2020}.
Since light can go through the asymptotic boundary in this setup,
we can probe the Hawking radiation from black holes by the entanglement entropy between the gravitational and bath systems.
This setup is sometimes called ``doubly holography''.
This is because we have three pictures, which are equivalent (see Figure \ref{fig:triality}).
By applying the AdS/CFT correspondence \cite{Maldacena1998} to the AdS${}_d$ gravity in this setup (left of Figure \ref{fig:triality}),
we can obtain a boundary conformal field theory, where the boundary is dual to the AdS${}_d$ gravity (right of Figure \ref{fig:triality} ).
We can obtain another picture by applying the AdS/BCFT correspondence \cite{Takayanagi2011,Fujita2011} to this BCFT (see bottom of Figure \ref{fig:triality}).
Here we apply the AdS/CFT correspondence ``twice'', so we call this holography the doubly holography.
The bottom picture in Figure \ref{fig:triality} is sometimes useful because we can evaluate ``quantum'' entanglement therein
by just calculating the holographic entanglement entropy (i.e. the area of Ryu-Takayanagi surface)  in the setup shown in bottom of Figure \ref{fig:triality}
(for example, see \cite{Ageev:2021ipd,Akal:2021foz,Chu:2021gdb,Bousso:2020kmy,Sully:2020pza,Akal2021,Akal2020,Miao2021,Geng2021, Geng2021a }).

On this background, it would be very interesting to explore the duality,
\footnote{
This duality can be obtained by (i) applying the AdS/CFT to the left of Figure \ref{fig:triality}, or (ii) applying the braneworld holography and the AdS/BCFT to the left of Figure \ref{fig:triality}.
It is still unclear whether this duality really holds.
Our investigations of BCFTs is partly motivated to clarify the consistency of this duality.
We hope to come back to this issue more explicitly in the future paper.
}
\begin{screen}
\begin{center}
$d$ dimensional boundary of CFT${}_{d+1}$ = $d+1$ dimensional braneworld
\end{center}
\end{screen}
In particular, it would be interesting to consider the question, ``can the boundary be dual to classical gravity?'' or ``which type of gravity is realized on the braneworld associated with the BCFT?''.
Nevertheless, there has been little knowledge of the CFT with this holographic boundary.
This is because there has been currently very little knowledge of irrational BCFTs.
\footnote{Except for the Liouville CFT, \cite{Teschner2000,Ponsot2002,Fateev2000}.
The Liouville CFT is not of interest in this paper,
since it is not a compact CFT, which seems to be quite different from the holographic CFT.
}
\footnote{When we were writing up this article, we noticed the interesting preprint \cite{Collier2021} on the arXiv, which has provided some results about irrational BCFTs by using the bootstrap equation at the fixed point. This is complementary to this article, which considers the bootstrap equation in the high-low temperature limit.}
For example, many techniques in CFTs with no boundaries have been generalized to CFTs with boundaries, and various constraints have been given by using the conformal bootstrap equation as shown in \cite{Cardy2004}. However, these results are given only when we restrict ourselves to rational conformal field theories (RCFTs).

Against this backdrop, it is very important to exhibit various universal formulae in irrational CFTs with boundaries.
For example, the universal formula for the entropy at high energy, known as the Cardy formula \cite{Cardy1986}, plays an important role in the understanding of quantum gravity. This entropy formula surprisingly matches with the black hole entropy \cite{Strominger1996}. This result may therefore answer the greatest question, ``Is black hole thermodynamics is really thermodynamics?''. The Cardy formula answers this question, ``Yes''. 
This accomplishment surely supports the high usefulness of discovering universal formulae in BCFTs with some assumptions that are needed for reproducing classical gravity.

There is a useful tool to identify unknown information in a given CFT, known as the conformal bootstrap.
The conformal bootstrap is just a consistency condition in CFTs, more precisely, the OPE associativity.
We can utilize this consistency condition to give non-trivial constraints on the spectrum and the OPE coefficients in a given CFT.
Some significant breakthroughs have been achieved by the numerical application (or the low-order approximation inspired by that) of the conformal bootstrap to investigate low-lying operators \cite{Rattazzi2008, Rychkov2009, Kos2014, SimmonsDuffin2015, Collier2018}, the upper bound on the gap from the vacuum \cite{Hellerman2011, Friedan2013, Collier2018, Hartman:2019pcd, AfkhamiJeddi2019, Besken:2021eps} and the uniqueness of Liouville CFT \cite{Collier2018a}. 
The conformal bootstrap equation can also be used to derive universal formulae in some asymptotic regions.
For example, one can derive the universal formula for the spectrum at high energy region \cite{Cardy1986,Mukhametzhanov2019,Pal2020}, the heavy-light-light OPE coefficients \cite{Das2018, Kusuki2018, Kusuki2018a,Das2021,Tsiares2020},
the heavy-heavy-light OPE coefficients \cite{Kraus2017, Hikida2018, RomeroBermudez2018, Brehm2018},
the heavy-heavy-heavy OPE coefficients \cite{Cardy2017}
and the OPE in the large spin limit \cite{Fitzpatrick2013, Komargodski2013,Alday2015,Kaviraj2015,Kaviraj2015a, Alday2017a, SimmonsDuffin2017,Alday2017,Sleight2018,Fitzpatrick2014}.

Recently, there is remarkable progress in solving the bootstrap equation in 2D CFTs.
The idea is to act the fusion transformation on the vacuum block,
which provides a new analytic method to access the CFT data.
The fusion matrix bootstrap has provided important progress in the light cone conformal bootstrap \cite{Kusuki2019a, Collier2018, Kusuki2019} and
the light-cone modular bootstrap \cite{Maxfield2019, Benjamin2019,Benjamin2020}.
Moreover, by making use of the fusion matrix bootstrap, all of the results in \cite{Das2018, Kusuki2018, Kusuki2018a, Das2021, Kraus2017, Hikida2018, RomeroBermudez2018, Brehm2018, Cardy2017} can be organized \cite{Brehm2020, Collier2020, Anous2021}.
For this reason, in this article, we investigate universal formulae in BCFTs by using the fusion matrix bootstrap.

\subsection{Summary}

One goal of this paper is to formulate the bootstrap equation in irrational BCFTs
and to clarify what we can learn from the bootstrap equation beyond rational BCFTs.
The techniques developed in this paper are very useful to provide a new understanding of the braneworld holography from the CFT side.
Although we only show the asymptotic universal formulae in this paper,
there are a lot of applications of our techniques in the context of the braneworld holography.
At the end of this paper, we propose some interesting questions that we can address by using our techniques.

Here we briefly summarize our results:
\begin{itemize}
\item Universality for OPE coefficients (Section \ref{sec:bootstrap})

The asymptotic behaviors of the OPE coefficients can be estimated by using the fusion matrix bootstrap.
In particular, we will show the universal formulae for
\begin{itemize}
\item boundary primary spectrum (i.e., open-string spectrum) at high energy,
\item boundary-boundary-boundary OPE coefficients for heavy-heavy-heavy weights,
\item boundary-boundary-boundary OPE coefficients for heavy-heavy-light weights,
\item boundary-boundary-boundary OPE coefficients for heavy-light-light weights,
\item bulk-boundary OPE coefficients for heavy-heavy weights,
\item bulk-boundary OPE coefficients for heavy-light weights,
\item bulk-boundary OPE coefficients for light-heavy weights.
\end{itemize}
For this purpose, we will consider various bootstrap equations on various manifolds.
Although in principle, we can fix all components in a BCFT by considering only six conformal bootstrap equations \cite{Lewellen1992},
it is useful to consider other bootstrap equations to fix the asymptotics in practice.

\item Light-cone bootstrap in BCFT (Section \ref{sec:light-cone})

We will show that in some setups, we can consider the large-spin asymptotics of the OPE coefficients even though BCFTs do not have the anti-holomorphic part.
It implies that it would be possible to consider an analog of the light-cone modular bootstrap \cite{Benjamin2019} in BCFTs.

\item Braneworld holography (Section \ref{sec:braneworld})

Although many recent works heavily make use of the braneworld holography,
there is very limited knowledge about this type of holography.
We believe that the braneworld should be realized on the boundary of the BCFT,
nevertheless, we do not know what gravity is realized as the boundary of the BCFT.
Some hints can be obtained from the asymptotics of the boundary ingredients, like the matching between the Cardy formula and the Bekenstein-Hawking entropy.
In some recent works, we consider the braneworld in the JT or Liouville gravity, which can have black hole solutions.
From this observation, we can expect that the boundary ingredients follow the Cardy formula, the eigenstate thermalization hypothesis (ETH), and some other chaotic nature, like the AdS${}_3$ gravity.
On this background, we will give some thoughts on the braneworld.
Here we do not give the explicit connection between the braneworld and the BCFT because our main purpose is to give a new tool to access the data of the boundary ingredients in irrational BCFTs as the first step. We hope that we will show the explicit connection by using our results obtained here in the near future paper.

\end{itemize}

Note that some of our results can also be straightforwardly applied to the one-dimensional conformal bootstrap.
This consideration could provide a generalization of the analytic bootstrap result in 1D CFT \cite{Mazac2016, Mazac2018} into a Virasoro version.
For example, our result (\ref{eq:bdybdybdy}) can be thought of as the Virasoro generalization of the asymptotic results in \cite{Mazac2016, Mazac2018}.
On the bulk side, the Virasoro generalization allows us to access information including graviton effects.
This is particularly useful to consider the black hole information paradox as one can see in \cite{Fitzpatrick2014, Fitzpatrick2015, Fitzpatrick2016}
since the exchange of the Virasoro descendants can create an effect of the presence of the BTZ black hole background, unlike the global conformal descendants.

\section{Review of Fusion Matrix Bootstrap} \label{sec:review}

In this section, we will review how to solve the conformal bootstrap equation by using the fusion matrix, first introduced in \cite{Kusuki2019a, Kusuki2019, Collier2019} and generalized in  \cite{Collier2020}.

The conformal bootstrap equation is the relation between different OPE coefficients $C_{ijk}$ in
the following form,
\begin{equation}
\begin{aligned}\label{eq:bootstrap}
&\sum_p C_{12p}C_{34p} \ca{F}^{21}_{34}(h_p|z)\overline{\ca{F}^{21}_{34}}(\bar{h}_p|\bar{z})\\
&=\sum_p C_{14p}C_{23p} \ca{F}^{23}_{14}(h_p|1-z)\overline{\ca{F}^{23}_{14}}(\bar{h}_p|1-\bar{z}),
\end{aligned}
\end{equation}
where we define the Virasoro conformal block as follows,
\newsavebox{\boxpc}
\sbox{\boxpc}{\includegraphics[width=120pt]{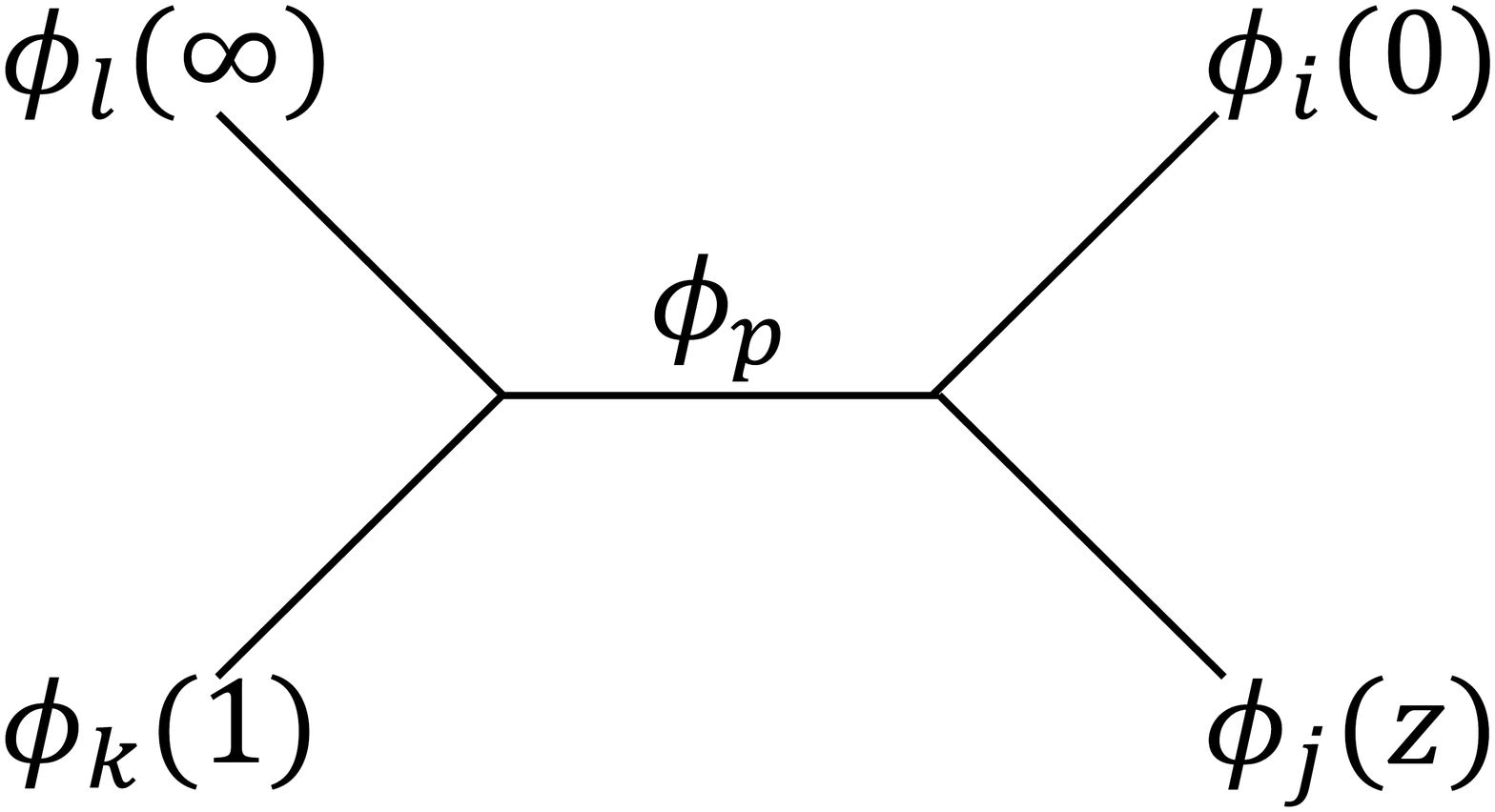}}
\newlength{\pcw}
\settowidth{\pcw}{\usebox{\boxpc}} 

\begin{equation}\label{eq:blockdef}
 \ca{F}^{ji}_{kl}(h_p|z) \equiv \parbox{\pcw}{\usebox{\boxpc}}.
\end{equation}
For simplicity, we focus on the case where $h_1=h_2=h_A$ and $h_3=h_4=h_B$.
Even though it is hard to solve this equation in general,
in some asymptotic regimes, we can analytically solve the bootstrap equation.
Here, we will show some examples.
In the limit as $z,\bar{z} \to 0$, we can approximate the bootstrap equation as
\footnote{
In fact, we do not need to take both $z \to 0$ and $\bar{z} \to 0$ limit if we assume our CFT to have no extended current.
The limit $z \to 0$ with $\bar{z}$ fixed leads to the same vacuum block approximation.
This technique is called the light cone bootstrap (see Section \ref{sec:light-cone} in more details).
Note that the same technique (including the fusion matrix approach as we show in the following) can also be useful to investigate the Lorentzian dynamics (see for example, \cite{He2019,Kusuki2019,Kusuki2019a}).
}
\begin{equation}
\begin{aligned}
 \ca{F}^{AA}_{BB}(0|z)\overline{\ca{F}^{AA}_{BB}}(0|\bar{z}) \simeq \sum_p C_{ABp}C_{ABp} \ca{F}^{AB}_{AB}(h_p|1-z)\overline{\ca{F}^{AB}_{AB}}(\bar{h}_p|1-\bar{z}),
\end{aligned}
\end{equation}
The key point is that we have invertible fusion transformations between the s and t-channel Virasoro conformal block.
\begin{equation}\label{eq:bootstrapAABB2}
\begin{aligned}
&\ca{F}^{AA}_{BB}(0|z)\\
&=
 \sum_{\substack{\a_{n,m}<\fr{Q}{2} \\ n,m \in \bb{Z}_{\geq0}}}\ \text{Res}\pa{   -2\pi i 
  {\bold F}_{0, \a_t} 
   \left[
    \begin{array}{cc}
    \a_A   & \a_A  \\
     \a_B  &   \a_B\\
    \end{array}
  \right]
  \ca{F}^{AB}_{AB}(h_t|1-z);\a_t=\a_{n,m}}\\
&+
\int_{\fr{Q}{2}+0}^{\fr{Q}{2}+i \infty} \dd \a_t {\bold F}_{0, \a_t} 
   \left[
    \begin{array}{cc}
    \a_A   & \a_A  \\
     \a_B  &   \a_B\\
    \end{array}
  \right]
  \ca{F}^{AB}_{AB}(h_t|1-z),
\end{aligned}
\end{equation}
where $\a_{n,m}\equiv\a_A+\a_B+mb+nb^{-1}$ and kernel $ {\bold F}_{\a_s, \a_t} $ denotes the {\it fusion matrix} (or {\it crossing matrix}), and we introduce the notation usually found in the Liouville CFT,
\begin{equation}
c=1+6Q^2, \ \ \ \ \ Q=b+\fr{1}{b}, \ \ \ \ \ h_i=\a_i(Q-\a_i), \ \ \ \ \ \bar{h}_i=\bar{\a}_i(Q-\bar{\a}_i).
\end{equation}
For simplicity, we define the degeneracy of primary states $\rho (\a,\bar{\a})$ as
\begin{equation}
\rho (\a,\bar{\a})=\sum_p D_p \delta(\a-\a_p) \delta(\bar{\a}-\bar{\a}_p),
\end{equation}
where the function $D_p$ denotes the degeneracy of primary operators with weights $(h_p,\bar{h}_p)$
and then define the averaged OPE coefficients as
\begin{equation}
\rho(\alpha_p,\bar{\alpha}_p) \overline{C_{ABp}C_{ABp}} = \sum_q C_{ABq}C_{ABq}\delta\pa{\alpha_p-\alpha_q} \delta\pa{\bar{\alpha}_p-\bar{\alpha}_q},
\end{equation}
where the average is taken over the states with the Liouville momentum $(\a_p, \bar{\a}_p)$.

With this notation, we obtain the following asymptotic bootstrap equation,
\begin{equation}
\begin{aligned}
&\int \dd \a_p \dd \bar{\a}_p \
{\bold F}_{0, \a_p} 
   \left[
    \begin{array}{cc}
    \a_A   & \a_A  \\
     \a_B  &   \a_B\\
    \end{array}
  \right]
{\bold F}_{0, \bar{\a}_p} 
   \left[
    \begin{array}{cc}
    \bar{\a}_A   & \bar{\a}_A  \\
     \bar{\a}_B  &   \bar{\a}_B\\
    \end{array}
  \right]
  \ca{F}^{AB}_{AB}(h_p|1-z)  \overline{\ca{F}^{AB}_{AB}}(\bar{h}_p|1-\bar{z})
 \\
&\simeq
\int \dd \a_p \dd \bar{\a}_p \
\rho(\alpha_p,\bar{\alpha}_p) \overline{C_{ABp}C_{ABp}}
\ca{F}^{AB}_{AB}(h_p|1-z)\overline{\ca{F}^{AB}_{AB}}(\bar{h}_p|1-\bar{z}).
\end{aligned}
\end{equation}
Thus, we can find the universal asymptotic formula for the OPE coefficients as
\begin{equation}\label{eq:bulk^3}
\rho(\alpha_p,\bar{\alpha}_p) \overline{C_{ABp}C_{ABp}} \simeq 
{\bold F}_{0, \a_p} 
   \left[
    \begin{array}{cc}
    \a_A   & \a_A  \\
     \a_B  &   \a_B\\
    \end{array}
  \right]
{\bold F}_{0, \bar{\a}_p} 
   \left[
    \begin{array}{cc}
    \bar{\a}_A   & \bar{\a}_A  \\
     \bar{\a}_B  &   \bar{\a}_B\\
    \end{array}
  \right], \ \ \ \ \ \ \ h_p, \bar{h}_p \to \infty.
\end{equation}
The explicit form of the fusion matrix is given in Appendix \ref{app:FM}.
In particular,  the asymptotic behavior of the fusion matrix is given by
\begin{equation}
{\bold F}_{0, \a_p} 
   \left[
    \begin{array}{cc}
    \a_A   & \a_A  \\
     \a_B  &   \a_B\\
    \end{array}
  \right]
\simeq
16^{-h_p} \ex{\pi\sqrt{\fr{c}{6}\pa{h_p-\fr{c}{24}}}}  \ \ \ \ \ \ \ h_p \to \infty.
\end{equation}
This asymptotic behavior is consistent with the results in \cite{Das2018, Kusuki2018, Kusuki2018a, Das2021},
which derive the asymptotic behavior of the OPE coefficients in a different way, using the inverse Laplace transformation.
Note that the above result (\ref{eq:bulk^3}) should be interpreted as the density of states averaged over a small window.
The dependence on the size of the window can be obtained by using the Tauberian theorem \cite{Mukhametzhanov2020, Das2021},
which show that a change of the size only gives $O(1)$ corrections.
In this paper, we will not attempt here to care about the size of the window.
It would be interesting to applying the Tauberian theorem on our results in a similar way as \cite{Mukhametzhanov2020, Das2021}.

The same approach can also be applied to the modular bootstrap equation.
The partition function can be expressed by
\begin{equation}
Z(\tau,\bar{\tau})=
\int \dd \a_p \dd \bar{\a}_p \
\rho (\a_p,\bar{\a}_p) \chi_{h_p}(\tau)\bar{\chi}_{\bar{h}_p}(\bar{\tau}),
\end{equation}
where $ \chi_{h_p}(\tau)$ is the Virasoro character.
The character has an analog of the fusion transformation, so-called the modular-S transformation,
\begin{equation}\label{eq:Smodular}
 \chi_{h_p}(\tau) = \int \dd \a_q \ S_{pq} \chi_{h_q} \pa{-\fr{1}{\tau}}.
\end{equation}
Thus, in the same way as the conformal bootstrap, we obtain the universal asymptotic formula for the density of states as
\begin{equation}\label{eq:Cardy}
\rho(\a_p, \bar{\a}_p) \simeq S_{0p} \bar{S}_{0p}
, \ \ \ \ \ \ \ h_p, \bar{h}_p \to \infty.
\end{equation}
Note that the explicit form of the modular-S matrix is given by
\begin{equation}
S_{0p} = -4\sqrt{2} \sin \pa{2\pi b \pa{\a-\fr{Q}{2}}}  \sin \pa{2\pi \fr{1}{b} \pa{\a-\fr{Q}{2}}}.
\end{equation}
The asymptotic behavior of the modular-S matrix is
\begin{equation}
S_{0p} \simeq \ex{2\pi\sqrt{\fr{c}{6}\pa{h_p-\fr{c}{24}}}},  \ \ \ \ \ \ \ h_p \to \infty.
\end{equation}
This completely matches the Cardy formula.
In other words, the result (\ref{eq:Cardy}) is just another derivation of the Cardy formula.

\section{Boundary Conformal Field Theory}
Let us move on to conformal field theories with boundaries.
As shown in \cite{Cardy2004},
if we restrict ourselves to boundary conditions satisfying
\begin{equation}\label{eq:T=T}
\left.
T(z) = \bar{T} (\bar{z})
\right|_{\text{bdy}} ,
\end{equation}
we can directly relate the analytic structure of correlators on the upper half plane to that on the full plane.
More precisely, we can reconstruct the Ward identity if we consider $ \bar{T} (\bar{z})$ to be the analytic continuation of $T(z)$ into the lower half plane.
It means that the Virasoro algebra can fix many ingredients in BCFTs, like in CFTs without boundaries as we have seen in Section \ref{sec:review}.

\begin{figure}[t]
 \begin{center}
  \includegraphics[width=8.0cm,clip]{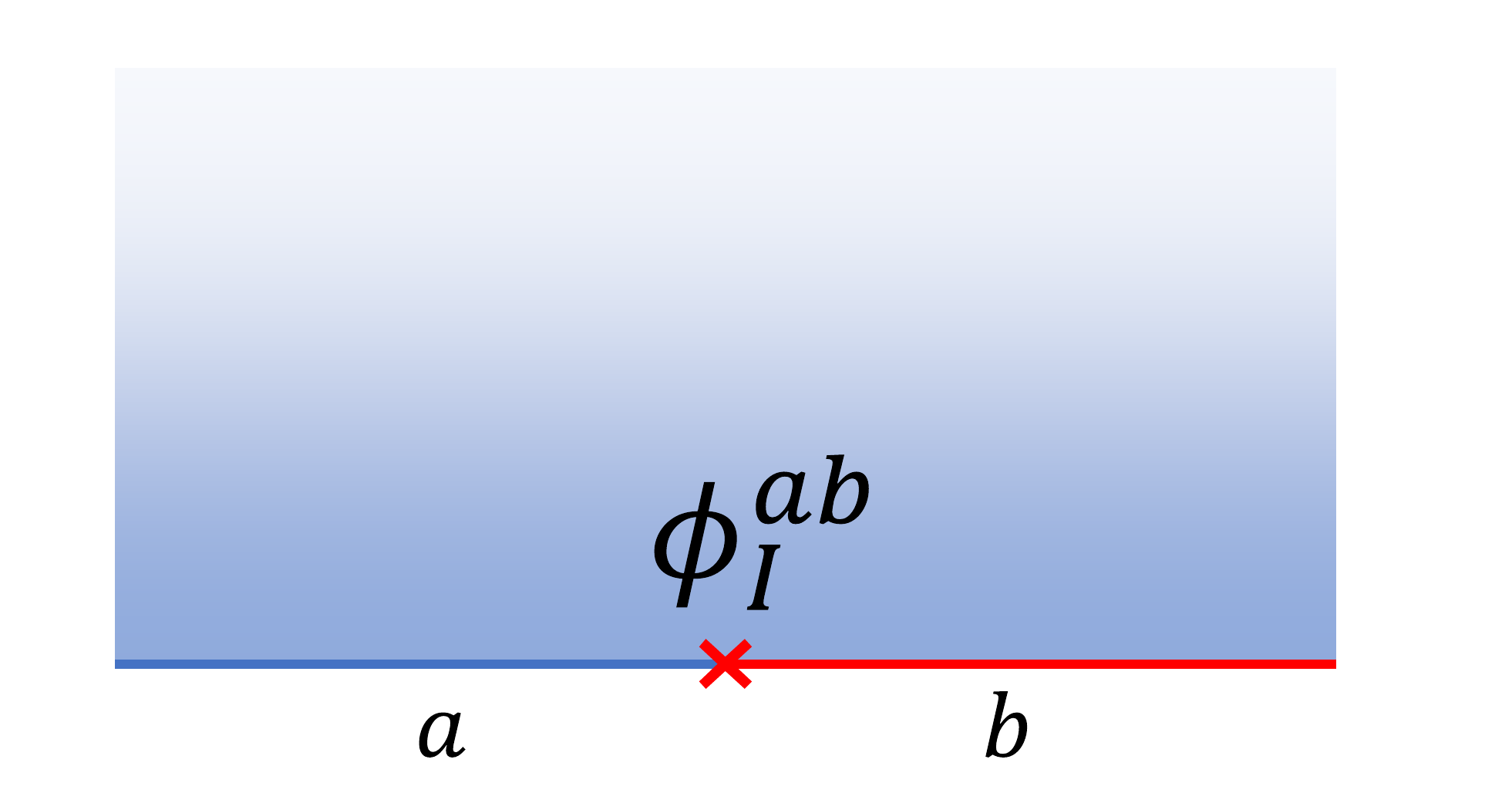}
 \end{center}
 \caption{ The definition of the boundary operator.
The index $I$ labels a primary field with the conformal dimension $h_I$ and the indices $a,b$ denote its boundary conditions, where the boundary condition $a$ is changed to $b$ by the operator $\phi^{ab}_I$.}
 \label{fig:bdyop}
\end{figure}

Let us consider what the new ingredients are, other than the bulk ingredients,
In addition to bulk primary operators $\phi_i$, BCFTs have boundary primary operators $\phi^{ab}_I (x)$, which live only on the boundary (see Figure \ref{fig:bdyop}).
The index $I$ labels a primary field with the conformal dimension $h_I$ and the indices $a,b$ denote its boundary conditions,
where the boundary condition $a$ is changed to $b$ by the operator $\phi^{ab}_I (x)$.
In the following, we denote the bulk primary operators by $\phi$ with lowercase letters $i$ and the boundary primary operators by $\phi$ with capital letter $I$.

The dynamics of a BCFT can be completely determined by the following information,
\begin{quote}
\begin{description}
\item[Disk partition function (i.e., boundary entropy)]\ \\
\begin{equation}\label{eq:Sbdy}
\braket{\mathbb{I}}_a = g^a,
\end{equation}

\item[Boundary-boundary-boundary OPE coefficients]\ \\
\begin{equation}
\phi^{ab}_I (0) \phi^{bc}_J (x) \sim \sum_K C^{abc}_{IJK} x^{h_K-h_I-h_J} \phi^{ac}_K(x)+\cdots,
\end{equation}

\item[Bulk-boundary OPE coefficients]\ \\
\begin{equation}
\phi_i (z) \sim \sum_I (2 \Im z)^{h_I-h_i-\bar{h}_i} C^a_{iI} \phi^{aa}_I(\Re z) + \cdots,
\end{equation}

\item[Bulk-bulk-bulk OPE coefficients]\ \\
\begin{equation}
\phi_i (0) \phi_j (z) \sim \sum_k C_{ijk} z^{h_k-h_i-h_j}  \bar{z}^{\bar{h}_k-\bar{h}_i-\bar{h}_j} \phi_k(z)+\cdots.
\end{equation}

\end{description}
\end{quote}
Note that the normalization $g^a$ in (\ref{eq:Sbdy}) cannot be freely chosen.
Let us consider an RCFT to see this explicitly.
From eq.(\ref{eq:T=T}),
one can show that possible boundary conditions can be expressed in terms of the Ishibashi states,
\begin{equation}
\ket{j}\rangle \equiv \sum_{N} \ket{j;N} \otimes   U \overline{\ket{j;N}},
\end{equation}
where $\ket{j;N}$ is a state in the Verma module $j$ labeled by $N$, and $U$ is an anti-unitary operator.
Moreover, the open-closed duality constraints the possible linear combinations of the Ishibashi states as
\begin{equation}
\ket{\ti{i}} = \sum_j \fr{S_{ij}}{\sqrt{S_{0j}}} \ket{j}\rangle,
\end{equation}
known as the Cardy state. From this expression , we can immediately find that the disk partition function is given by
\begin{equation}
\braket{\mathbb{I}}_a = g^a = \fr{S_{a0}}{\sqrt{S_{00}}}.
\end{equation}
In a similar way, we can also fix the bulk-boundary OPE coefficients with the vacuum boundary operator as \cite{Cardy1991}
\begin{equation}
C^a_{i \mathbb{I}} = \fr{S_{ai} \sqrt{S_{00}}}{S_{a0}\sqrt{S_{0i}}}
\end{equation}

Our interest is the asymptotic formula for the boundary ingredients where the boundary condition does not matter since the universal feature is only fixed by the Virasoro symmetry (i.e., the properties of Virasoro blocks with no boundary).
Therefore, we only focus on a special case where all boundary conditions are the same (i.e., identical boundaries), for simplicity.
In the following, we also omit the index $a$ that specifies the boundary condition to avoid cumbersome expressions.

Let us consider a natural normalization for a two point function,
\begin{equation}
\braket{\phi_I (0) \phi_J (x) } =  \delta_{IJ} x^{-2h_I} \a_I.
\end{equation}
Naively, we could set $\a_I$ freely by suitable normalization of the boundary primary fields $\phi_I$.
However, we have the non-trivial normalization for the vacuum amplitude, $g$,
which leads to a natural normalization $\a_I = g$.
Note that, for example in \cite{Cardy1991}, the normalization is set to be $\a_I = 1$.
In this case, one can find that $C_{\mathbb{I} \mathbb{I}} \neq 1$, which seems to be unnatural.
For this reason, we set $\a_I = g$ in the following.
With this normalization, we have
\begin{equation}
    C_{\mathbb{I} \mathbb{I}}=1,
\end{equation}
\begin{equation}
    C_{\mathbb{I} I I}=1,
\end{equation}

\newsavebox{\boxdisk}
\sbox{\boxdisk}{\includegraphics[width=60pt]{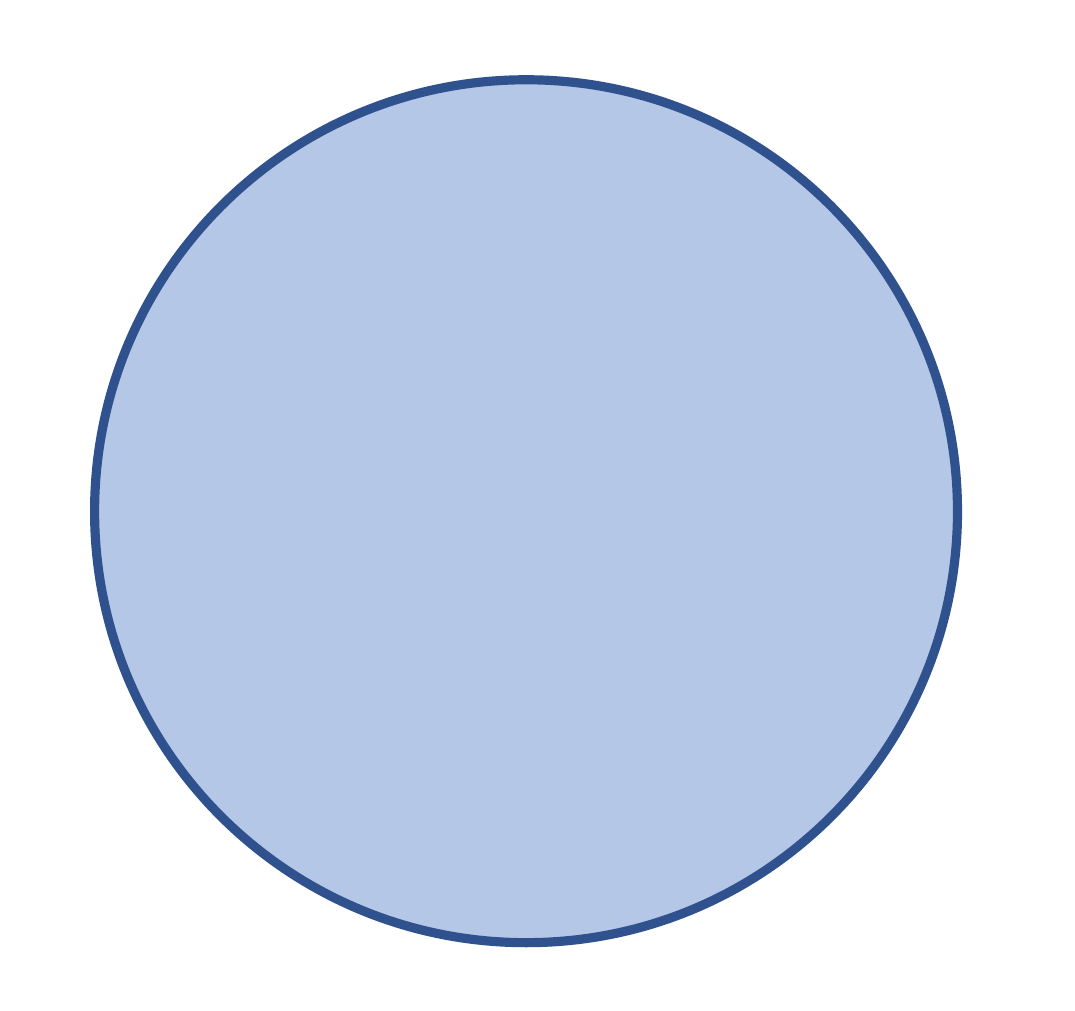}}
\newlength{\boxdiskw}
\settowidth{\boxdiskw}{\usebox{\boxdisk}} 

\newsavebox{\boxa}
\sbox{\boxa}{\includegraphics[width=60pt]{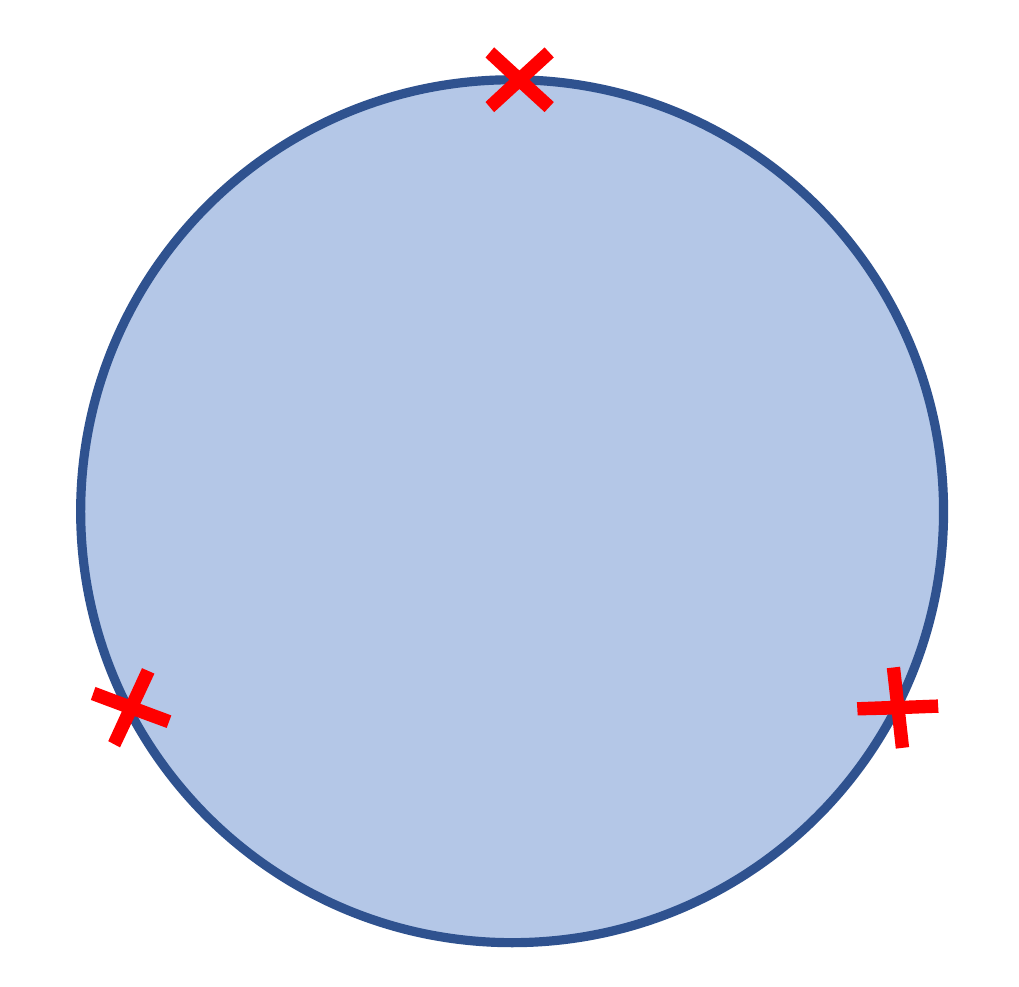}}
\newlength{\boxwa}
\settowidth{\boxwa}{\usebox{\boxa}} 

\newsavebox{\boxb}
\sbox{\boxb}{\includegraphics[width=60pt]{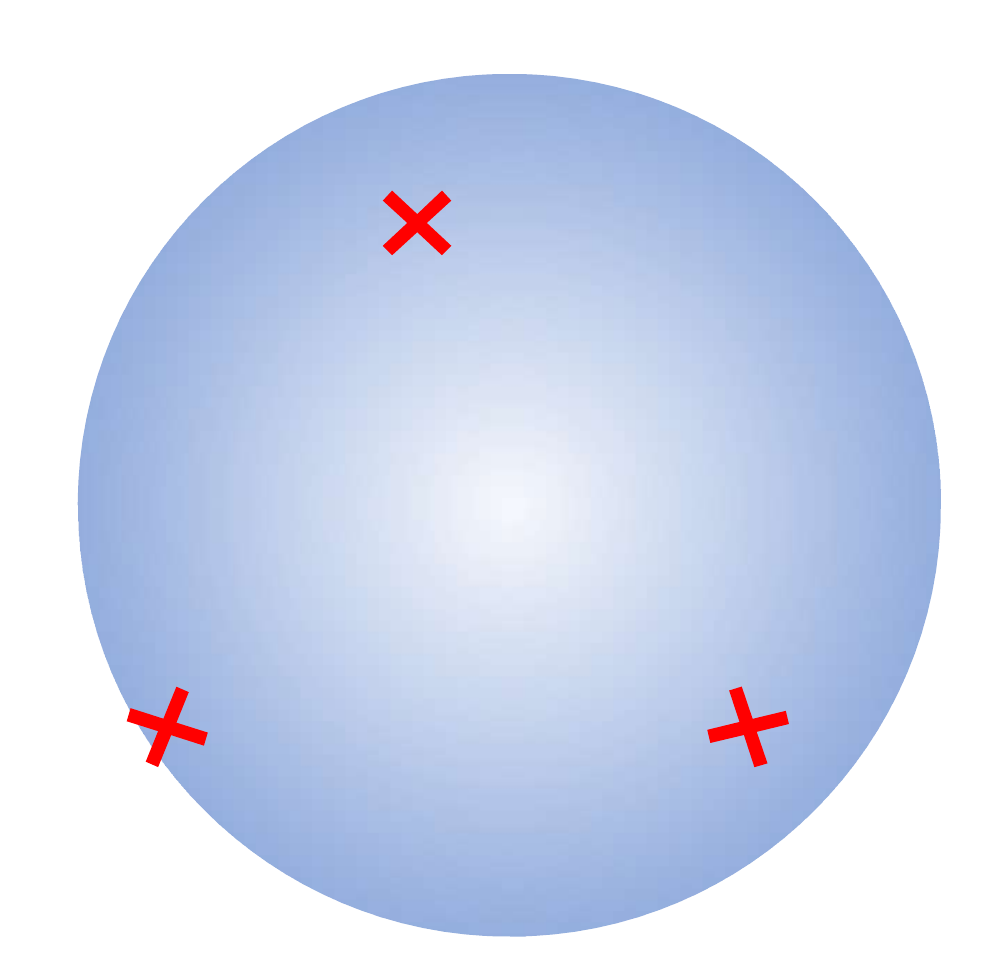}}
\newlength{\boxwb}
\settowidth{\boxwb}{\usebox{\boxb}} 

\newsavebox{\boxc}
\sbox{\boxc}{\includegraphics[width=60pt]{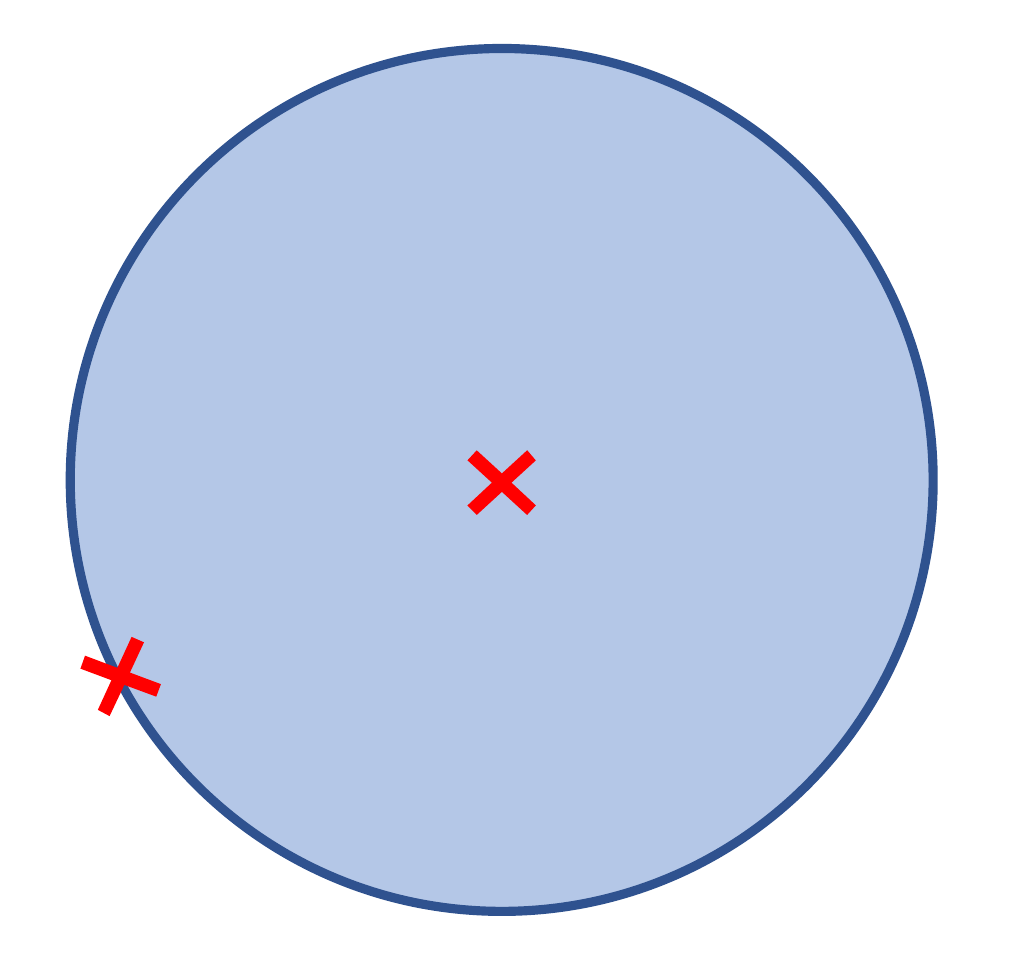}}
\newlength{\boxwc}
\settowidth{\boxwc}{\usebox{\boxc}}

Now we have four simplified components in a BCFT,
\begin{quote}
\begin{description}

\item[Disk partition function(i.e., boundary entropy)]\ \\
\begin{equation}
\parbox{\boxdiskw}{\usebox{\boxdisk}}  \equiv g.
\end{equation}

\item[Boundary-boundary-boundary OPE coefficients]\ \\
\begin{equation}
 \parbox{\boxwa}{\usebox{\boxa}} \equiv C_{IJK}.
\end{equation}

\item[Bulk-boundary OPE coefficients]\ \\
\begin{equation}
\parbox{\boxwc}{\usebox{\boxc}} \equiv C_{iI} .
\end{equation}

\item[Bulk-bulk-bulk OPE coefficients]\ \\
\begin{equation}\label{eq:bulk3}
 \parbox{\boxwb}{\usebox{\boxb}} \equiv C_{ijk}.
\end{equation}

\end{description}
\end{quote}
We sketch a sphere without boundaries by a gradient color disk (\ref{eq:bulk3}).

With this information, we can completely evaluate any correlation function.
For example, let us consider a two point function on a disk.
\begin{figure}[H]
 \begin{center}
  \includegraphics[width=3.0cm,clip]{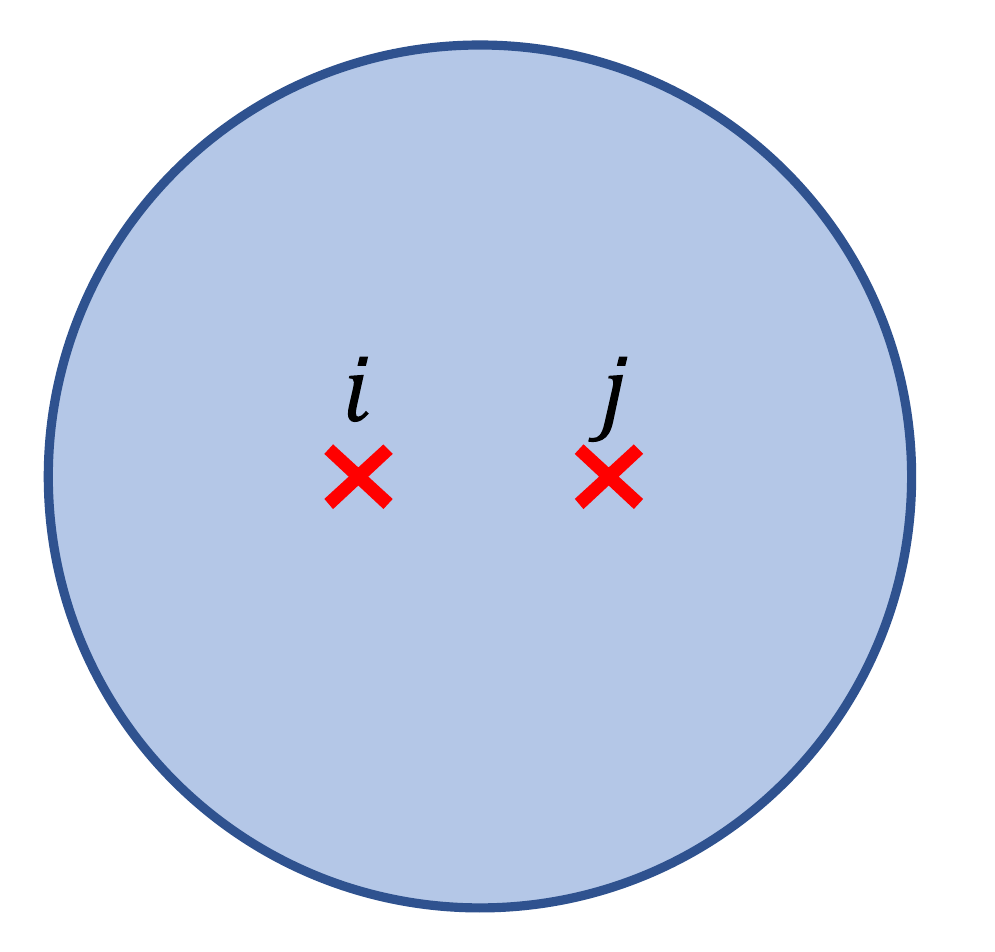}
 \end{center}
\end{figure}
\noindent
This can be decomposed into two parts by inserting the bulk identity $\sum_p \ket{p}\bra{p}$ along the red line in the following picture,
\begin{figure}[H]
 \begin{center}
  \includegraphics[width=3.0cm,clip]{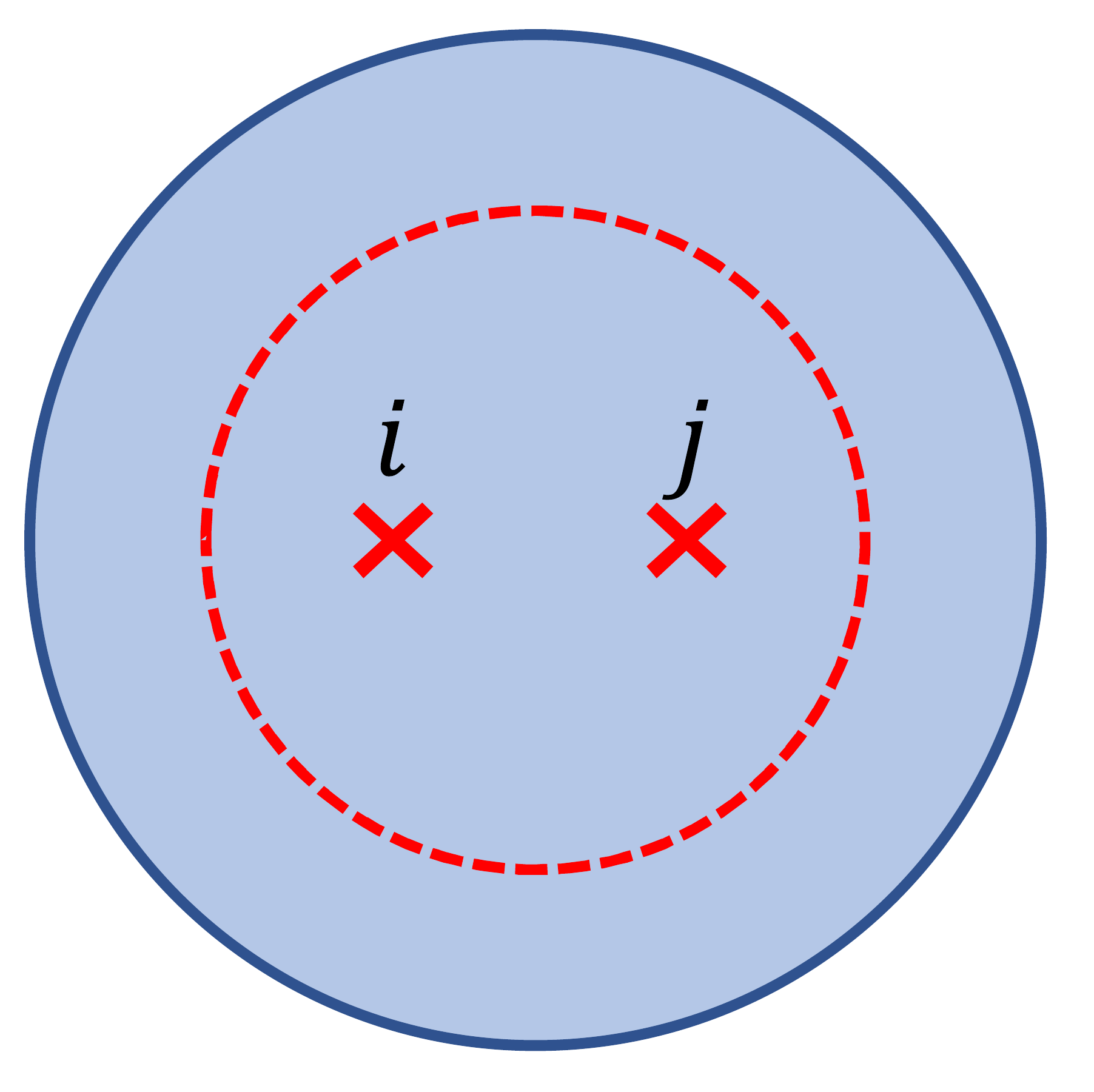}
 \end{center}
\end{figure}

\newsavebox{\boxd}
\sbox{\boxd}{\includegraphics[width=150pt]{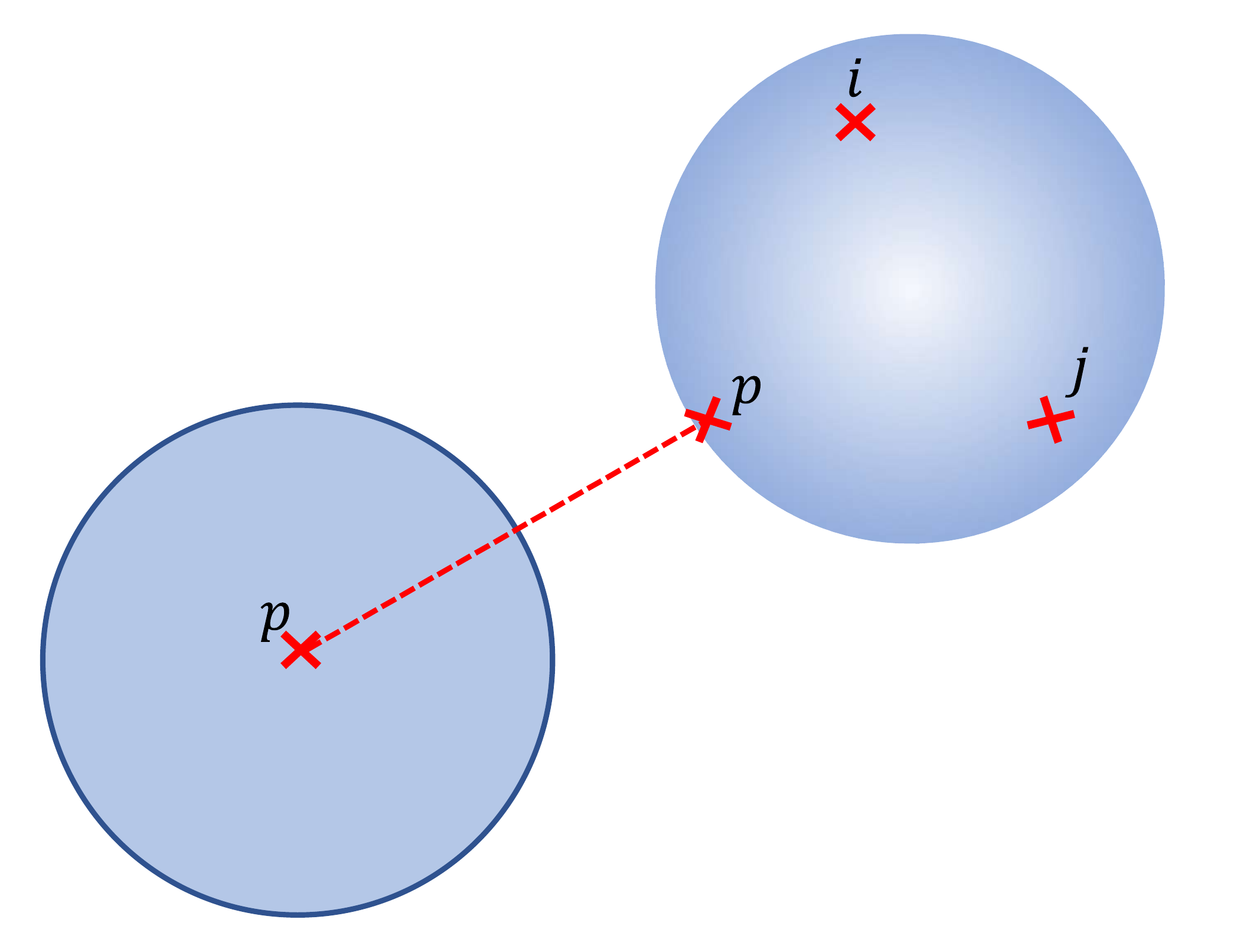}}
\newlength{\boxwd}
\settowidth{\boxwd}{\usebox{\boxd}} 

\newsavebox{\boxijji}
\sbox{\boxijji}{\includegraphics[width=120pt]{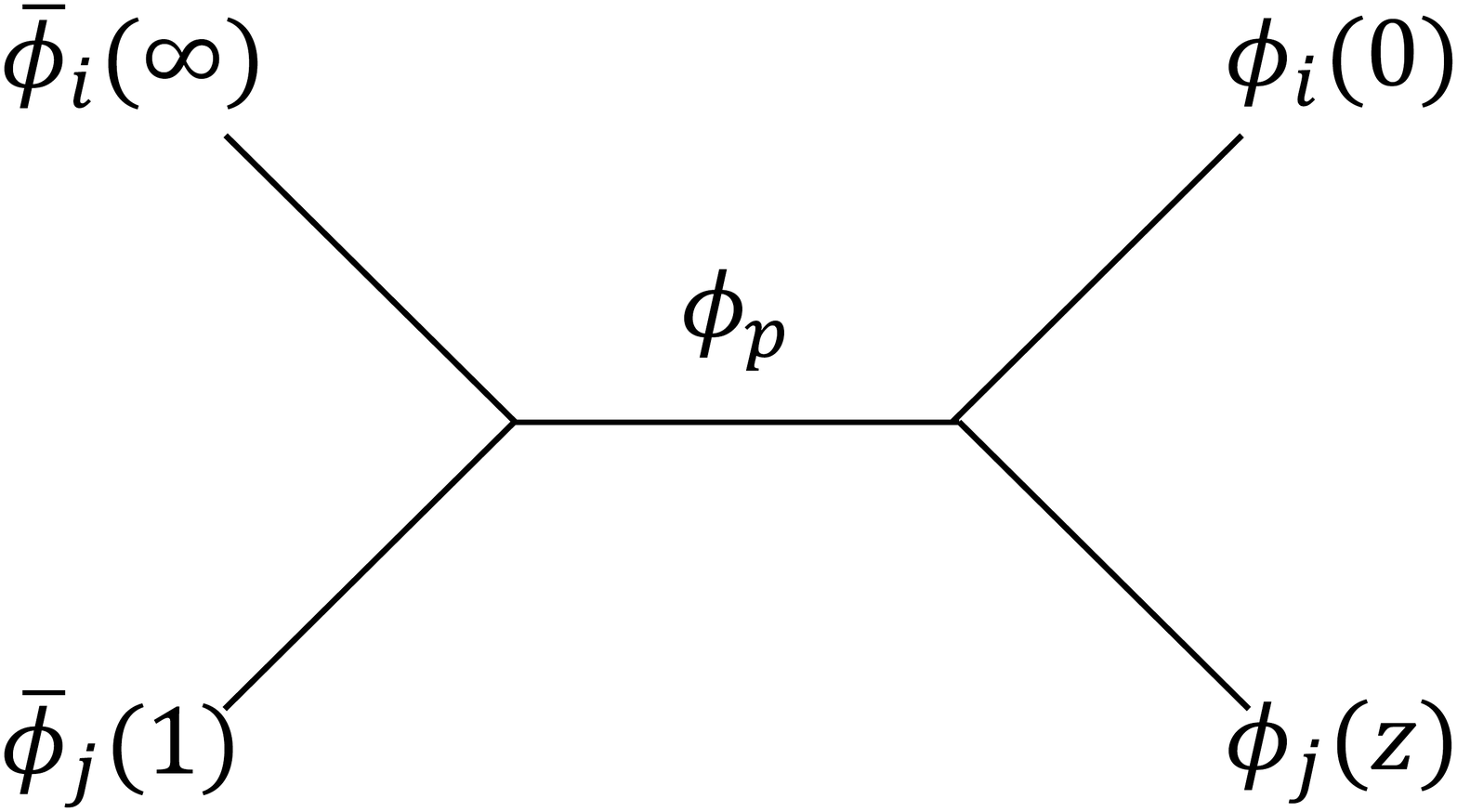}}
\newlength{\boxwijji}
\settowidth{\boxwijji}{\usebox{\boxijji}}

\newsavebox{\boxexa}
\sbox{\boxexa}{\includegraphics[width=200pt]{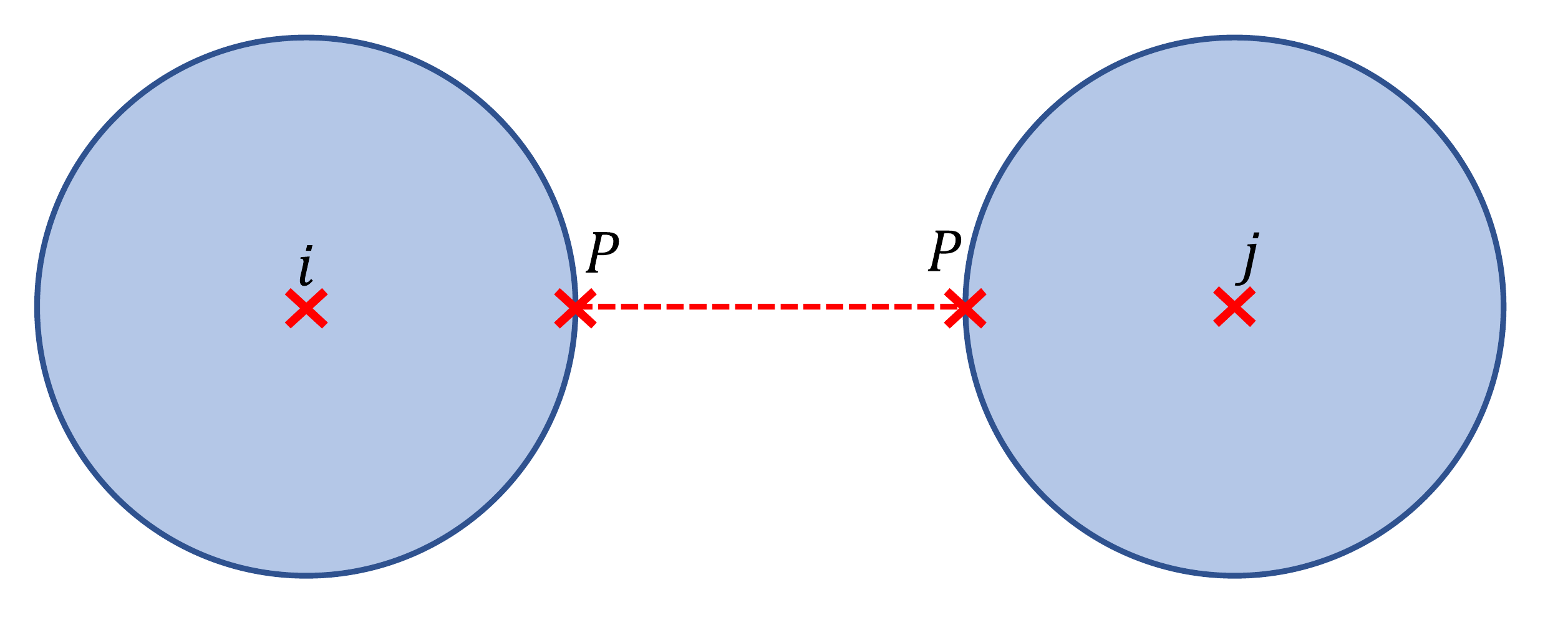}}
\newlength{\boxexaw}
\settowidth{\boxexaw}{\usebox{\boxexa}}

\newsavebox{\boxexb}
\sbox{\boxexb}{\includegraphics[width=120pt]{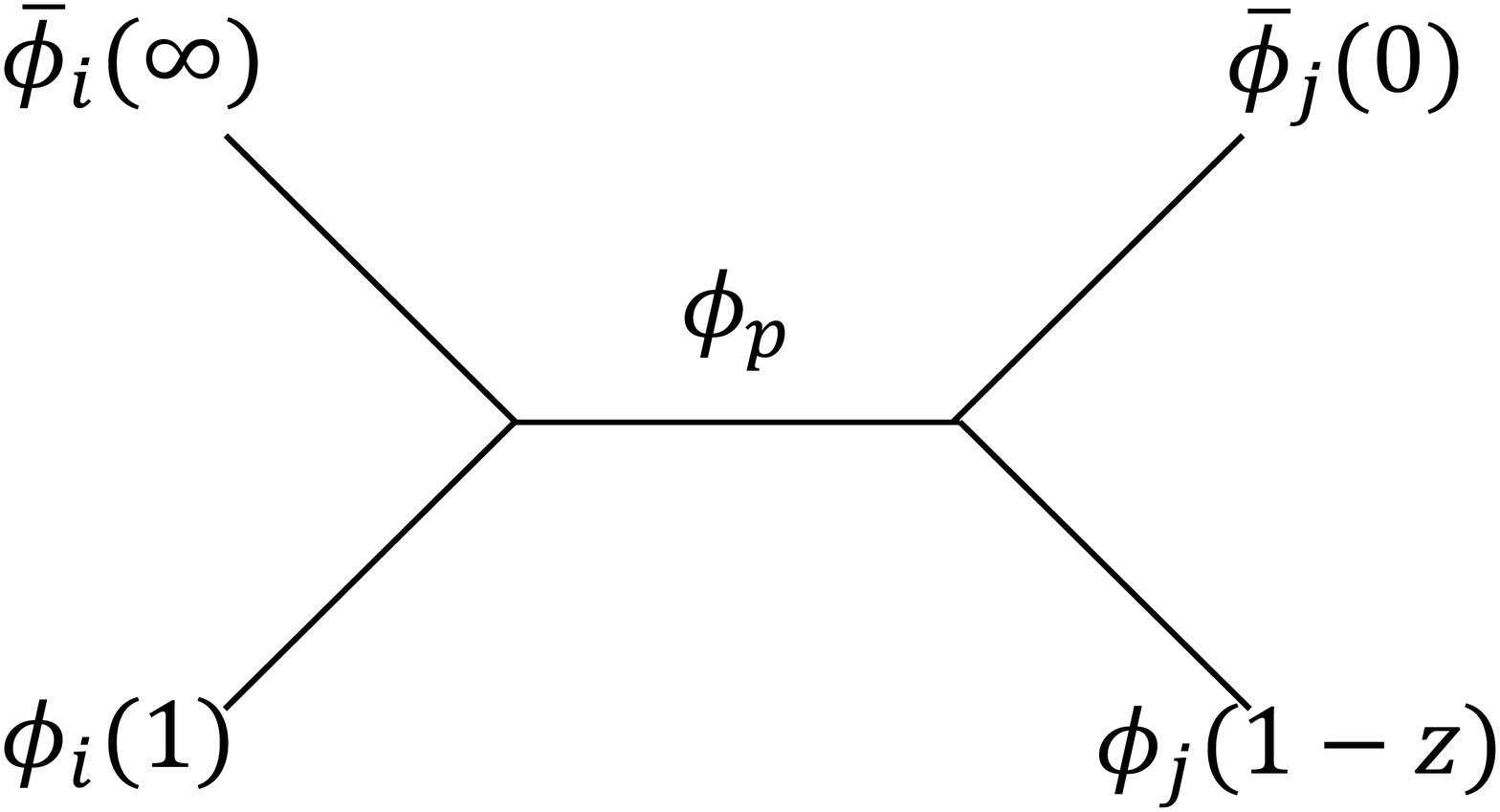}}
\newlength{\boxexbw}
\settowidth{\boxexbw}{\usebox{\boxexb}}

\noindent
Thus, we obtain 
\begin{equation}
\parbox{\boxwd}{\usebox{\boxd}} = g\sum_p C_{p\mathbb{I}}C_{ijp} \parbox{\boxwijji}{\usebox{\boxijji}},
\end{equation}
where the diagram is the same as the standard Virasoro conformal block.
The prefactor $g$ comes from the bulk-boundary OPE,
\begin{equation}
\braket{\phi_p}_{\text{disk}} =  \sum_Q C_{pQ} \braket{\phi_Q} = g \  C_{p \mathbb{I}}.
\end{equation}
We can obtain another channel expansion by inserting the boundary identity $\fr{1}{g}\sum_P \ket{P}\bra{P}$ along the red line in the following picture,
where the prefactor $\fr{1}{g}$ comes from the normalization of the boundary two-point function $\braket{P|P}=g$.
\begin{figure}[H]
 \begin{center}
  \includegraphics[width=3.0cm,clip]{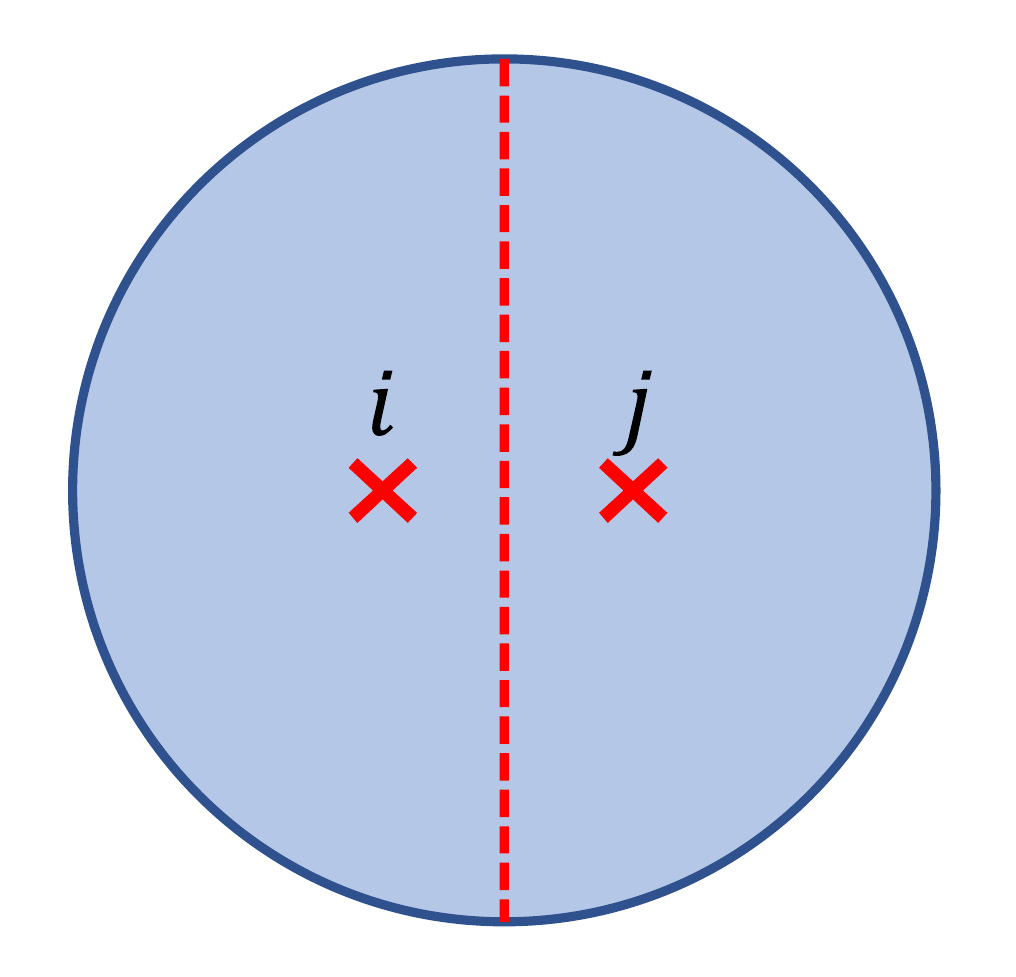}
 \end{center}
\end{figure}
\noindent
Thus, we obtain
\begin{equation}
\parbox{\boxexaw}{\usebox{\boxexa}} =
g
\sum_P
C_{iP}C_{jP}
\parbox{\boxexbw}{\usebox{\boxexb}}.
\end{equation}
What we would like to emphasize here is that the sewing and cutting procedure in the full plane  \cite{Sonoda1988, Moore1989a, Moore1989, Moore1988} also work in the upper half-plane.
More details can be found in the reference \cite{Lewellen1992}.
The point is that any correlation function in a BCFT can also be decomposed in terms of the standard Virasoro block.
Therefore, in a similar way as the procedure shown in Section \ref{sec:review},
the information about the boundary ingredients at asymptotically large dimensions can be fixed by the fusion matrix.
We will show them in the next section.

\section{Bootstrap in BCFT}\label{sec:bootstrap}
In this section, we will solve various bootstrap equations in a BCFT.
In the following, we assume our CFT to be a unitary compact CFT with  $c>1$.

\subsection{Zero-point on a cylinder: spectrum of boundary primaries}

\newsavebox{\boxaa}
\sbox{\boxaa}{\includegraphics[width=70pt]{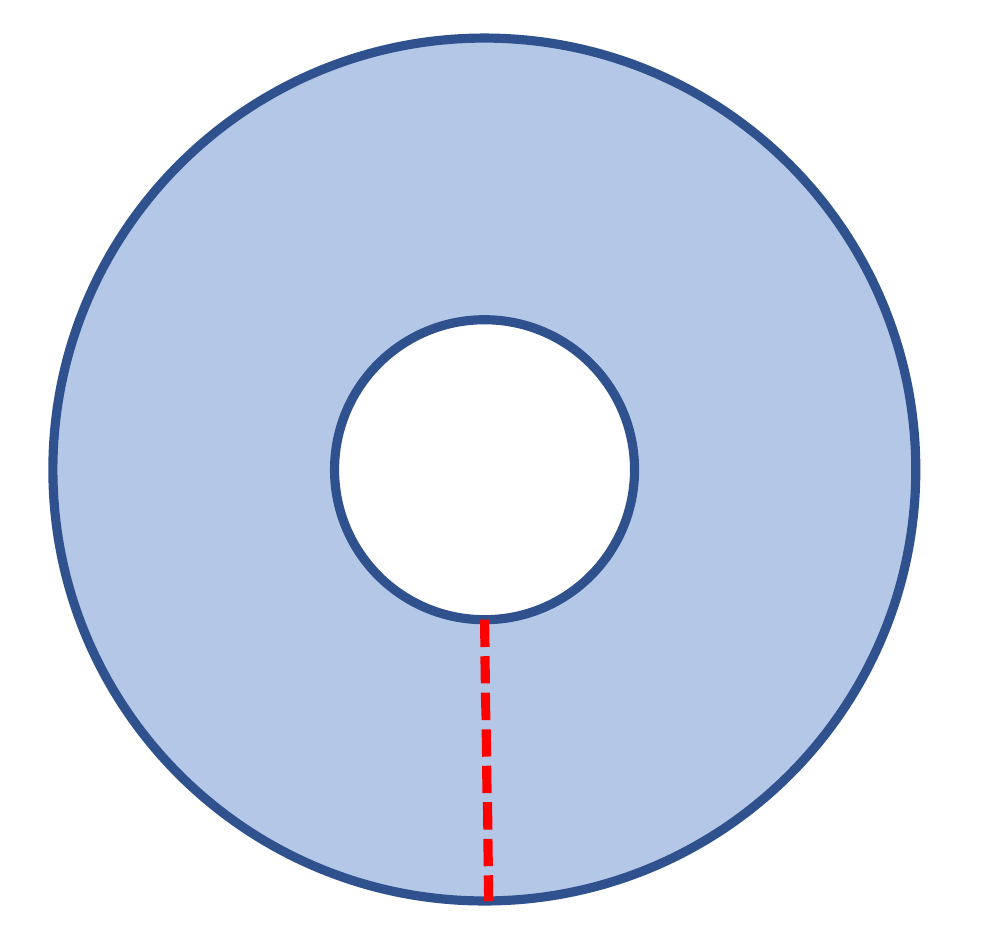}}
\newlength{\boxwaa}
\settowidth{\boxwaa}{\usebox{\boxaa}} 

\newsavebox{\boxab}
\sbox{\boxab}{\includegraphics[width=70pt]{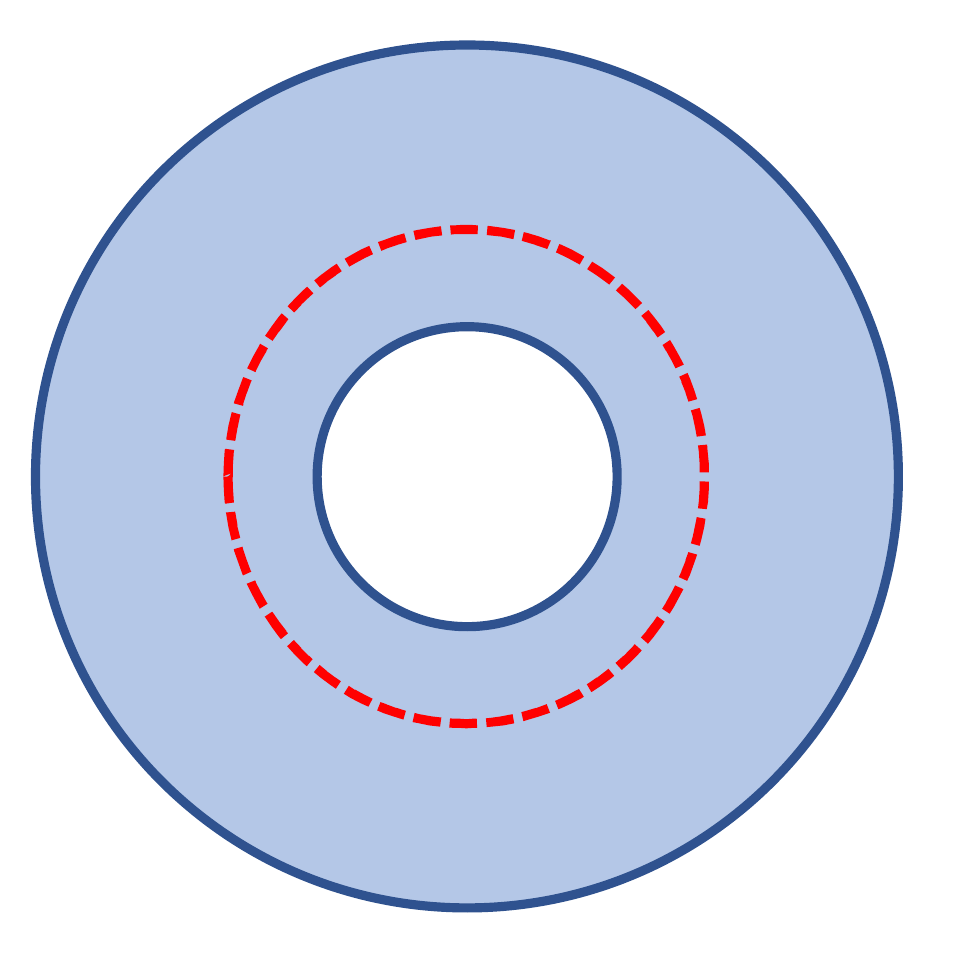}}
\newlength{\boxwab}
\settowidth{\boxwab}{\usebox{\boxab}}

\newsavebox{\boxtoruss}
\sbox{\boxtoruss}{\includegraphics[width=60pt]{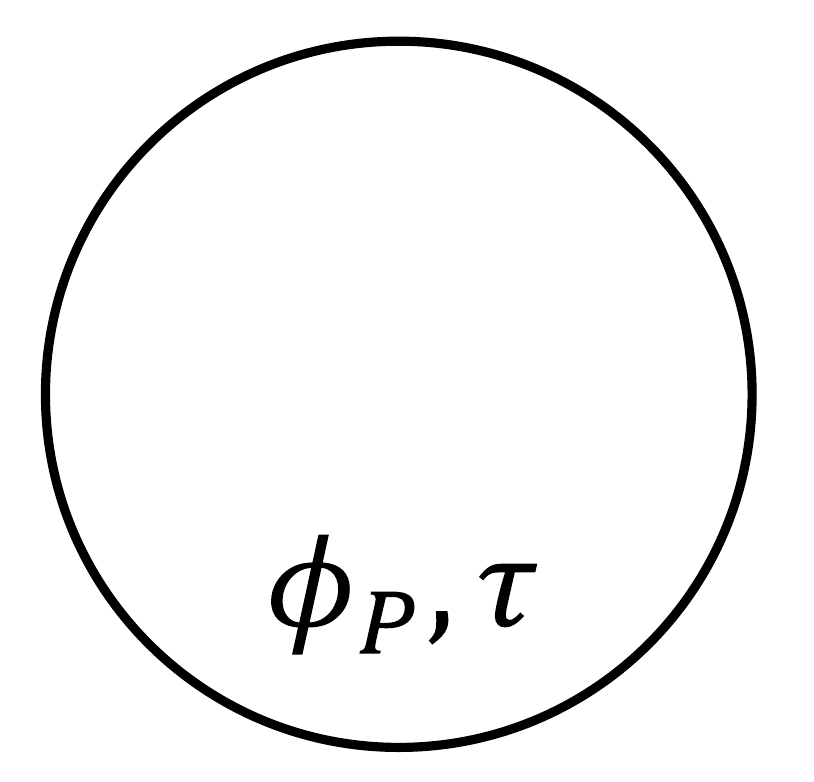}}
\newlength{\boxtorussw}
\settowidth{\boxtorussw}{\usebox{\boxtoruss}} 

\newsavebox{\boxtorust}
\sbox{\boxtorust}{\includegraphics[width=60pt]{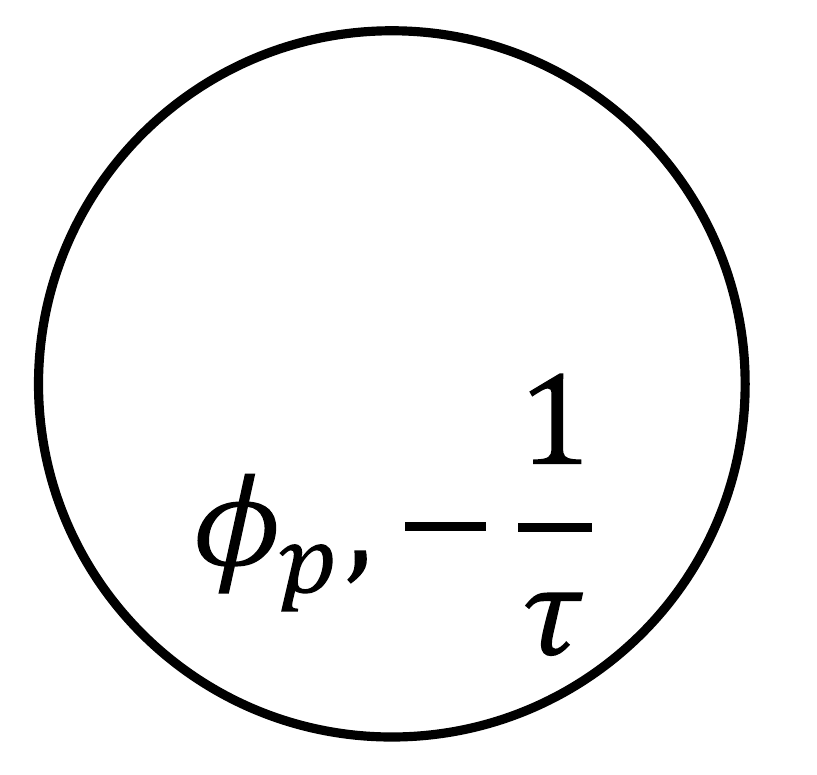}}
\newlength{\boxtorustw}
\settowidth{\boxtorustw}{\usebox{\boxtorust}} 

\newsavebox{\boxtorusI}
\sbox{\boxtorusI}{\includegraphics[width=60pt]{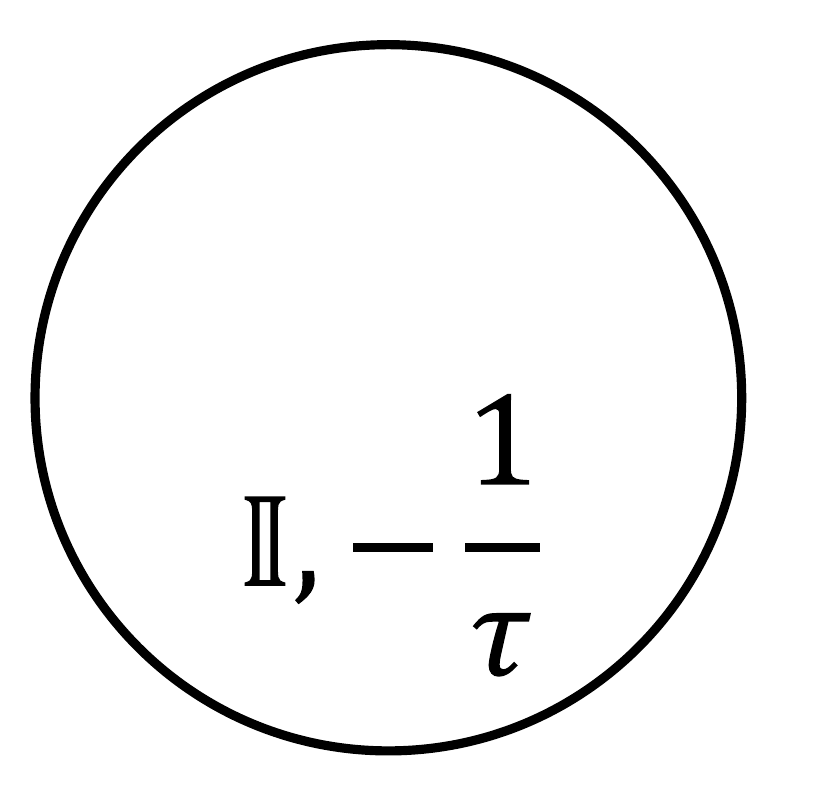}}
\newlength{\boxtorusIw}
\settowidth{\boxtorusIw}{\usebox{\boxtorusI}} 

For a zero-point function on a cylinder, we have the following two choices of how to cut,
\begin{equation}
\parbox{\boxwaa}{\usebox{\boxaa}} = \parbox{\boxwab}{\usebox{\boxab}}.
\end{equation}
The corresponding bootstrap equation is given by
\begin{equation}\label{eq:cylinder}
\begin{aligned}
\int \dd \a_P \ 
\rho (\a_P)
\parbox{\boxtorussw}{\usebox{\boxtoruss}}
=
g^2
\int \dd \a_p \ 
\rho (\a_p)
\overline{C_{p\mathbb{I}}  C_{p\mathbb{I}} }
\parbox{\boxtorustw}{\usebox{\boxtorust}},
\end{aligned}
\end{equation}
where the diagram represents the Virasoro character $\chi_p(\tau)$ (which depends only on the conformal dimension $h_p$ and independent of whether the field lives on the boundary or the bulk),
and $\rho (\a_P)$ represents the density of the ``boundary'' primary states, which is different from the density of the ``bulk'' primary states  $\rho (\a_p) \equiv \rho(\a_p,\a_p) $ with lowercase letters.
The right hand side can be approximated by the vacuum block if we take the $\tau \to i0$ limit.
\footnote{Here we assume that the boundary satisfies the condition $g \neq 0$.
In general, this condition would be satisfied since D-branes usually have non-zero tension. }
As a result, we obtain
\begin{equation}\label{eq:torusbootstrap}
\begin{aligned}
\int \dd \a_P \ 
\rho (\a_P)
\parbox{\boxtorussw}{\usebox{\boxtoruss}}
&\simeq
g^2
\parbox{\boxtorusIw}{\usebox{\boxtorusI}}\\
&=
g^2
\int \dd \a_P \ 
S_{0P}
\parbox{\boxtorussw}{\usebox{\boxtoruss}}
.
\end{aligned}
\end{equation}
In the second equation, we use the modular-S transformation (\ref{eq:Smodular}).
Thus, the universal formula for the density of the boundary primary states is given by
\begin{equation}\label{eq:open}
\rho(\a_P) \simeq 
g^2 S_{0P}, \ \ \ \ \ \ h_P \to \infty.
\end{equation}
This is similar to the Cardy formula for the spectrum of bulk primaries.
Indeed, the same result was derived in \cite{Hikida2018} by the traditional approach using the inversion Laplace transformation.
Note that although we retain the non-universal factor $g$,
this factor is subleading in the large $h_p$ asymptotics.
This is just a constant in the conformal dimension $h_p$.

One can take another limit $\tau \to i \infty$, where the left hand side can be approximated by the vacuum block.
In a similar way as (\ref{eq:torusbootstrap}), we obtain
\begin{equation}
\begin{aligned}
g^2
\int \dd \a_p \ 
\rho (\a_p)
\overline{C_{p\mathbb{I}} C_{p\mathbb{I}} }
\parbox{\boxtorustw}{\usebox{\boxtorust}}
&\simeq
\int \dd \a_p \ 
S_{0p}
\parbox{\boxtorustw}{\usebox{\boxtorust}}
.
\end{aligned}
\end{equation}
This equation implies
\begin{equation}
g^2 \rho(\a_p) \overline{C_{p\mathbb{I}} C_{p\mathbb{I}} } = S_{0p}, \ \ \ \ \ \ h_p \to \infty.
\end{equation}
As show  in (\ref{eq:Cardy}),
the density of the bulk primary states is given by
\begin{equation}
\rho(\a_p) \simeq S_{0p} S_{0p}  \ \ \ \ \ \ h_p \to \infty,
\end{equation}
therefore, we can simplify this result as
\begin{equation}
\overline{C_{p\mathbb{I}} C_{p\mathbb{I}} } \simeq g^{-2} S_{0p}^{-1}, \ \ \ \ \ \ h_p \to \infty.
\end{equation}

\subsection{Bulk two-point on a disk: bulk-boundary OPE coefficients}
\newsavebox{\boxba}
\sbox{\boxba}{\includegraphics[width=70pt]{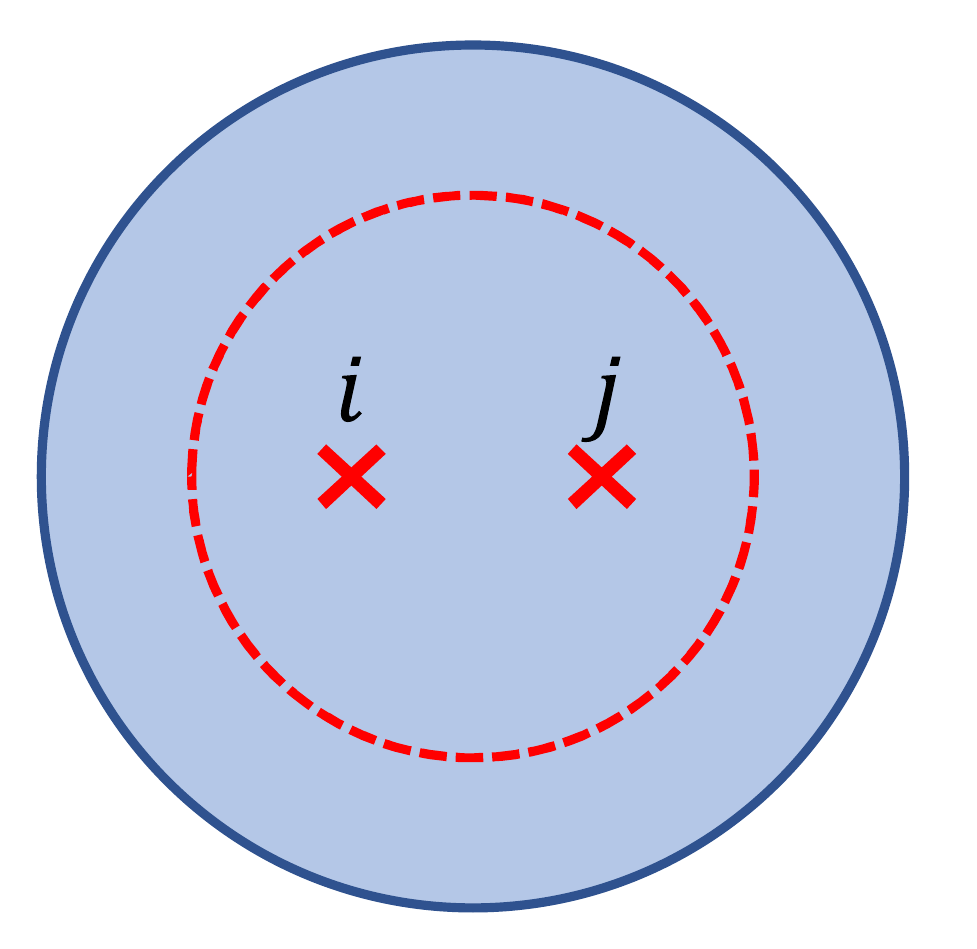}}
\newlength{\boxbaw}
\settowidth{\boxbaw}{\usebox{\boxba}} 

\newsavebox{\boxbb}
\sbox{\boxbb}{\includegraphics[width=70pt]{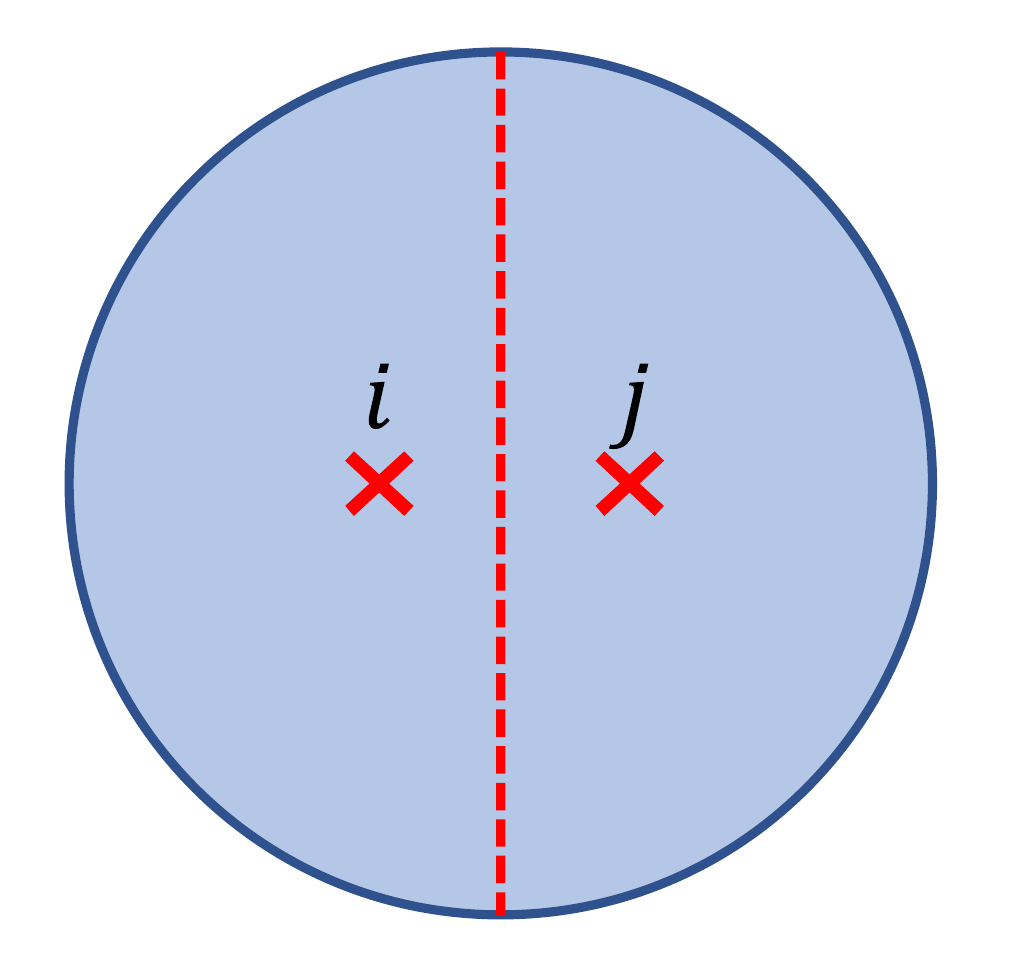}}
\newlength{\boxbbw}
\settowidth{\boxbbw}{\usebox{\boxbb}} 

\newsavebox{\boxtwoa}
\sbox{\boxtwoa}{\includegraphics[width=100pt]{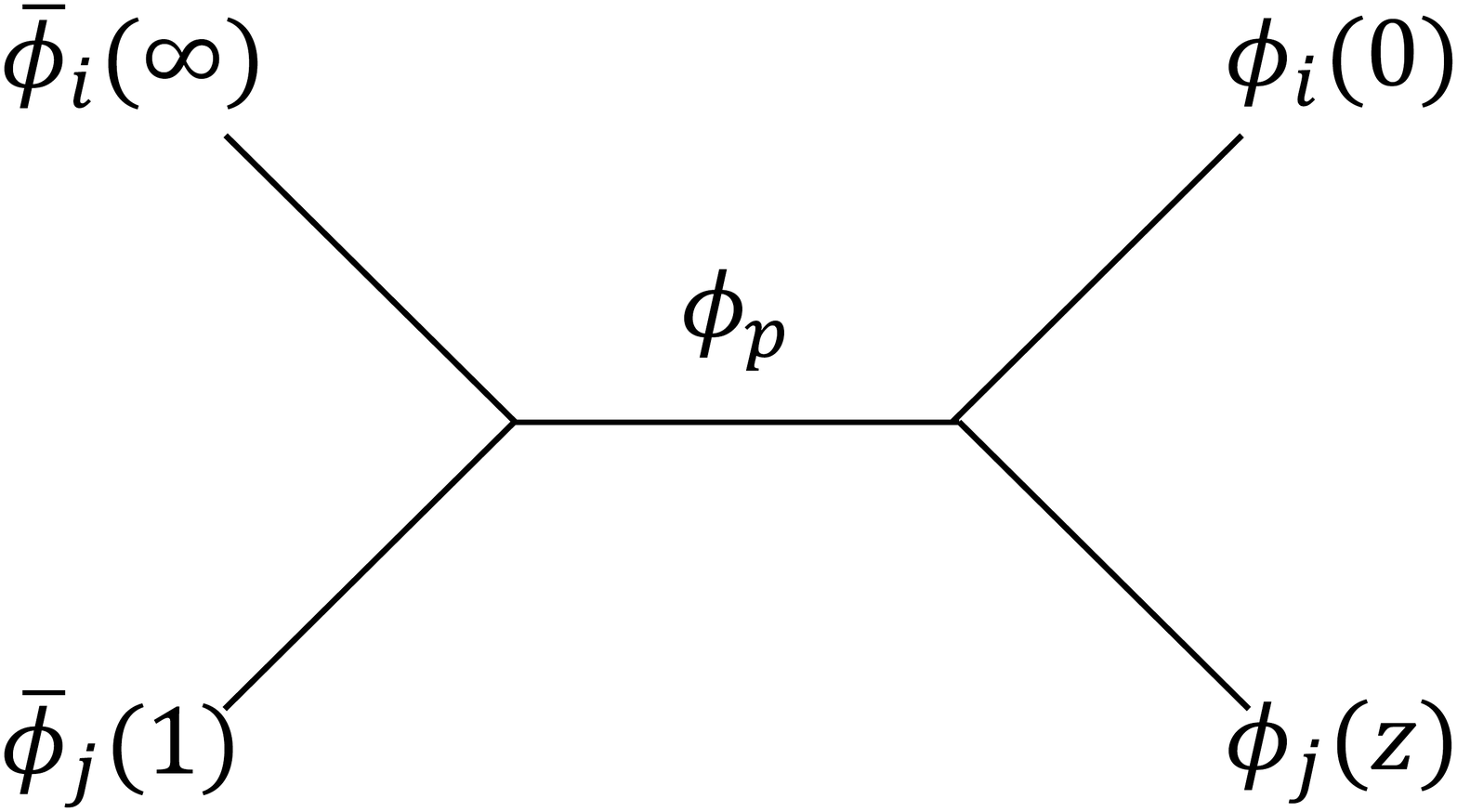}}
\newlength{\boxtwoaw}
\settowidth{\boxtwoaw}{\usebox{\boxtwoa}} 

\newsavebox{\boxtwob}
\sbox{\boxtwob}{\includegraphics[width=100pt]{two-b.pdf}}
\newlength{\boxtwobw}
\settowidth{\boxtwobw}{\usebox{\boxtwob}} 

\newsavebox{\boxtwoc}
\sbox{\boxtwoc}{\includegraphics[width=100pt]{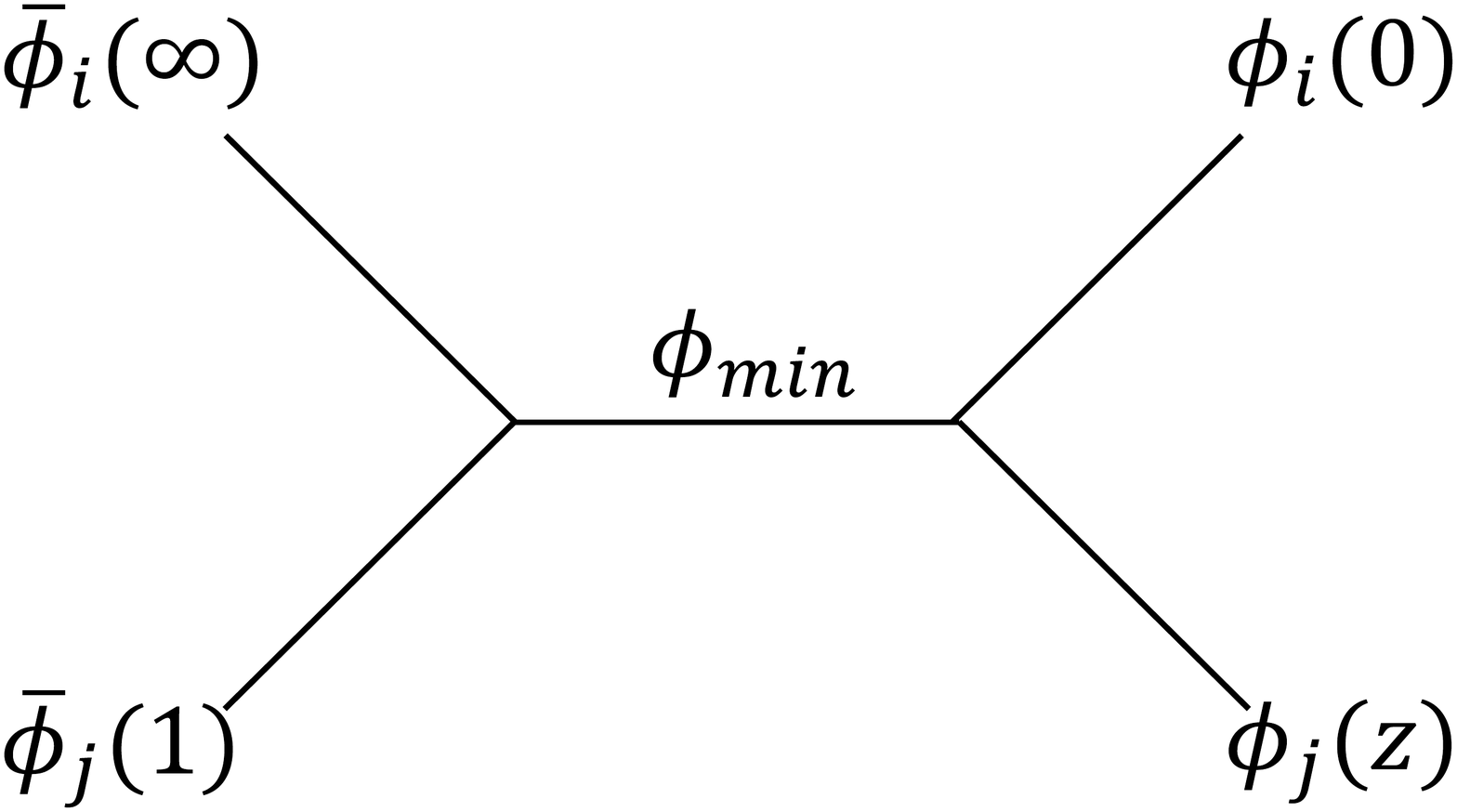}}
\newlength{\boxtwocw}
\settowidth{\boxtwocw}{\usebox{\boxtwoc}} 

For a bulk two-point function on a disk, we have the following two choices of how to cut,
\begin{equation}
\parbox{\boxbaw}{\usebox{\boxba}} = \parbox{\boxbbw}{\usebox{\boxbb}}.
\end{equation}
In the following, we assume $h_i=\bar{h}_i$ and  $h_j=\bar{h}_j$, which can simplify some phase factors to be one.
The corresponding bootstrap equation is given by
\footnote{
We have a mathematical subtlety involving positivity.
On the left-hand side of the bootstrap equation, the coefficients can be negative.
It implies that the asymptotic formula is more sensitive to the size of the averaging window than other cases with positive coefficients.
Nevertheless, it would be expected that the result is true for a suitable size of the averaging window.}
\begin{equation}
\begin{aligned}
\int \dd \a_p \ 
\rho (\a_p)
\overline{C_{ijp}C_{p\mathbb{I}}}
\parbox{\boxtwoaw}{\usebox{\boxtwoa}}
=
\int \dd \a_P \ 
\rho (\a_P)
\overline{C_{iP}C_{jP}}
\parbox{\boxtwobw}{\usebox{\boxtwob}}.
\end{aligned}
\end{equation}
Let us take the limit $z \to 0$, where the left-hand side can be approximated by a single block with the conformal dimension $h_{p_{min}}$, which is the lowest dimension in the OPE between $\phi_i$ and $\phi_j$.
By making use of the fusion transformation , we obtain the following equation,
\begin{equation}
\begin{aligned}
&\int \dd \a_P \ 
\rho (\a_P)
\overline{C_{iP}C_{jP}}
\parbox{\boxtwobw}{\usebox{\boxtwob}}
\simeq
\rho (\a_{p_{min}})
\overline{C_{ijp_{min}}C_{p_{min}\mathbb{I}}}
\parbox{\boxtwocw}{\usebox{\boxtwoc}}\\
&=
\rho (\a_{p_{min}})
\overline{C_{ijp_{min}}C_{p_{min}\mathbb{I}}}
\int \dd \a_P \ 
{\bold F}_{\a_{p_{min}}, \a_P} 
   \left[
    \begin{array}{cc}
    \a_j   & \a_i  \\
     \a_j  &   \a_i\\
    \end{array}
  \right]
\parbox{\boxtwobw}{\usebox{\boxtwob}}.
\end{aligned}
\end{equation}
From this equation, we obtain the universal formula for the bulk-boundary OPE coefficients as
\begin{equation}
\rho (\a_P)
\overline{C_{iP}C_{jP}}
\simeq
\rho (\a_{p_{min}})
\overline{C_{ijp_{min}}C_{p_{min}\mathbb{I}}}
{\bold F}_{\a_{p_{min}}, \a_P} 
   \left[
    \begin{array}{cc}
    \a_j   & \a_i  \\
     \a_j  &   \a_i\\
    \end{array}
  \right],
\ \ \ \ \ \ h_P \to \infty.
\end{equation}
In particular, in the case $i=j$, we have a simple formula,
\begin{equation}
\rho (\a_P)
\overline{C_{iP}C_{iP}}
\simeq
{\bold F}_{0, \a_P} 
   \left[
    \begin{array}{cc}
    \a_i   & \a_i  \\
     \a_i  &   \a_i\\
    \end{array}
  \right],
\ \ \ \ \ \ h_P \to \infty.
\end{equation}
By using (\ref{eq:open}), we have
\begin{equation}
\overline{C_{iP}C_{iP}}
\simeq
g^{-2}
S_{0P}^{-1}
{\bold F}_{0, \a_P} 
   \left[
    \begin{array}{cc}
    \a_i   & \a_i  \\
     \a_i  &   \a_i\\
    \end{array}
  \right]
,
\ \ \ \ \ \ h_P \to \infty.
\end{equation}

Note that the Virasoro block expansion of the bulk two-point function on a disk often appears in the evaluation of the entanglement entropy in BCFTs.
For example, one can provide constraints on the bulk-boundary OPE coefficients by considering the holographic entanglement entropy in BCFTs \cite{Sully2021}.
It would be interesting to consider a combination of the constraints from the bootstrap equation and the holographic entanglement entropy.

\subsection{Boundary four-point on a disk: boundary-boundary-boundary OPE coefficients}
\newsavebox{\boxca}
\sbox{\boxca}{\includegraphics[width=70pt]{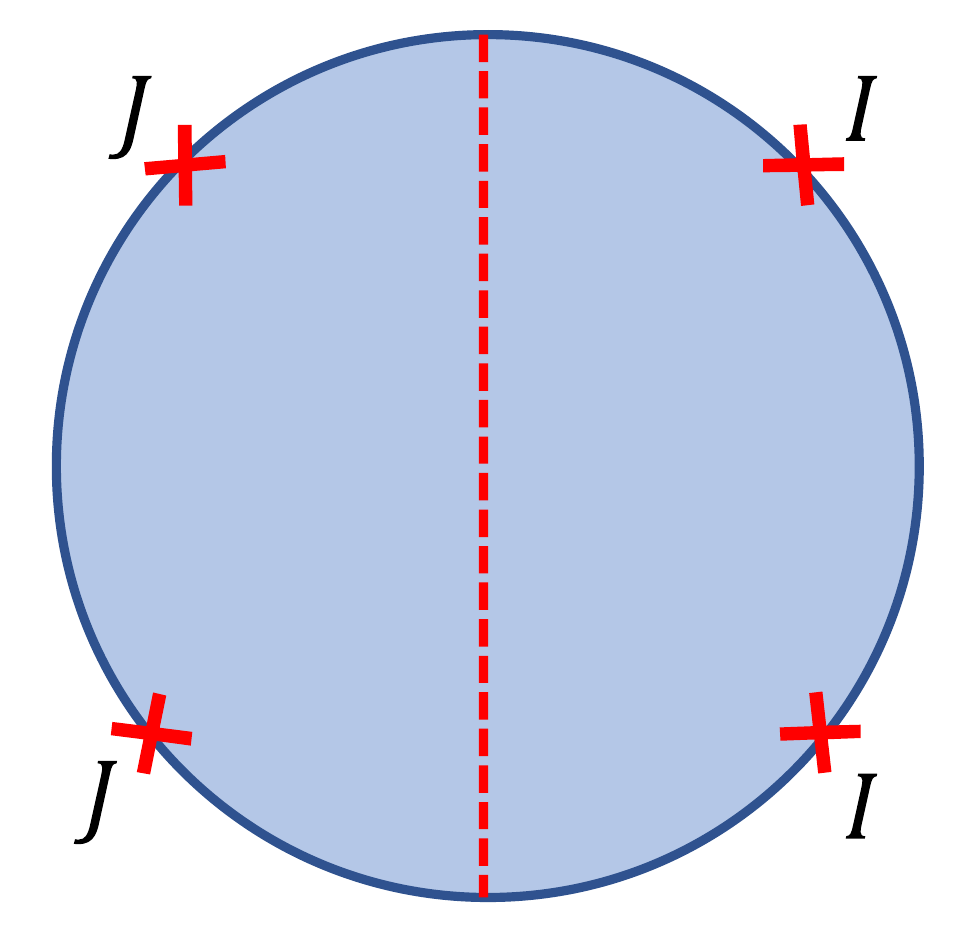}}
\newlength{\boxcaw}
\settowidth{\boxcaw}{\usebox{\boxca}} 

\newsavebox{\boxcb}
\sbox{\boxcb}{\includegraphics[width=70pt]{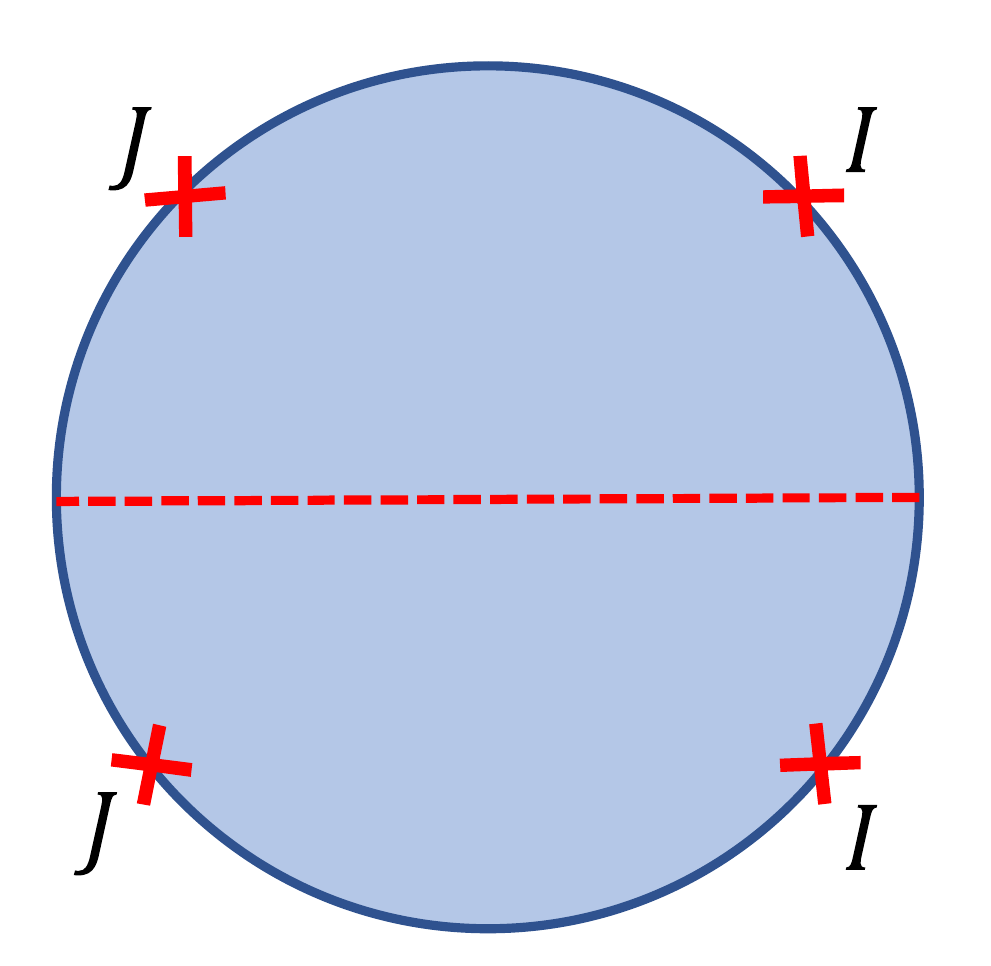}}
\newlength{\boxcbw}
\settowidth{\boxcbw}{\usebox{\boxcb}} 

\newsavebox{\boxfoura}
\sbox{\boxfoura}{\includegraphics[width=100pt]{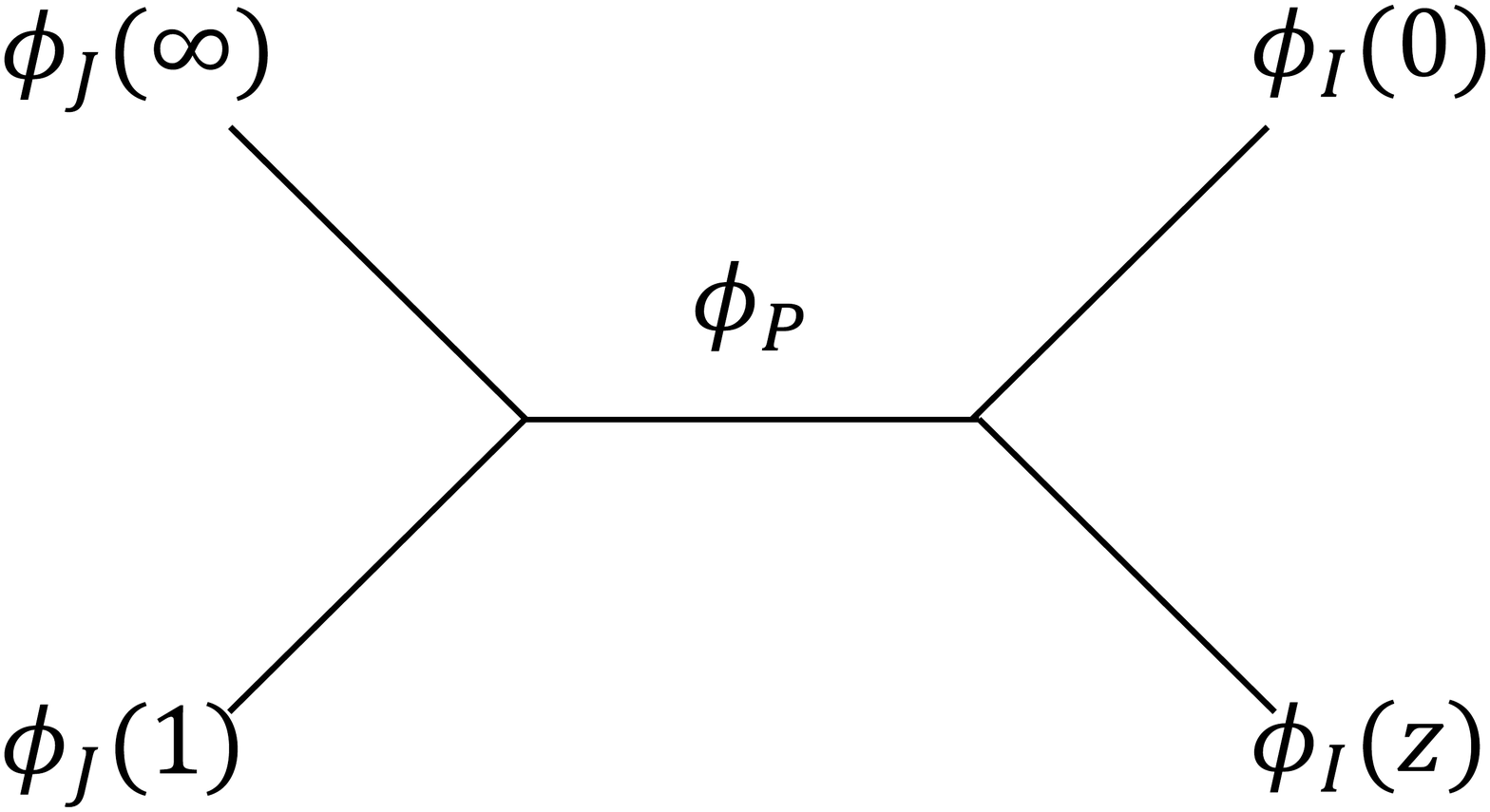}}
\newlength{\boxfouraw}
\settowidth{\boxfouraw}{\usebox{\boxfoura}} 

\newsavebox{\boxfourb}
\sbox{\boxfourb}{\includegraphics[width=100pt]{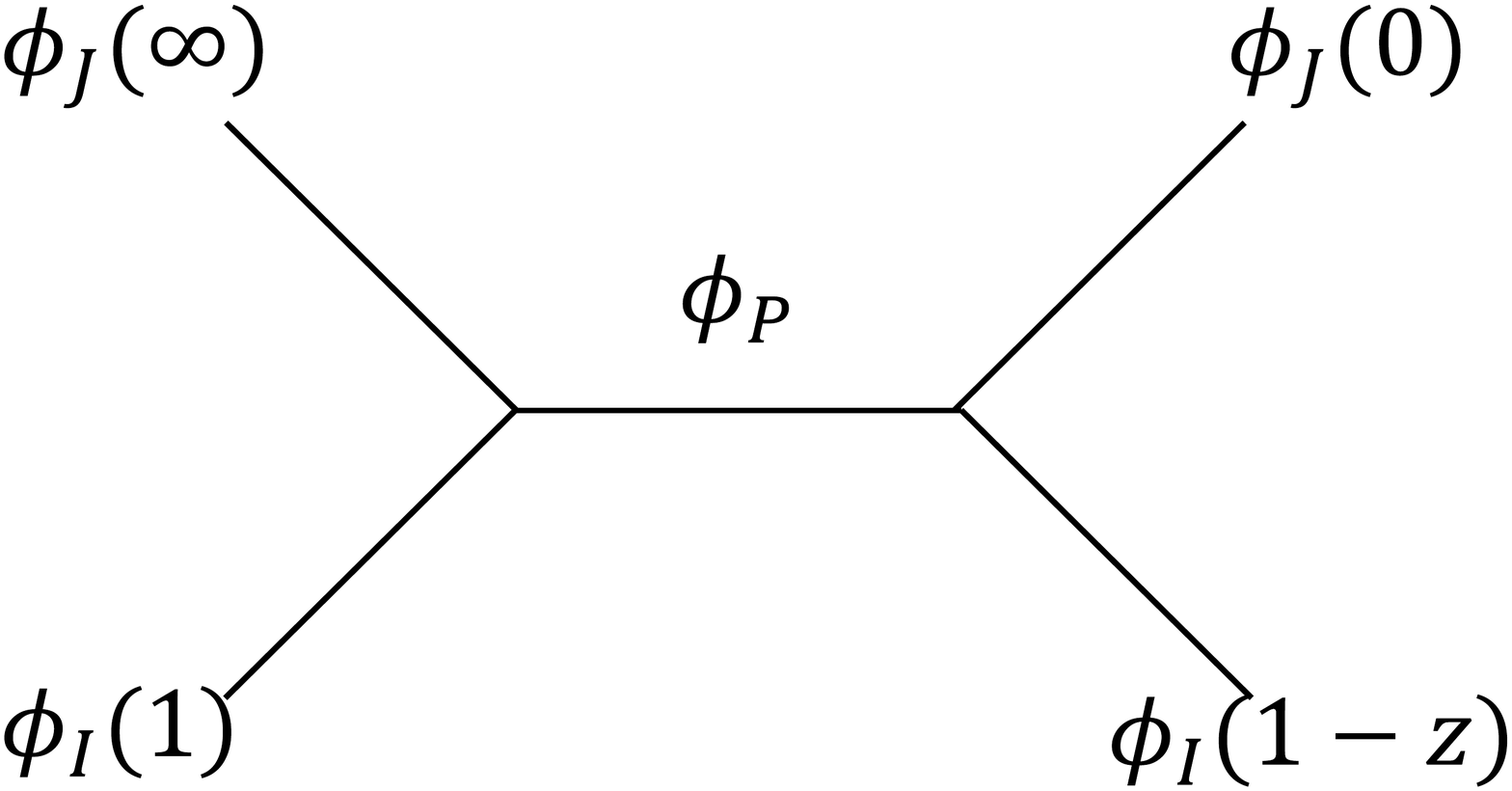}}
\newlength{\boxfourbw}
\settowidth{\boxfourbw}{\usebox{\boxfourb}} 

\newsavebox{\boxfourc}
\sbox{\boxfourc}{\includegraphics[width=100pt]{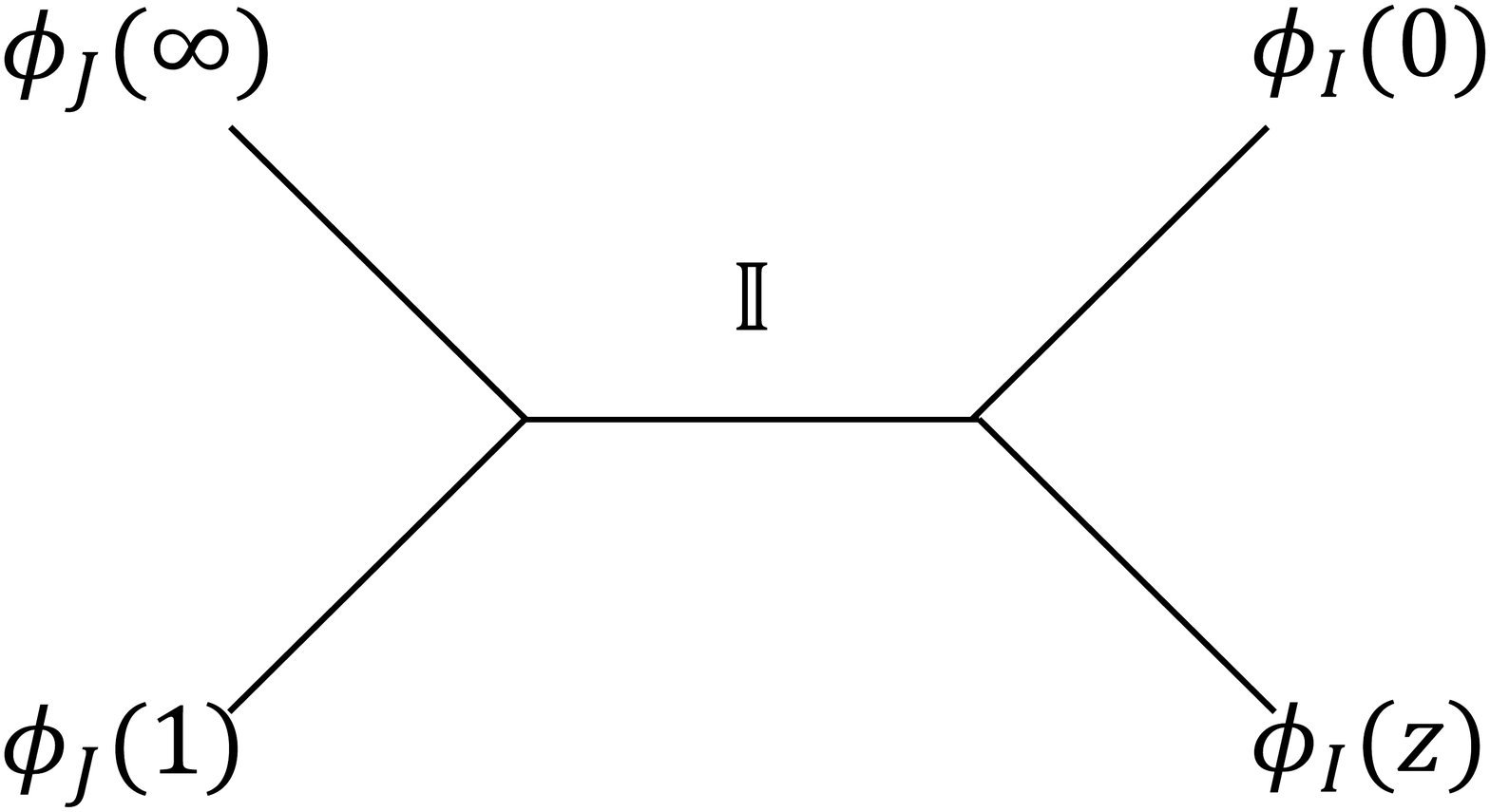}}
\newlength{\boxfourcw}
\settowidth{\boxfourcw}{\usebox{\boxfourc}}

For a boundary four-point function on a disk, we have the following two choices of how to cut,
\begin{equation}
\parbox{\boxcaw}{\usebox{\boxca}} = \parbox{\boxcbw}{\usebox{\boxcb}}.
\end{equation}
The corresponding bootstrap equation is given by
\begin{equation}
\begin{aligned}
\int \dd \a_P \ 
\rho (\a_P)
\overline{C_{IIP}C_{JJP}}
\parbox{\boxfouraw}{\usebox{\boxfoura}}
=
\int \dd \a_P \ 
\rho (\a_P)
\overline{C_{IJP}C_{IJP}}
\parbox{\boxfourbw}{\usebox{\boxfourb}}.
\end{aligned}
\end{equation}
In the $z \to 0$ limit, the left hand side can be approximated by the vacuum block,
which leads to the following approximated bootstrap equation,
\begin{equation}
\begin{aligned}
\int \dd \a_P \ 
\rho (\a_P)
\overline{C_{IJP}C_{IJP}}
\parbox{\boxfourbw}{\usebox{\boxfourb}}
&\simeq
\parbox{\boxfourcw}{\usebox{\boxfourc}}\\
&=
\int \dd \a_P \ 
{\bold F}_{0, \a_P} 
   \left[
    \begin{array}{cc}
    \a_I   & \a_I  \\
     \a_J  &   \a_J \\
    \end{array}
  \right]
\parbox{\boxfourbw}{\usebox{\boxfourb}}.
\end{aligned}
\end{equation}
Therefore, the boundary-boundary-boundary OPE coefficients follow the universal asymptotic formula,
\begin{equation}
\rho (\a_P)
\overline{C_{IJP}C_{IJP}}
\simeq
{\bold F}_{0, \a_P} 
   \left[
    \begin{array}{cc}
    \a_I   & \a_I  \\
     \a_J  &   \a_J \\
    \end{array}
  \right],
\ \ \ \ \ \ h_P \to \infty.
\end{equation}
By using (\ref{eq:open}), we have
\begin{equation}
\overline{C_{IJP}C_{IJP}}
\simeq
g^{-2}S_{0P}^{-1}
{\bold F}_{0, \a_P} 
   \left[
    \begin{array}{cc}
    \a_I   & \a_I  \\
     \a_J  &   \a_J \\
    \end{array}
  \right],
\ \ \ \ \ \ h_P \to \infty.
\end{equation}
This asymptotic formula is similar to that of the bulk-bulk-bulk OPE coefficients (\ref{eq:bulk^3}),
\begin{equation}
\rho (\a_p, \bar{\a}_p)
\overline{C_{ijp}C_{ijp}}
\simeq
{\bold F}_{0, \a_p} 
   \left[
    \begin{array}{cc}
    \a_i   & \a_i  \\
     \a_j  &   \a_j \\
    \end{array}
  \right]
{\bold F}_{0, \bar{\a}_p} 
   \left[
    \begin{array}{cc}
    \bar{\a}_i   & \bar{\a}_i  \\
     \bar{\a}_j  &   \bar{\a}_j \\
    \end{array}
  \right]
,
\ \ \ \ \ \ h_p, \bar{h}_p \to \infty.
\end{equation}

\subsection{Boundary one-point on a cylinder: boundary-boundary-boundary OPE coefficients}
\newsavebox{\boxfa}
\sbox{\boxfa}{\includegraphics[width=70pt]{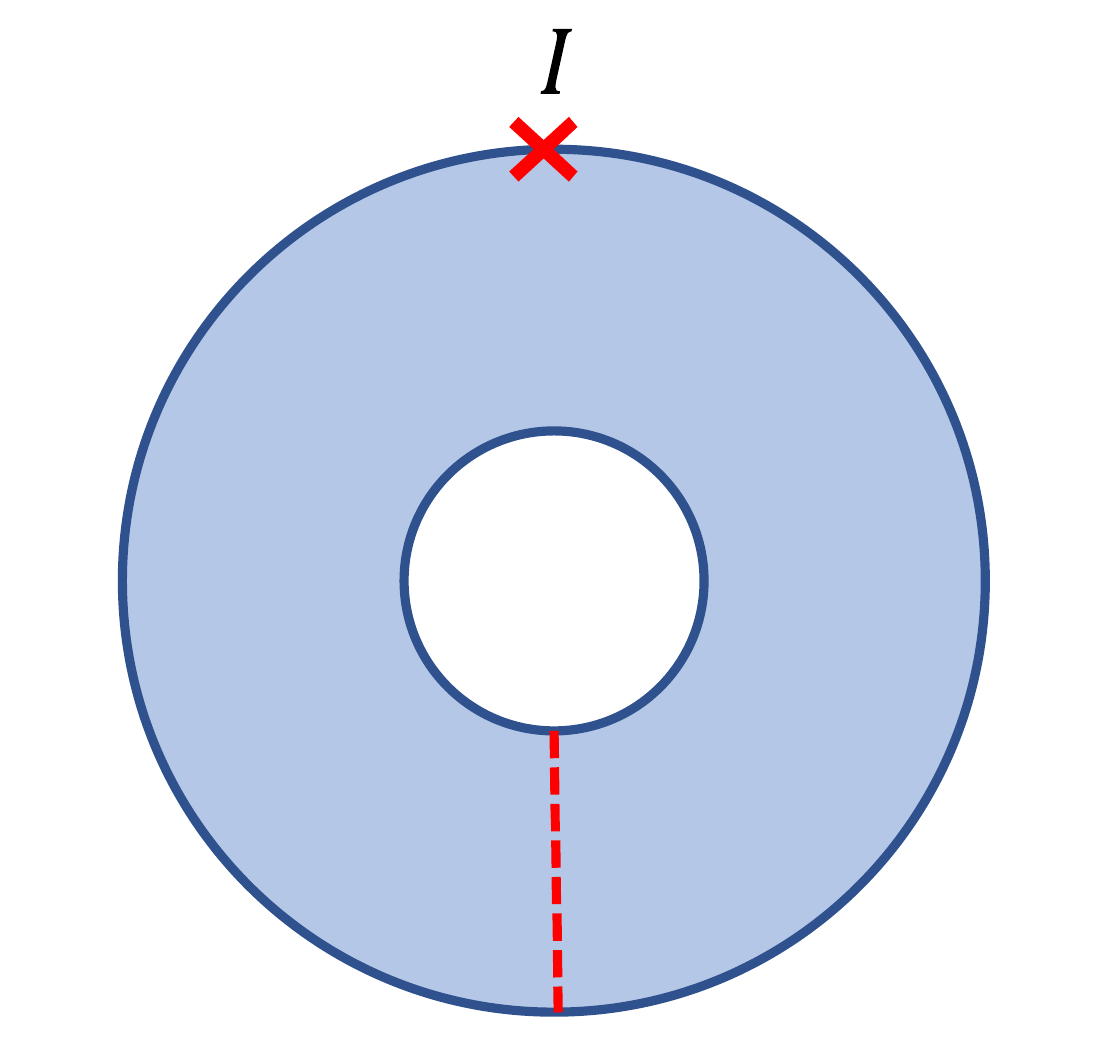}}
\newlength{\boxfaw}
\settowidth{\boxfaw}{\usebox{\boxfa}} 

\newsavebox{\boxfb}
\sbox{\boxfb}{\includegraphics[width=70pt]{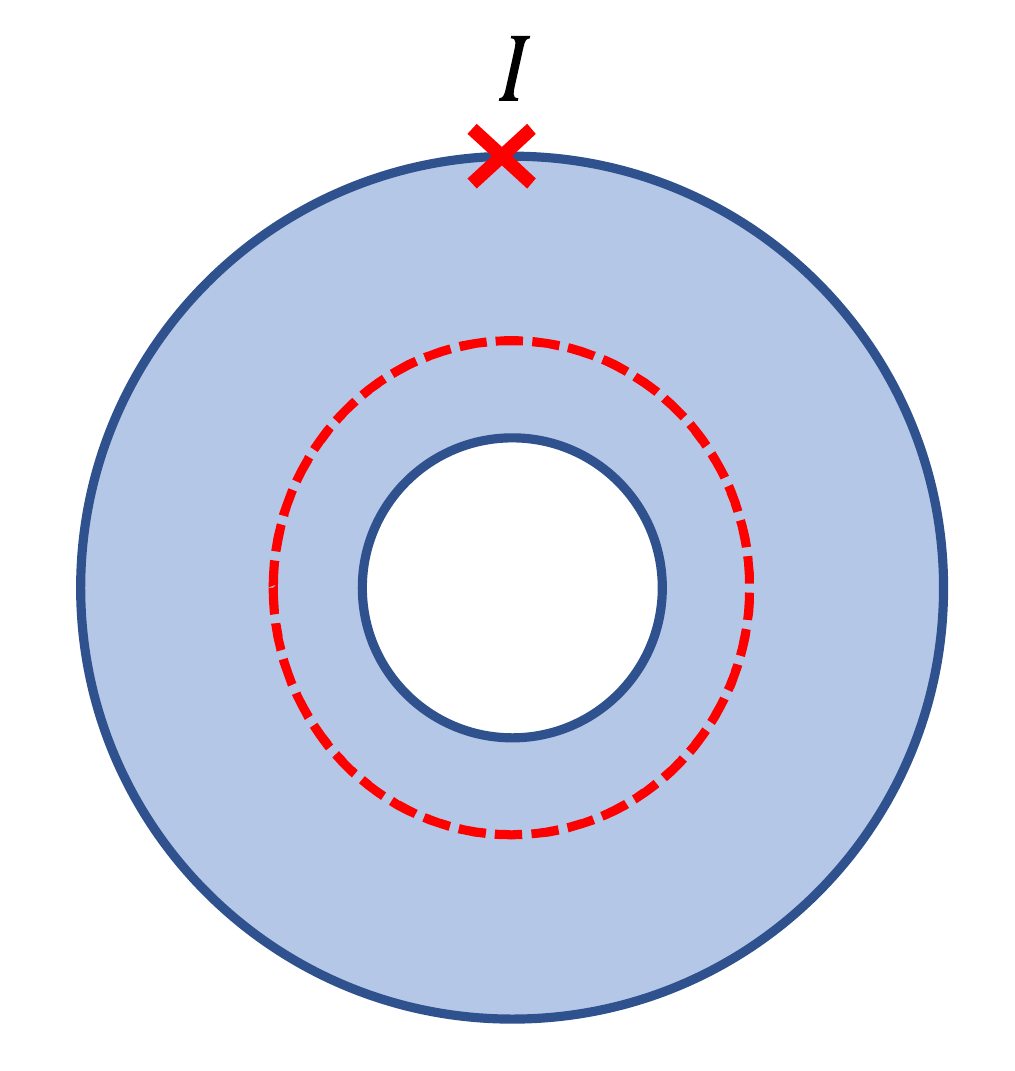}}
\newlength{\boxfbw}
\settowidth{\boxfbw}{\usebox{\boxfb}} 

\newsavebox{\boxonetorusa}
\sbox{\boxonetorusa}{\includegraphics[width=45pt]{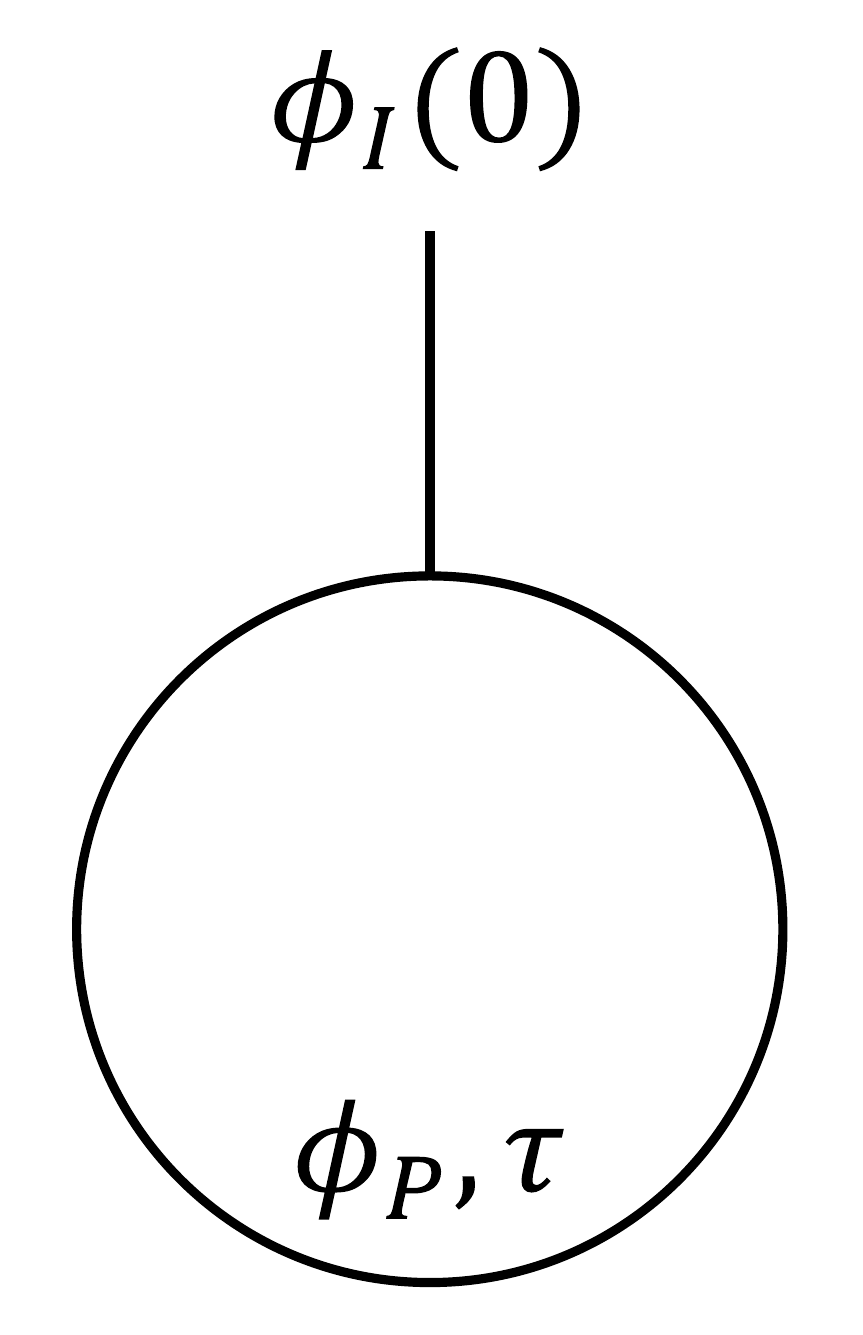}}
\newlength{\boxonetorusaw}
\settowidth{\boxonetorusaw}{\usebox{\boxonetorusa}} 

\newsavebox{\boxonetorusb}
\sbox{\boxonetorusb}{\includegraphics[width=45pt]{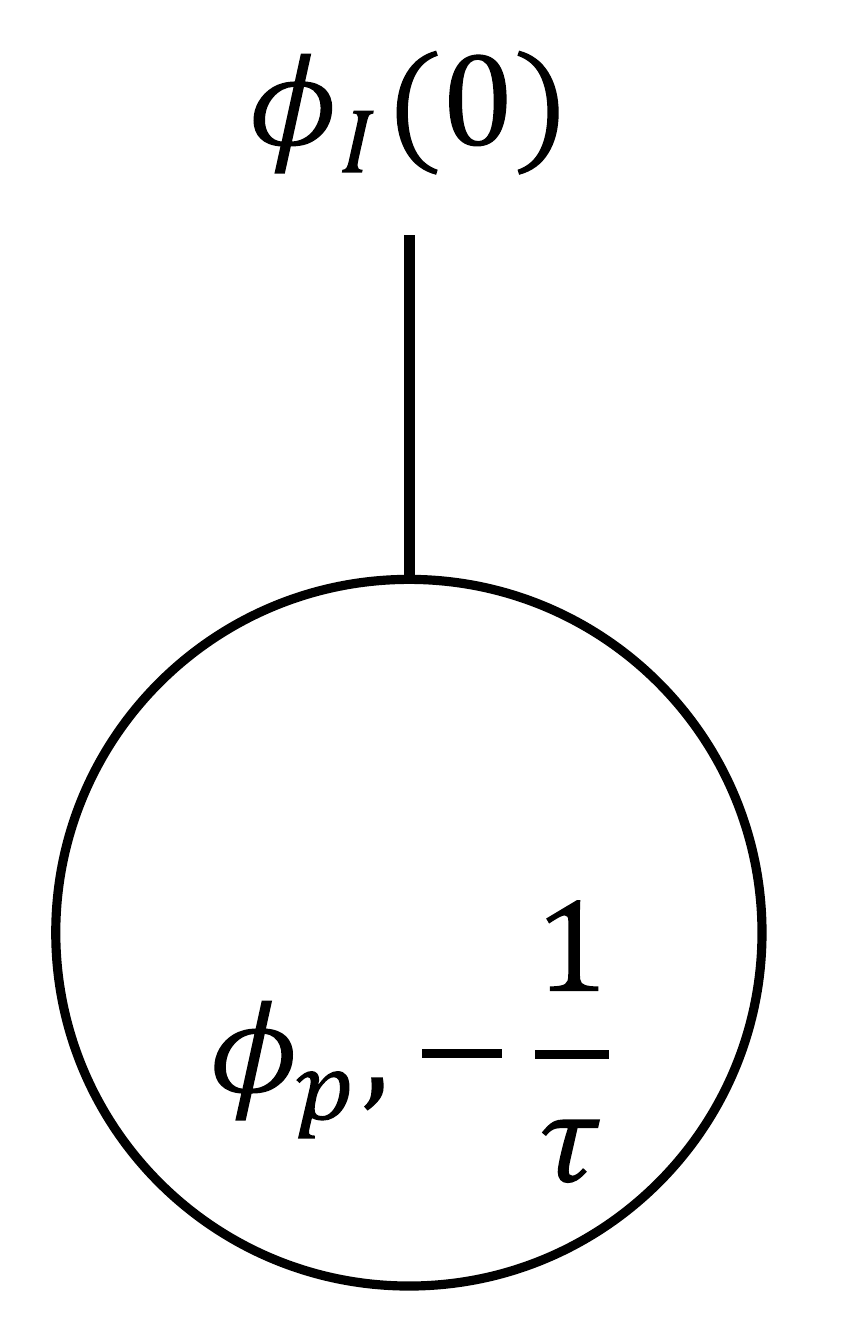}}
\newlength{\boxonetorusbw}
\settowidth{\boxonetorusbw}{\usebox{\boxonetorusb}} 

\newsavebox{\boxonetorusc}
\sbox{\boxonetorusc}{\includegraphics[width=45pt]{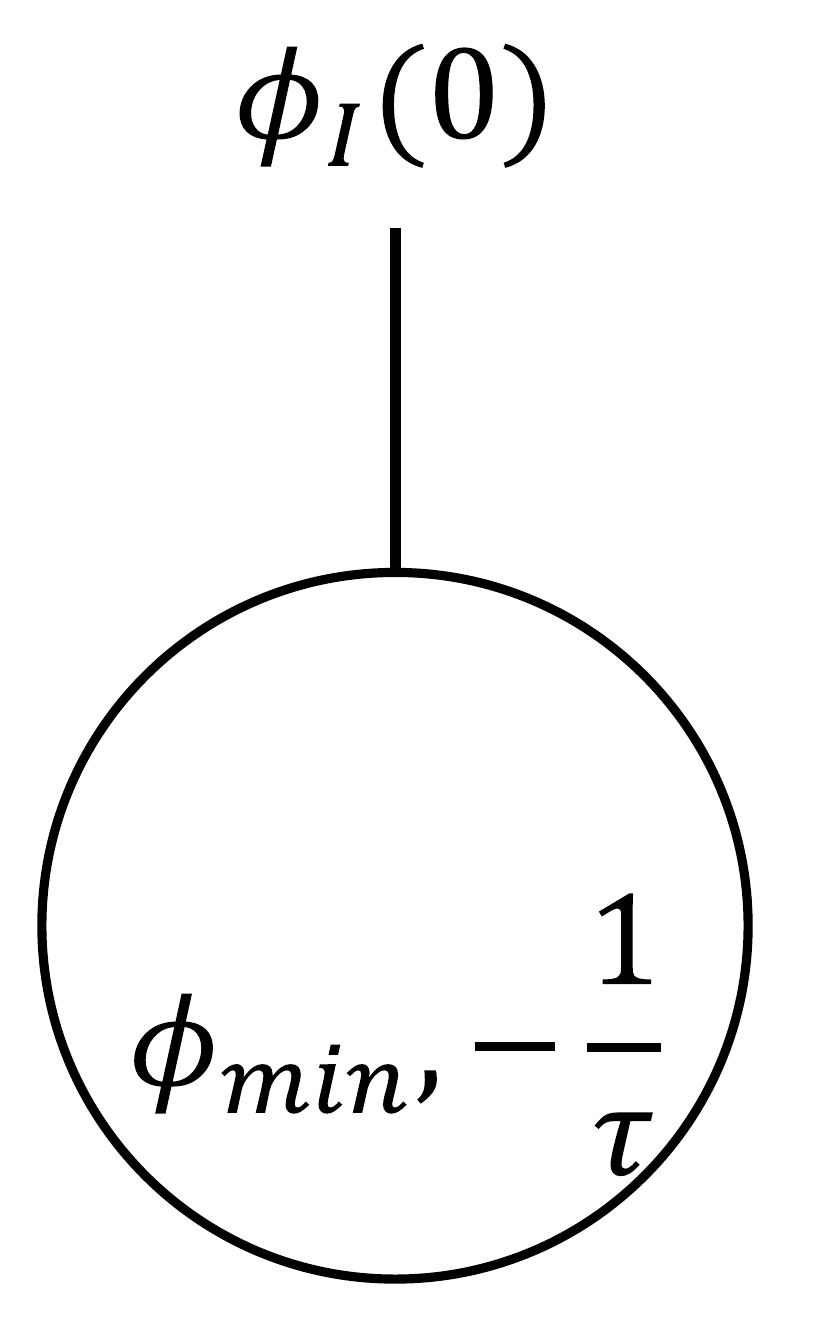}}
\newlength{\boxonetoruscw}
\settowidth{\boxonetoruscw}{\usebox{\boxonetorusc}}

For a boundary one-point function on a cylinder, we have the following two choices of how to cut,
\begin{equation}
\parbox{\boxfaw}{\usebox{\boxfa}} = \parbox{\boxfbw}{\usebox{\boxfb}}.
\end{equation}
The corresponding bootstrap equation is given by
\footnote{
In the standard convention of the conformal block,
we have conformal factors under this modular transformation.
For simplicity, we absorb these factors into the block.
In fact, this simplification does not change our results because we come back to the same channel by using the fusion transformation, which cancels the factors.
This simplification can also be found in \cite{Collier2020}. 
}
\begin{equation}
\begin{aligned}
\int \dd \a_P \ 
\rho (\a_P)
\overline{C_{IPP}}
\parbox{\boxonetorusaw}{\usebox{\boxonetorusa}}
=
g^2
\int \dd \a_p \ 
\rho (\a_p)
\overline{C_{pI}C_{p\mathbb{I}}}
\parbox{\boxonetorusbw}{\usebox{\boxonetorusb}}.
\end{aligned}
\end{equation}
In the $\tau \to i0$ limit, the right hand side can be approximated by a single block with the conformal dimension $h_{p_{min}}$, which is the lowest dimension that satisfies $C_{p_{min} I} \neq 0$.
Thus, the approximated bootstrap equation is given by
\begin{equation}
\begin{aligned}
\int \dd \a_P \ 
\rho (\a_P)
\overline{C_{IPP}}
\parbox{\boxonetorusaw}{\usebox{\boxonetorusa}}
&\simeq
g^2
\int \dd \a_p \ 
\rho (\a_{p_{min}})
\overline{C_{p_{min}I}C_{p_{min}\mathbb{I}}}
\parbox{\boxonetorusbw}{\usebox{\boxonetorusc}}\\
&=
g^2 \ 
\rho (\a_{p_{min}})
\overline{C_{p_{min}I}C_{p_{min}\mathbb{I}}}
\int \dd \a_P \ 
S_{p_{min}P}[I]
\parbox{\boxonetorusaw}{\usebox{\boxonetorusa}},
\end{aligned}
\end{equation}
where the matrix $S_{pq}[I]$ represents the kernel of the modular-S transformation for the torus one-point block \cite{Teschner2003} (see also \cite{Collier2020}).
As a result, we obtain the universal formula for the boundary-boundary-boundary OPE coefficients as
\begin{equation}
\rho (\a_P)
\overline{C_{IPP}}
\simeq
g^2 \ 
\rho (\a_{p_{min}})
\overline{C_{p_{min}I}C_{p_{min}\mathbb{I}}}
S_{p_{min}P}[I],
\ \ \ \ \ \ h_P \to \infty,
\end{equation}
or equivalently, by using (\ref{eq:open}), we have
\begin{equation}
\overline{C_{IPP}}
\simeq
S_{0P}^{-1} \ 
\rho (\a_{p_{min}})
\overline{C_{p_{min}I}C_{p_{min}\mathbb{I}}}
S_{p_{min}P}[I],
\ \ \ \ \ \ h_P \to \infty.
\end{equation}

\subsection{Boundary two-point on a cylinder: boundary-boundary-boundary OPE coefficients}
\newsavebox{\boxea}
\sbox{\boxea}{\includegraphics[width=70pt]{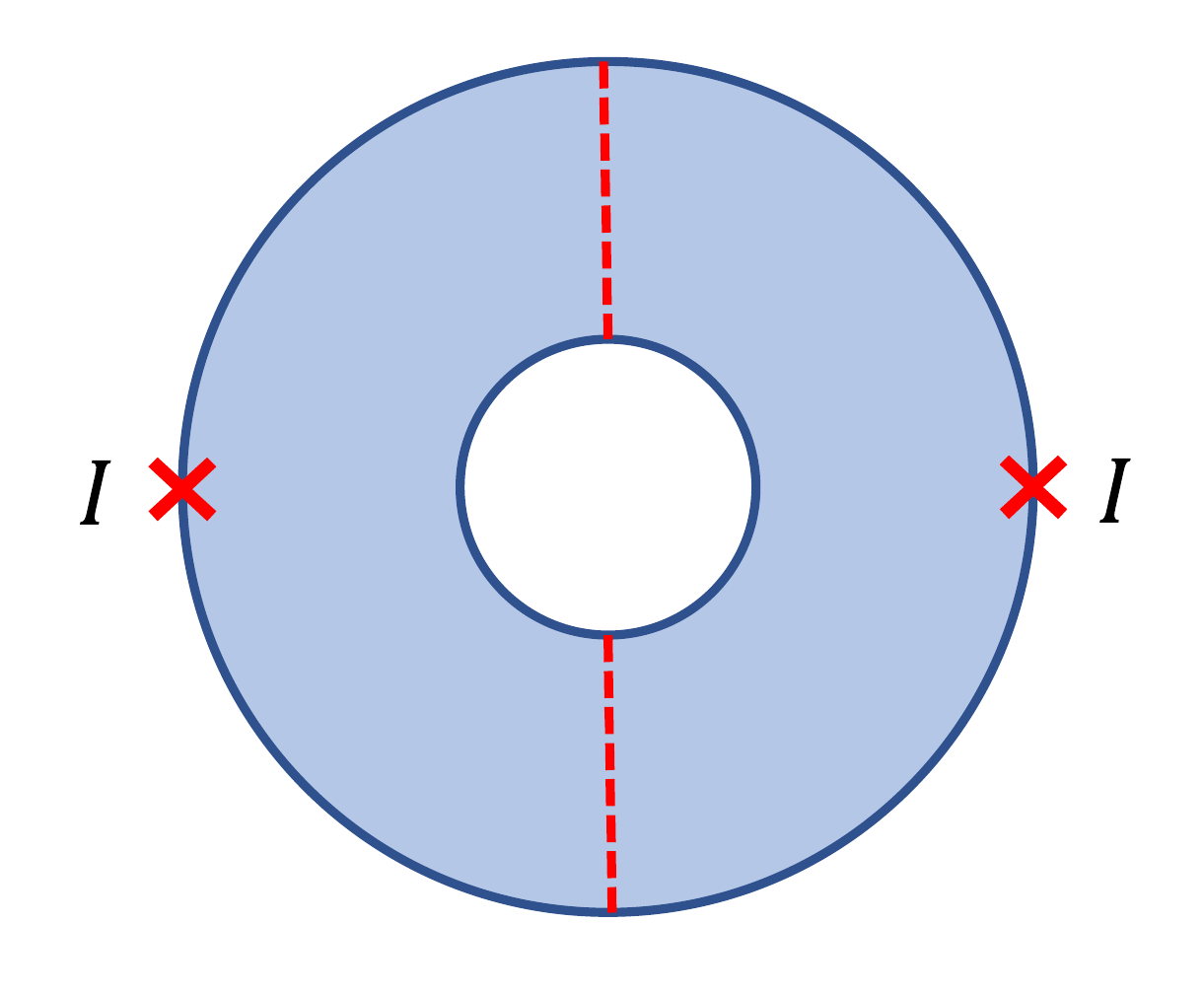}}
\newlength{\boxeaw}
\settowidth{\boxeaw}{\usebox{\boxea}} 

\newsavebox{\boxeb}
\sbox{\boxeb}{\includegraphics[width=70pt]{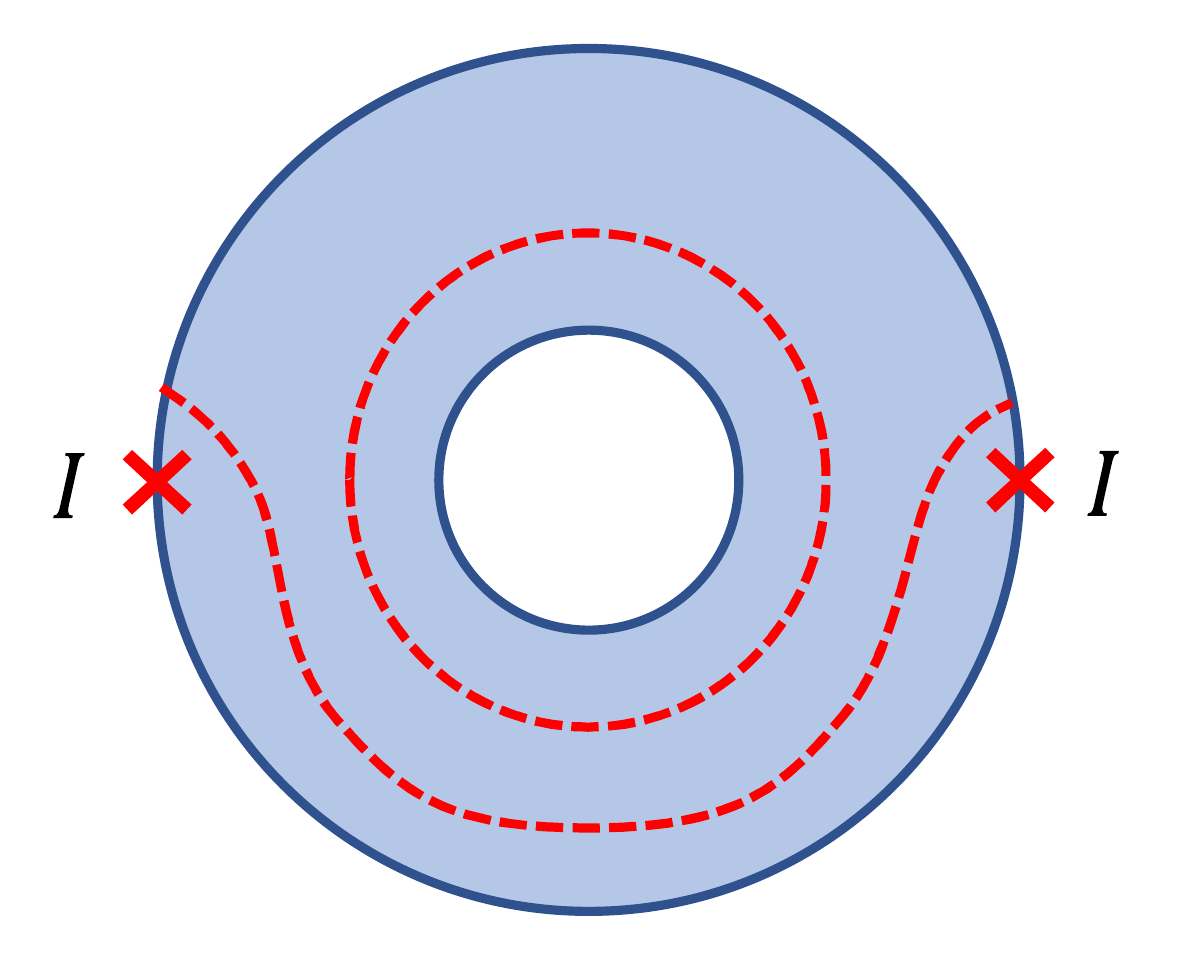}}
\newlength{\boxebw}
\settowidth{\boxebw}{\usebox{\boxeb}} 

\newsavebox{\boxtwotorusa}
\sbox{\boxtwotorusa}{\includegraphics[width=70pt]{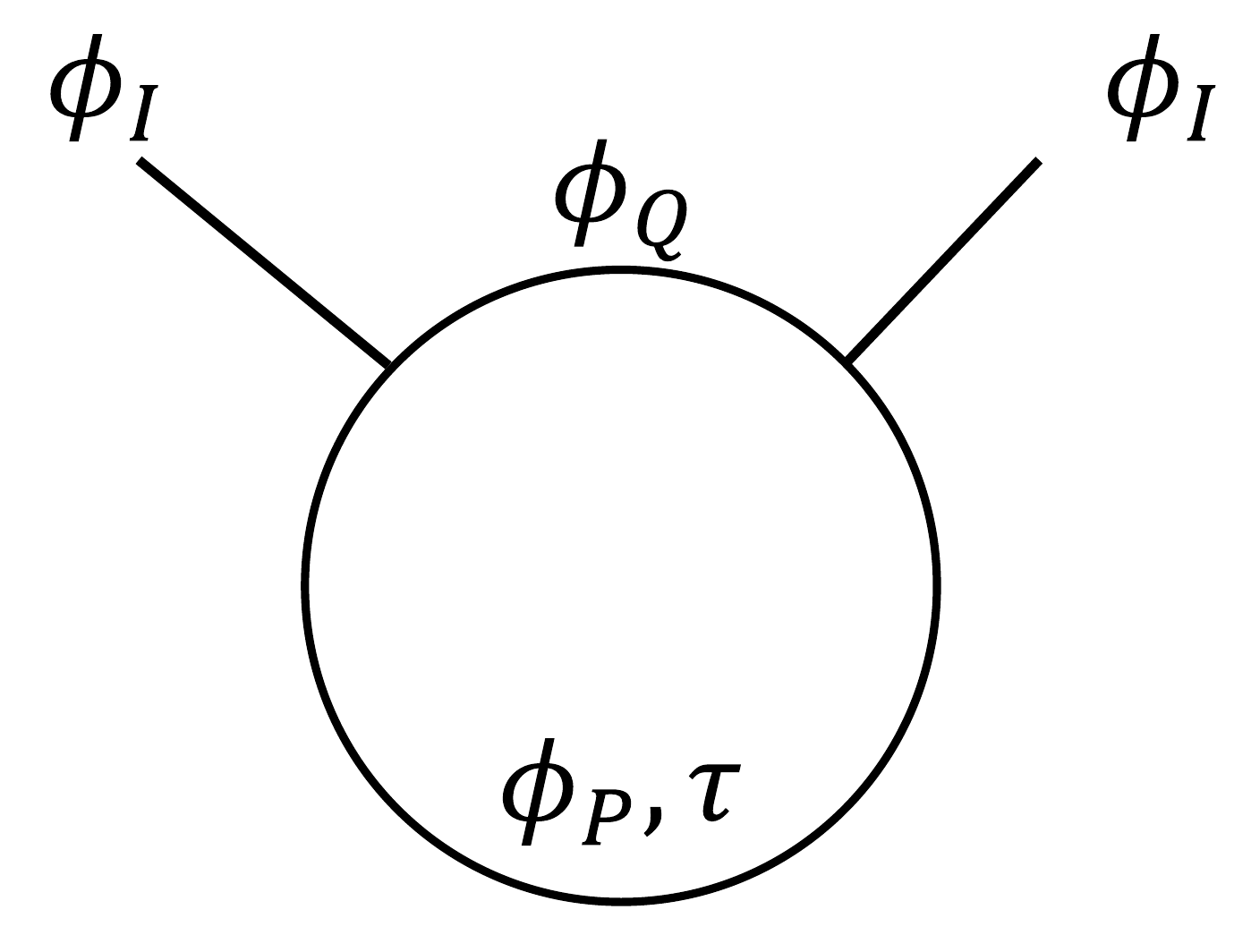}}
\newlength{\boxtwotorusaw}
\settowidth{\boxtwotorusaw}{\usebox{\boxtwotorusa}} 

\newsavebox{\boxtwotorusb}
\sbox{\boxtwotorusb}{\includegraphics[width=50pt]{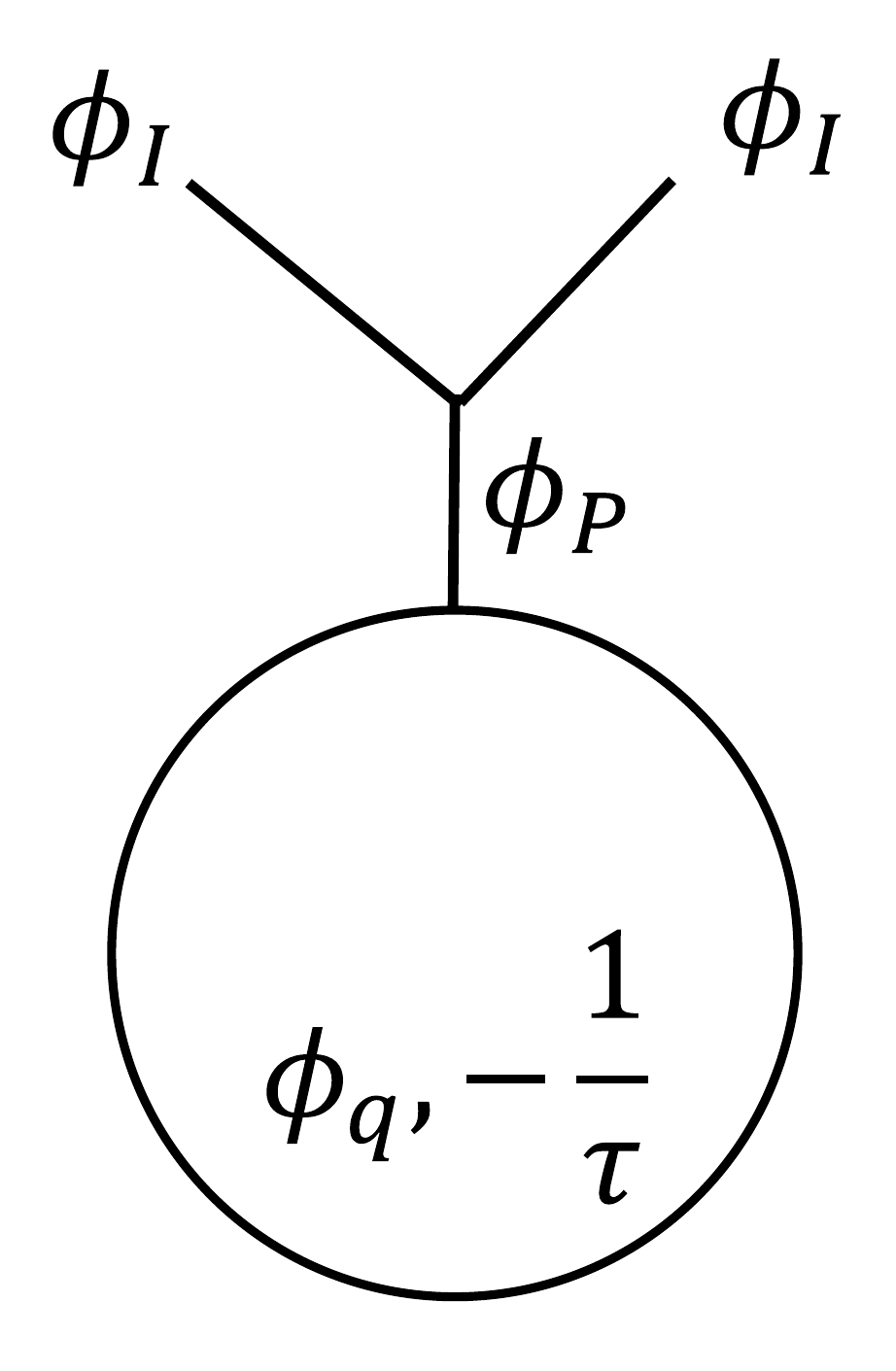}}
\newlength{\boxtwotorusbw}
\settowidth{\boxtwotorusbw}{\usebox{\boxtwotorusb}} 

\newsavebox{\boxtwotorusc}
\sbox{\boxtwotorusc}{\includegraphics[width=50pt]{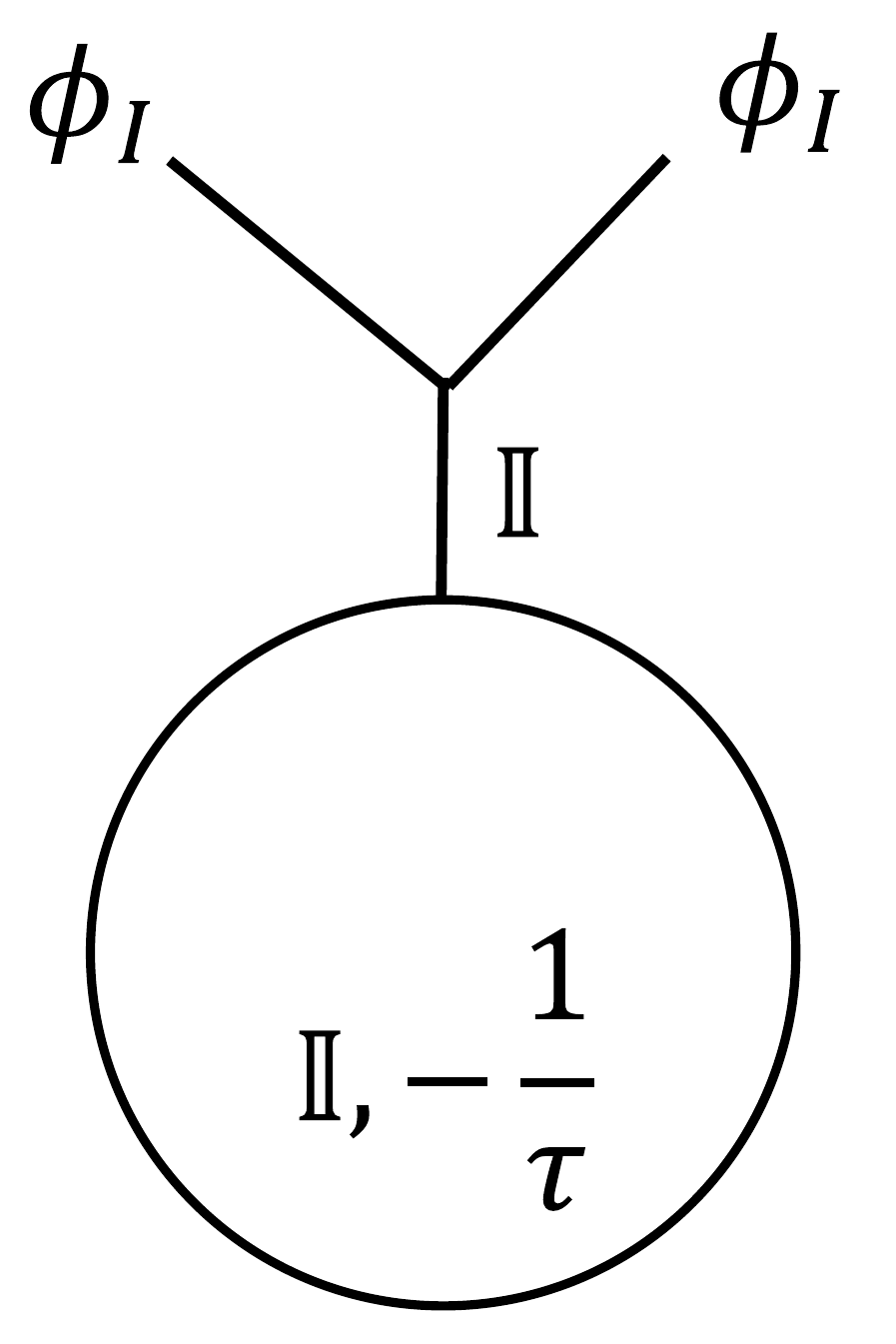}}
\newlength{\boxtwotoruscw}
\settowidth{\boxtwotoruscw}{\usebox{\boxtwotorusc}} 

\newsavebox{\boxtwotorusd}
\sbox{\boxtwotorusd}{\includegraphics[width=50pt]{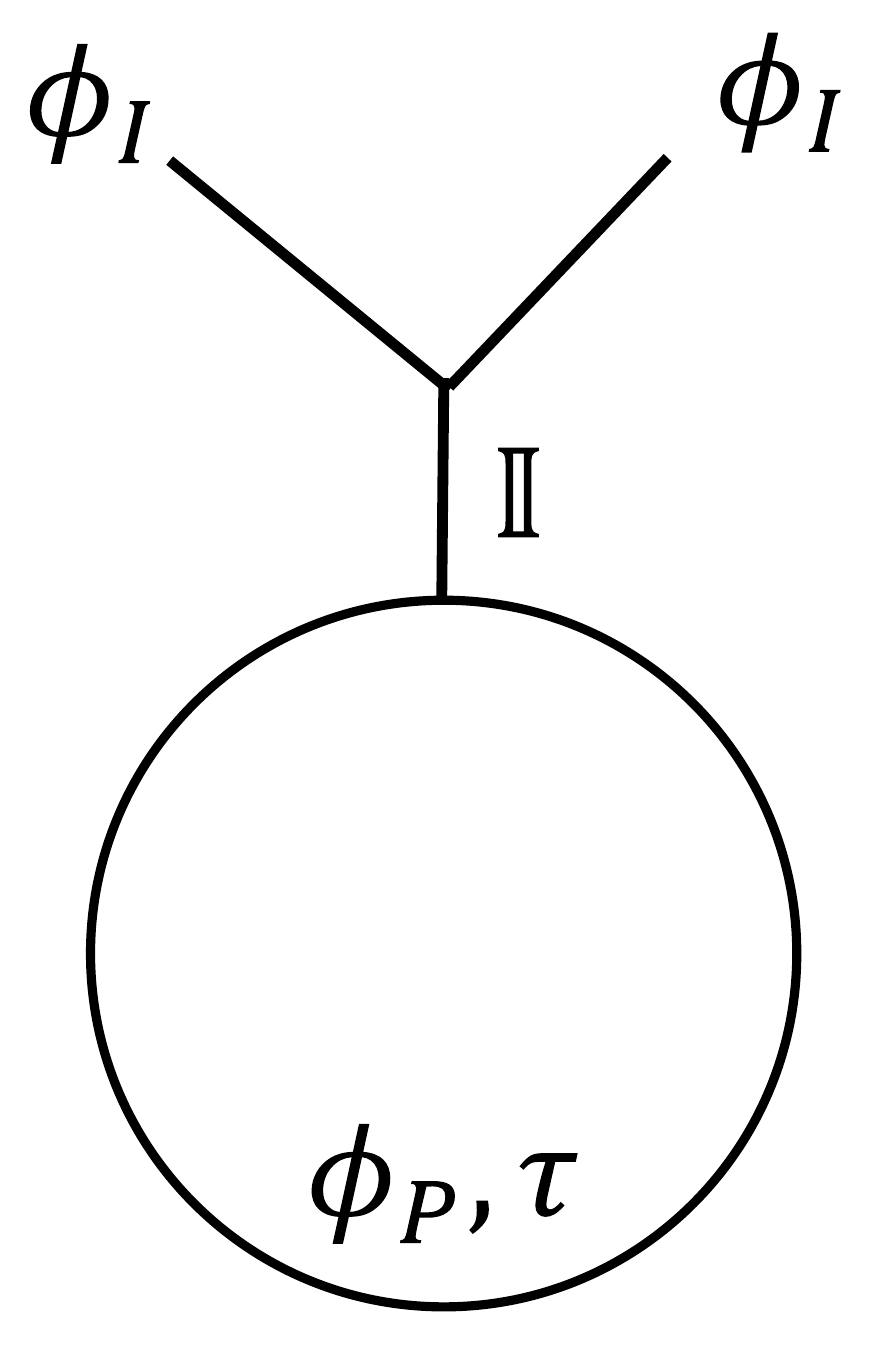}}
\newlength{\boxtwotorusdw}
\settowidth{\boxtwotorusdw}{\usebox{\boxtwotorusd}} 

For a boundary two-point function on a cylinder, we can consider the following two choices of how to cut,
\begin{equation}
\parbox{\boxeaw}{\usebox{\boxea}} = \parbox{\boxebw}{\usebox{\boxeb}}.
\end{equation}
The corresponding bootstrap equation is given by
\begin{equation}
\begin{aligned}
&
\int \dd \a_P \ 
\int \dd \a_Q \ 
\rho (\a_P)
\rho (\a_Q)
\overline{C_{IPQ}C_{IPQ}}
\parbox{\boxtwotorusaw}{\usebox{\boxtwotorusa}}\\
&=
g^2
\int \dd \a_P \ 
\int \dd \a_q \ 
\rho (\a_P)
\rho (\a_q)
\overline{C_{IIP}C_{qP}C_{q\mathbb{I}}}
\parbox{\boxtwotorusbw}{\usebox{\boxtwotorusb}} .
\end{aligned}
\end{equation}
Let us consider the limit $\tau \to i0$ and the short distance limit between two boundary operators.
In this limit, the right-hand side can be approximated by the vacuum block and then we obtain the following equation,
\begin{equation}
\begin{aligned}
&
\int \dd \a_P \ 
\int \dd \a_Q \ 
\rho (\a_P)
\rho (\a_Q)
\overline{C_{IPQ}C_{IPQ}}
\parbox{\boxtwotorusaw}{\usebox{\boxtwotorusa}}\\
&\simeq
g^2
\parbox{\boxtwotoruscw}{\usebox{\boxtwotorusc}}\\
&=
g^2
\int \dd \a_P \ 
S_{0P}
\parbox{\boxtwotorusdw}{\usebox{\boxtwotorusd}}\\
&=
g^2
\int \dd \a_P \ 
\int \dd \a_Q \ 
S_{0P}
{\bold F}_{0, \a_Q} 
   \left[
    \begin{array}{cc}
    \a_I   & \a_I  \\
     \a_P  &   \a_P \\
    \end{array}
  \right]
\parbox{\boxtwotorusaw}{\usebox{\boxtwotorusa}}.
\end{aligned}
\end{equation}
Thus, the universal formula for the boundary-boundary-boundary OPE coefficients is given by
\begin{equation}
\rho (\a_P)
\rho (\a_Q)
\overline{C_{IPQ}C_{IPQ}}
\simeq
g^2
S_{0P}
{\bold F}_{0, \a_Q} 
   \left[
    \begin{array}{cc}
    \a_I   & \a_I  \\
     \a_P  &   \a_P \\
    \end{array}
  \right],
  \ \ \ \ \ \ h_P, h_Q \to \infty.
\end{equation}
We can re-express it by (\ref{eq:open}) as
\begin{equation}
\overline{C_{IPQ}C_{IPQ}}
\simeq
g^{-2}
S_{0Q}^{-1}
{\bold F}_{0, \a_Q} 
   \left[
    \begin{array}{cc}
    \a_I   & \a_I  \\
     \a_P  &   \a_P \\
    \end{array}
  \right],
  \ \ \ \ \ \ h_P, h_Q \to \infty.
\end{equation}

\subsection{Sphere with three holes: boundary-boundary-boundary OPE coefficients}
\newsavebox{\boxda}
\sbox{\boxda}{\includegraphics[width=70pt]{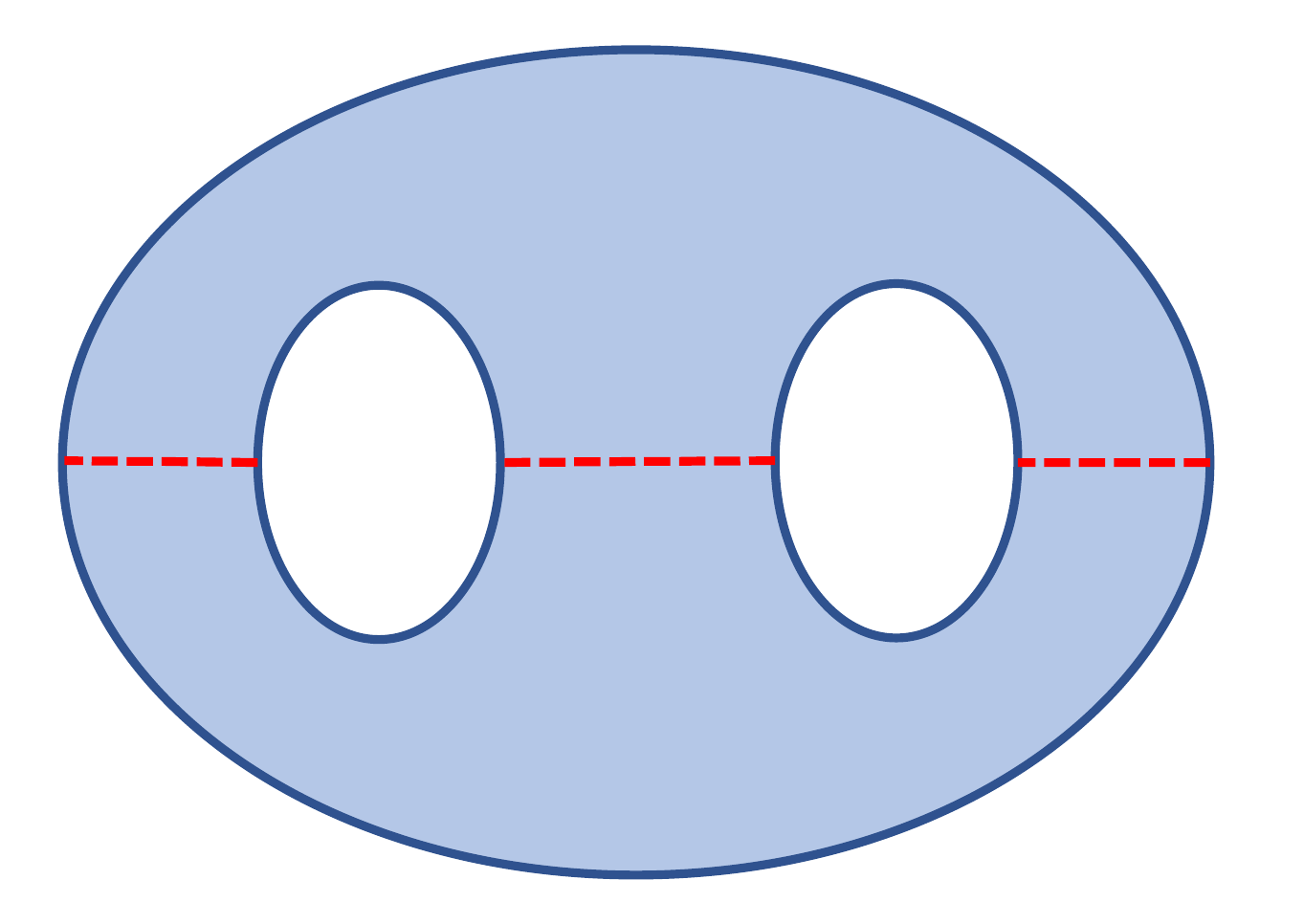}}
\newlength{\boxdaw}
\settowidth{\boxdaw}{\usebox{\boxda}} 

\newsavebox{\boxdb}
\sbox{\boxdb}{\includegraphics[width=70pt]{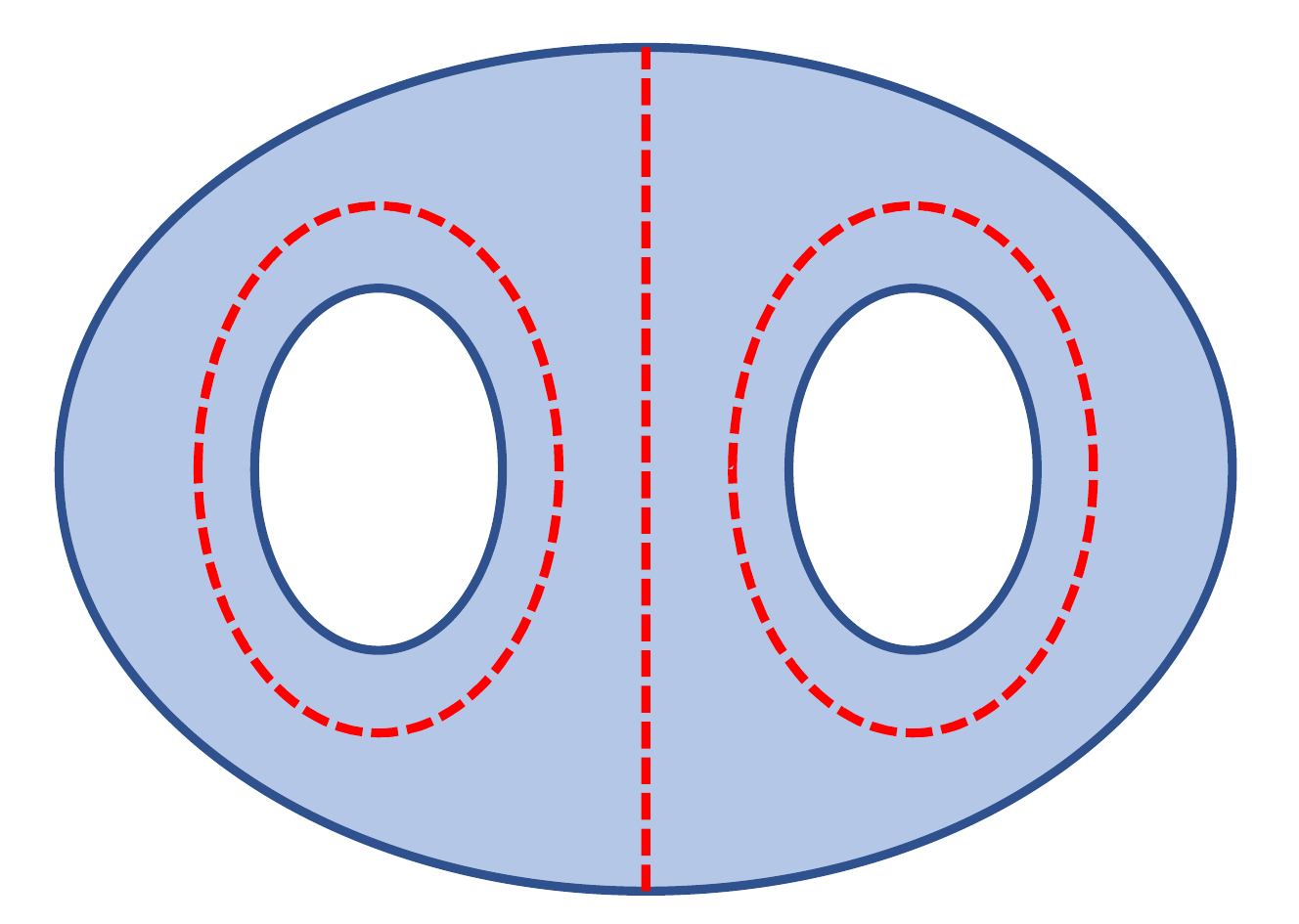}}
\newlength{\boxdbw}
\settowidth{\boxdbw}{\usebox{\boxdb}} 

\newsavebox{\boxgenusa}
\sbox{\boxgenusa}{\includegraphics[width=70pt]{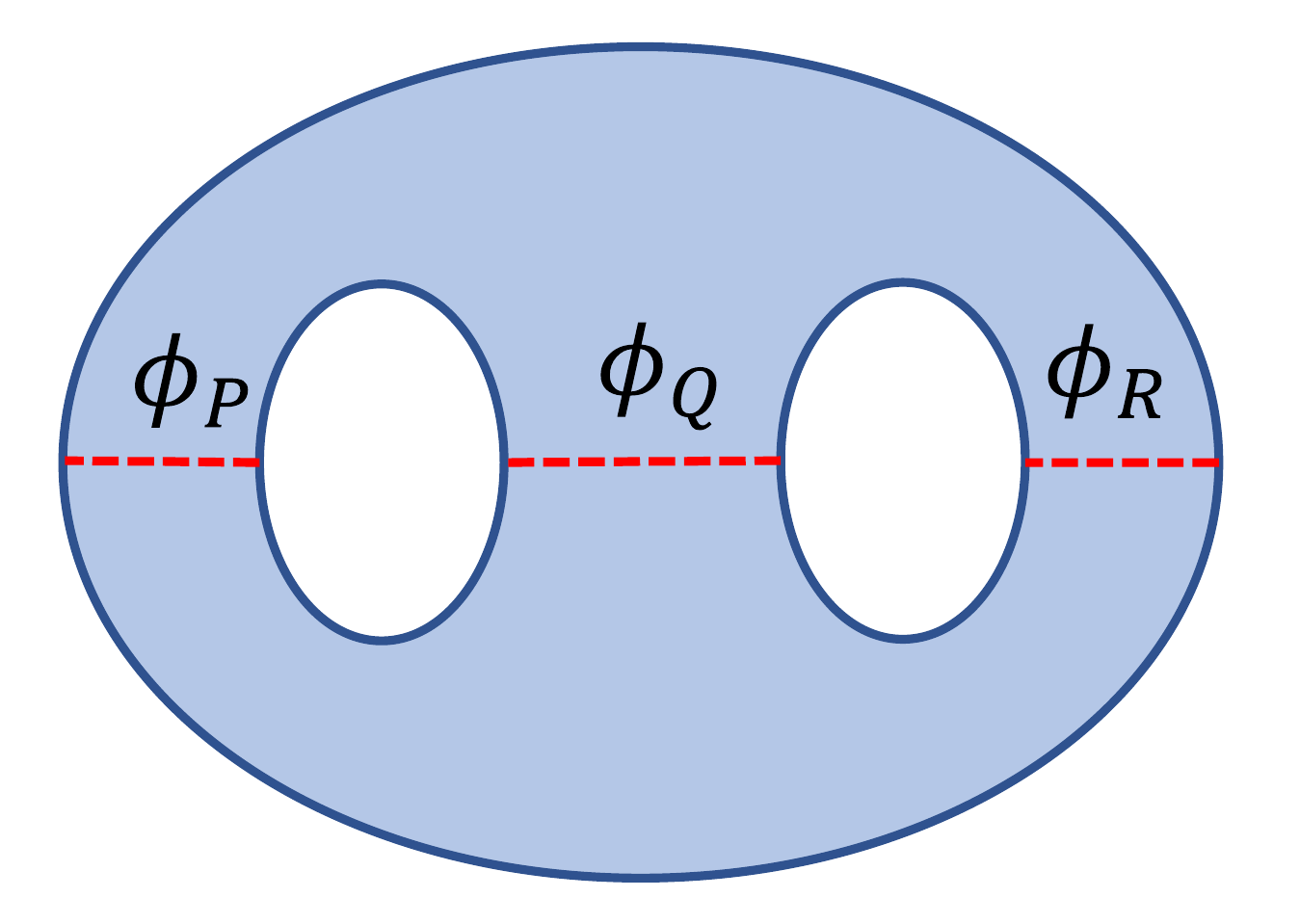}}
\newlength{\boxgenusaw}
\settowidth{\boxgenusaw}{\usebox{\boxgenusa}} 

\newsavebox{\boxgenusb}
\sbox{\boxgenusb}{\includegraphics[width=70pt]{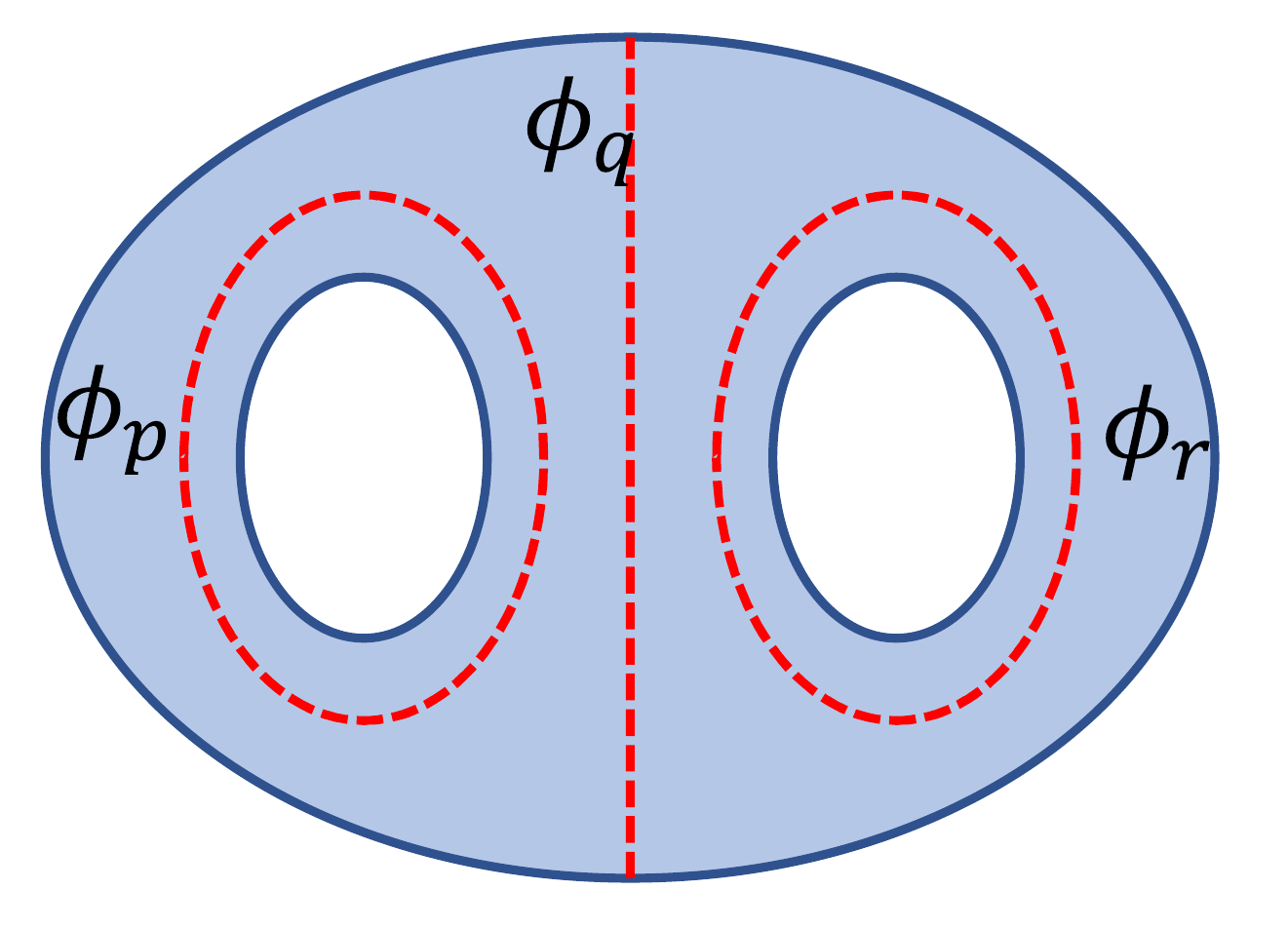}}
\newlength{\boxgenusbw}
\settowidth{\boxgenusbw}{\usebox{\boxgenusb}} 

\newsavebox{\boxgenusc}
\sbox{\boxgenusc}{\includegraphics[width=70pt]{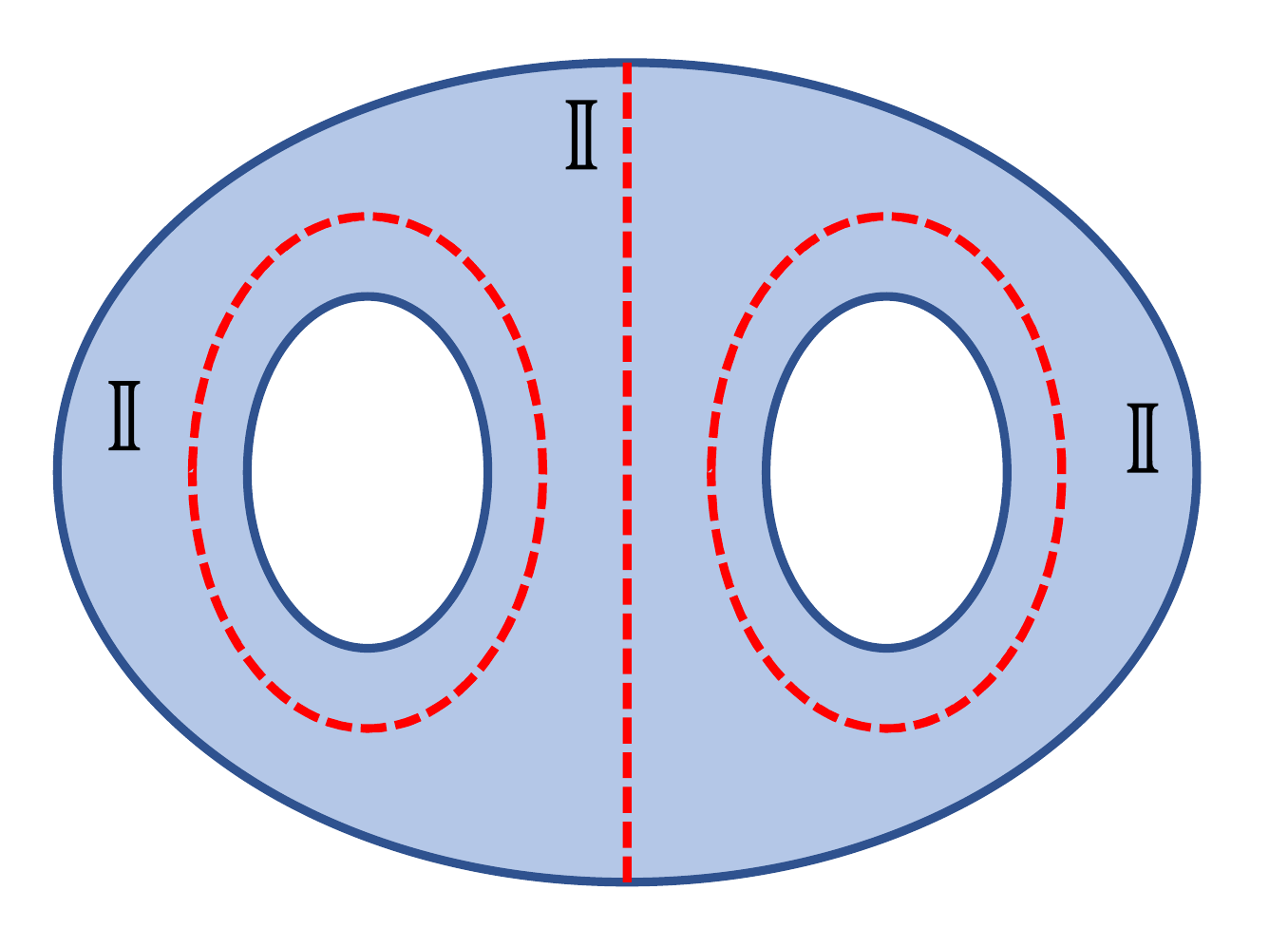}}
\newlength{\boxgenuscw}
\settowidth{\boxgenuscw}{\usebox{\boxgenusc}} 

\newsavebox{\boxgenusd}
\sbox{\boxgenusd}{\includegraphics[width=70pt]{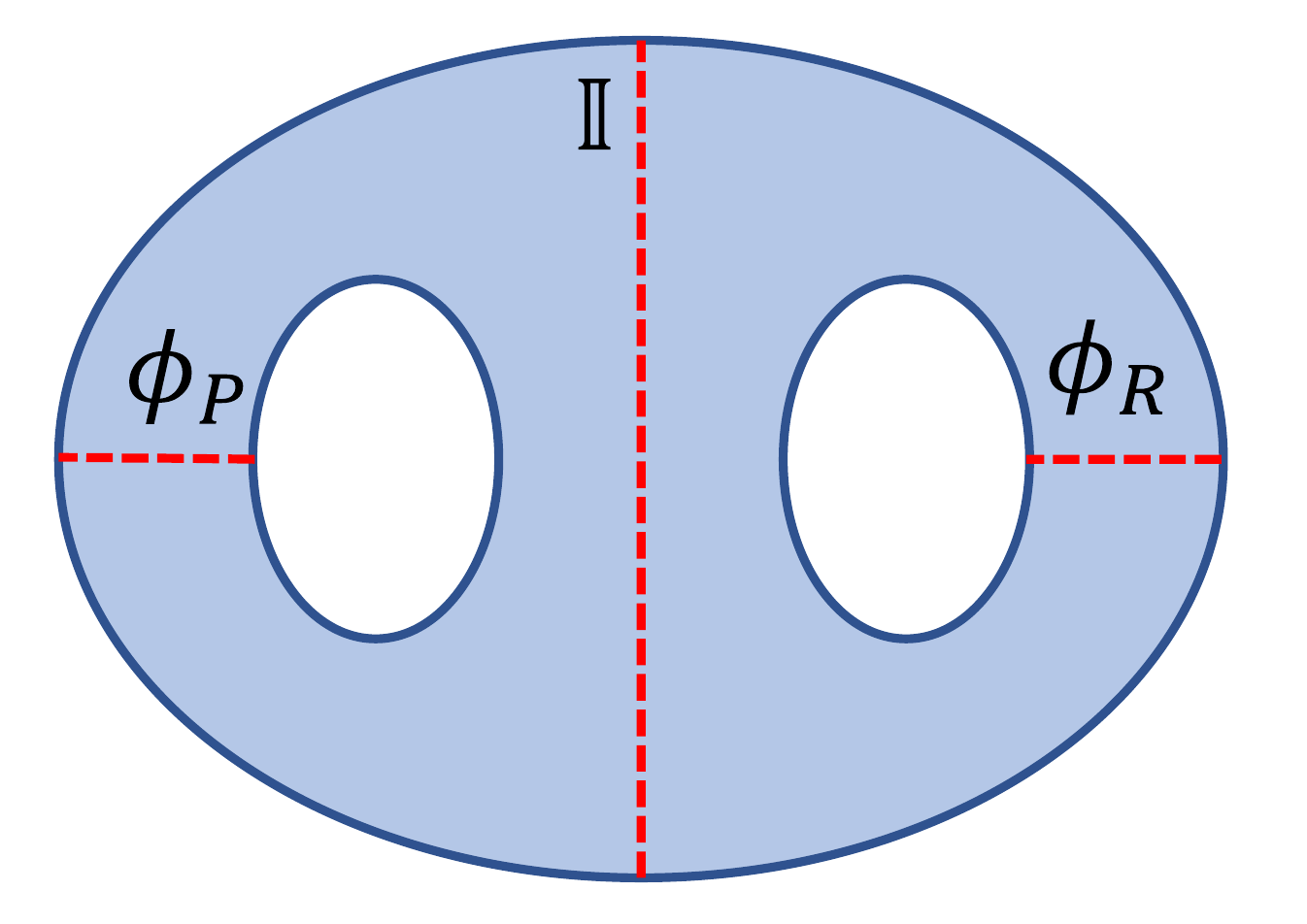}}
\newlength{\boxgenusdw}
\settowidth{\boxgenusdw}{\usebox{\boxgenusd}} 

For a sphere partition function with three holes, we can consider the following two choices of how to cut,
\begin{equation}
\parbox{\boxdaw}{\usebox{\boxda}} = \parbox{\boxdbw}{\usebox{\boxdb}}.
\end{equation}
In general, partition functions on manifolds with genus $g\geq2$ have complicated moduli spaces.
To avoid cumbersome expressions, we represent the conformal block with simple sketches instead of explicit forms.
This makes sense because we do not need the exact values of the modulus to show the universal formula, unlike the traditional approach making use of the inverse Laplace transformation (as shown in \cite{Cardy2017}).
We only use the fusion transformation.

With this notational simplification, the corresponding bootstrap equation can be expressed by
\begin{equation}
\begin{aligned}
&
\int \dd \a_P \ 
\int \dd \a_Q \ 
\int \dd \a_R \ \ 
\rho (\a_P)
\rho (\a_Q)
\rho (\a_R)
\overline{C_{PQR}C_{PQR}}
\parbox{\boxgenusaw}{\usebox{\boxgenusa}}\\
&=
g^4
\int \dd \a_p \ 
\int \dd \a_q \ 
\int \dd \a_r \ \ 
\rho (\a_p)
\rho (\a_q)
\rho (\a_r)
\overline{C_{pqr} C_{p\mathbb{I}} C_{q\mathbb{I}} C_{r\mathbb{I}} }
\parbox{\boxgenusbw}{\usebox{\boxgenusb}}.
\end{aligned}
\end{equation}
In an appropriate limit of the moduli parameters, the right-hand side can be approximated by the vacuum block.
By using the fusion transformation, we obtain
\begin{equation}
\begin{aligned}
&
\int \dd \a_P \ 
\int \dd \a_Q \ 
\int \dd \a_R \ \ 
\rho (\a_P)
\rho (\a_Q)
\rho (\a_R)
\overline{C_{PQR}C_{PQR}}
\parbox{\boxgenusaw}{\usebox{\boxgenusa}}\\
&\simeq
g^4
\parbox{\boxgenuscw}{\usebox{\boxgenusc}}\\
&=
g^4
\int \dd \a_P \ 
\int \dd \a_R \ \ 
S_{0P}
S_{0R}
\parbox{\boxgenusdw}{\usebox{\boxgenusd}}\\
&=
g^4
\int \dd \a_P \ 
\int \dd \a_Q \ 
\int \dd \a_R \ \ 
S_{0P}
S_{0R}
{\bold F}_{0, \a_Q} 
   \left[
    \begin{array}{cc}
    \a_P   & \a_P  \\
     \a_R  &   \a_R \\
    \end{array}
  \right]
\parbox{\boxgenusaw}{\usebox{\boxgenusa}}.
\end{aligned}
\end{equation}
Thus, the universal formula for the boundary-boundary-boundary OPE coefficients is given by
\begin{equation}
\rho (\a_P)
\rho (\a_Q)
\rho (\a_R)
\overline{C_{PQR}C_{PQR}}
\simeq
g^4
S_{0P}
S_{0R}
{\bold F}_{0, \a_Q} 
   \left[
    \begin{array}{cc}
    \a_P   & \a_P  \\
     \a_R  &   \a_R \\
    \end{array}
  \right],
  \ \ \ \ \ \ h_P,h_Q,h_R \to \infty,
\end{equation}
or equivalently, we have
\begin{equation}\label{eq:bdybdybdy}
\overline{C_{PQR}C_{PQR}}
\simeq
g^{-2}
S_{0Q}^{-1}
{\bold F}_{0, \a_Q} 
   \left[
    \begin{array}{cc}
    \a_P   & \a_P  \\
     \a_R  &   \a_R \\
    \end{array}
  \right],
  \ \ \ \ \ \ h_P,h_Q,h_R \to \infty.
\end{equation}

\section{Light-Cone Bootstrap in BCFT}\label{sec:light-cone}
Besides the high-low temperature limit, we can analytically solve the bootstrap equation in the light-cone limit.
As shown in  \cite{Kusuki2019a, Collier2018, Kusuki2019}, we can derive the universal feature for large-spin states in this limit.
Here we have a natural question: how we can derive the large-spin universality in a BCFT?
In this section, we will provide simple examples of the light-cone bootstrap in BCFTs.

\subsection{Torus partition function with a hole: bulk-boundary OPE coefficients}\label{sec:hole}
\newsavebox{\boxga}
\sbox{\boxga}{\includegraphics[width=120pt]{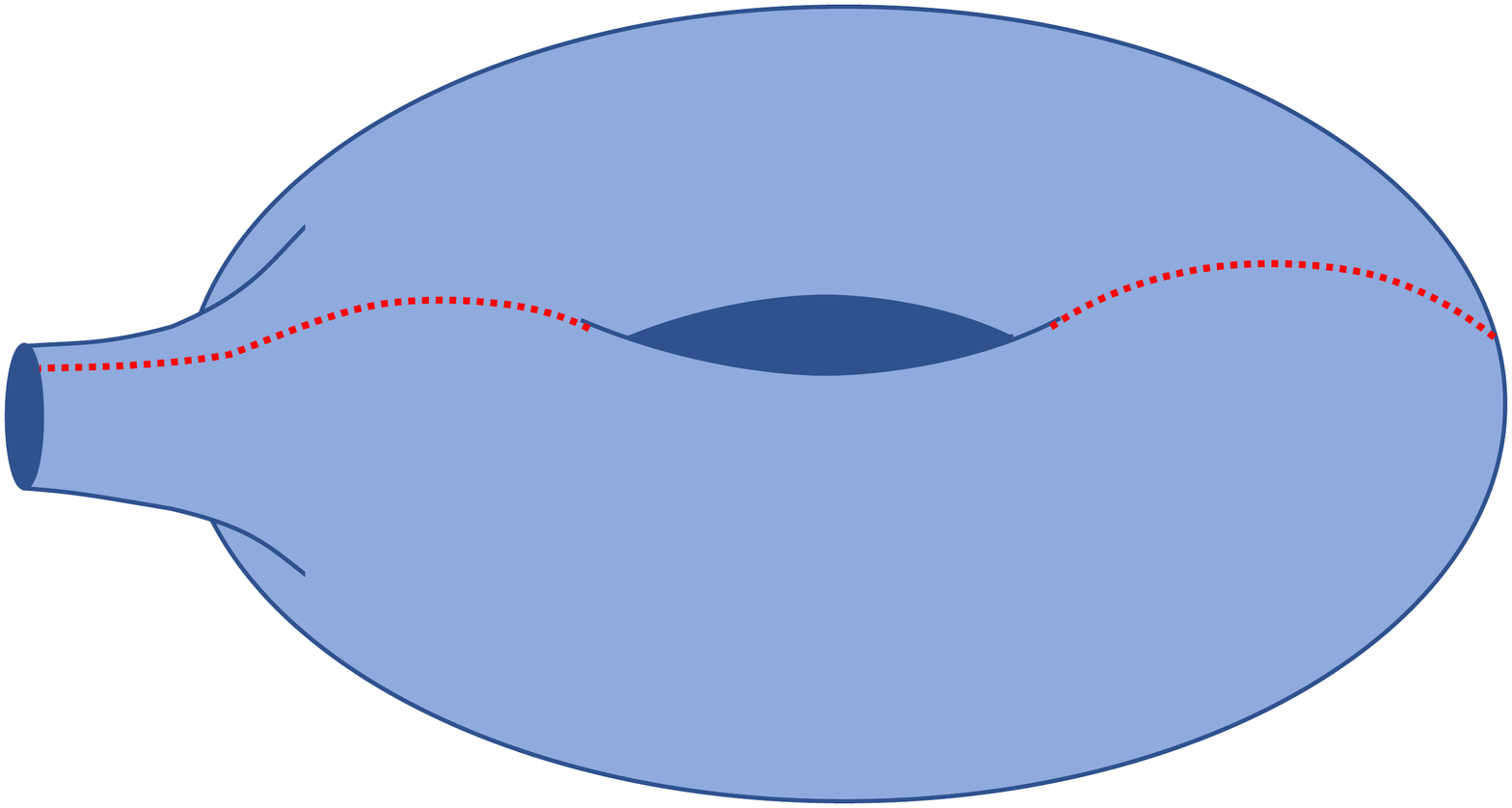}}
\newlength{\boxgaw}
\settowidth{\boxgaw}{\usebox{\boxga}} 

\newsavebox{\boxgb}
\sbox{\boxgb}{\includegraphics[width=120pt]{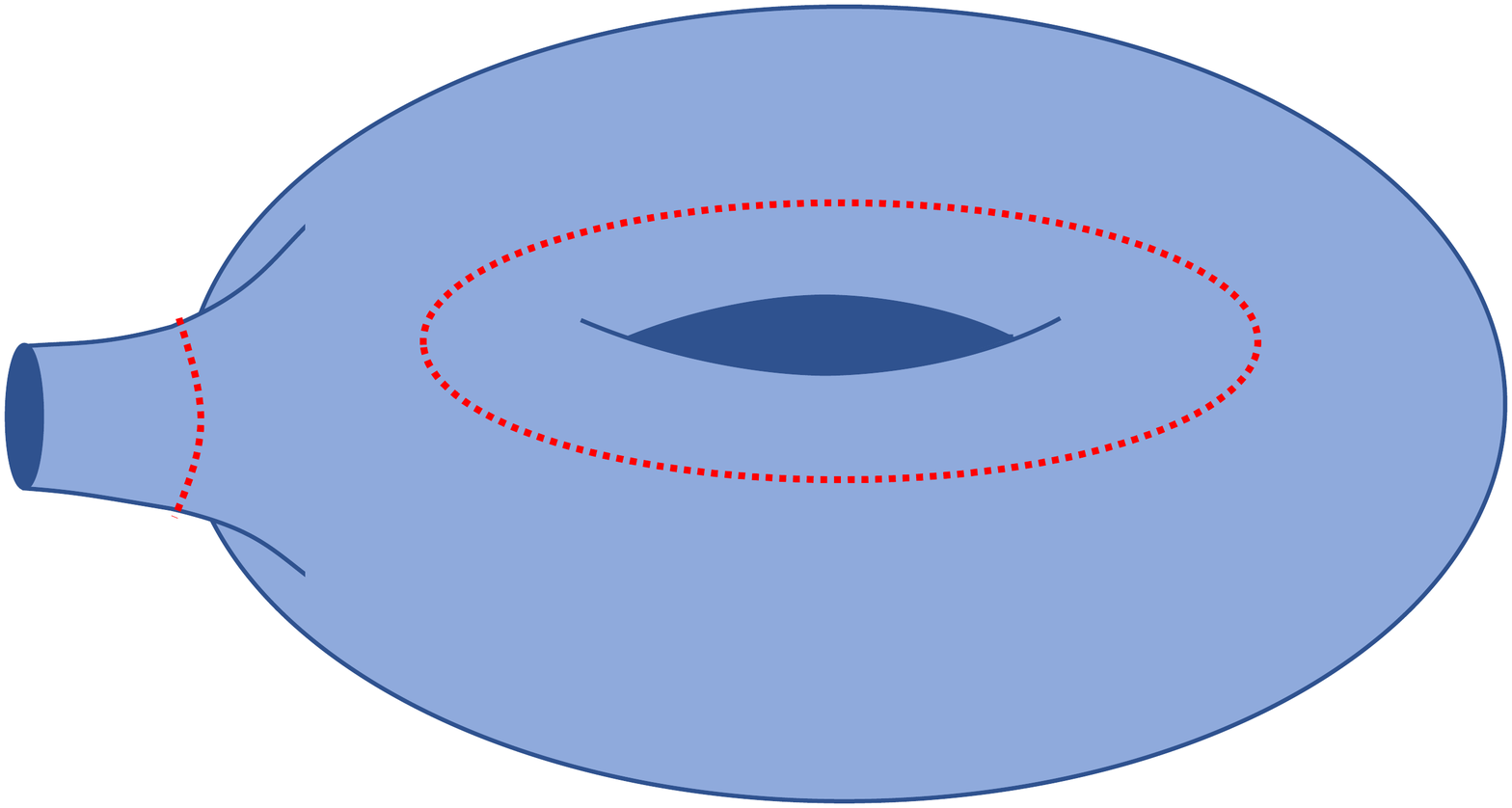}}
\newlength{\boxgbw}
\settowidth{\boxgbw}{\usebox{\boxgb}} 

\newsavebox{\boxholea}
\sbox{\boxholea}{\includegraphics[width=120pt]{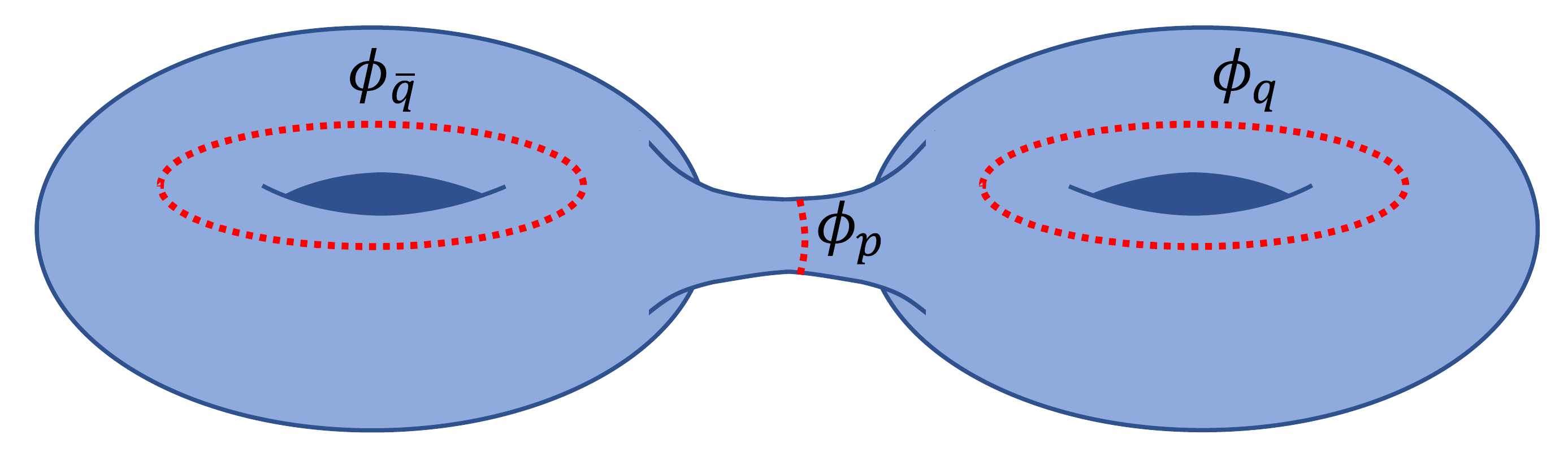}}
\newlength{\boxholeaw}
\settowidth{\boxholeaw}{\usebox{\boxholea}} 

\newsavebox{\boxholeb}
\sbox{\boxholeb}{\includegraphics[width=120pt]{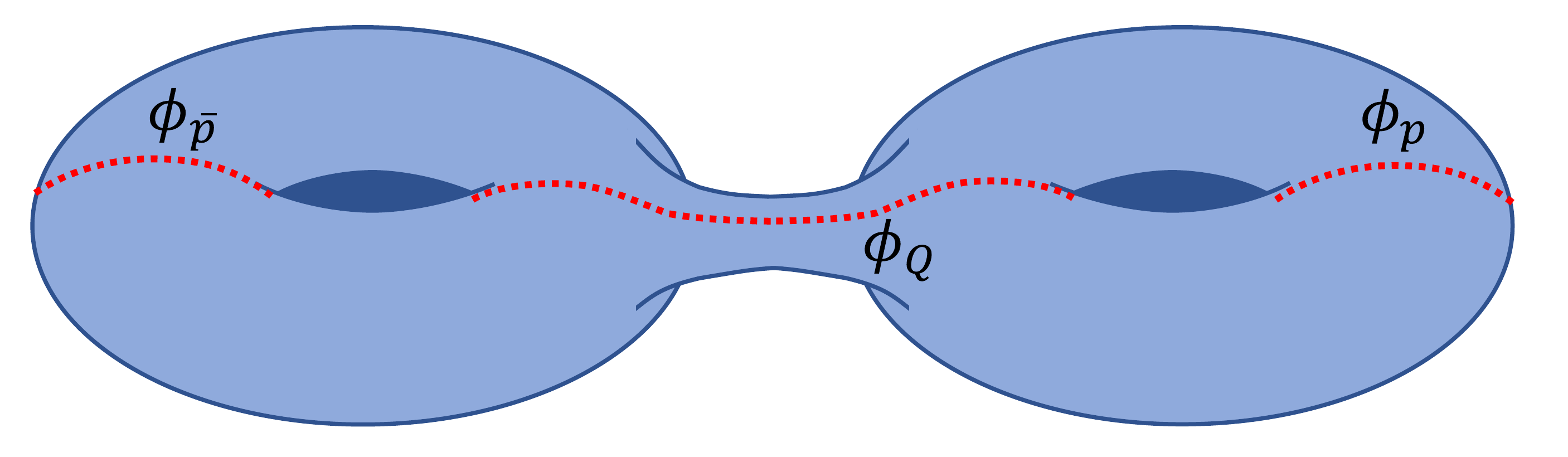}}
\newlength{\boxholebw}
\settowidth{\boxholebw}{\usebox{\boxholeb}} 

\newsavebox{\boxholec}
\sbox{\boxholec}{\includegraphics[width=120pt]{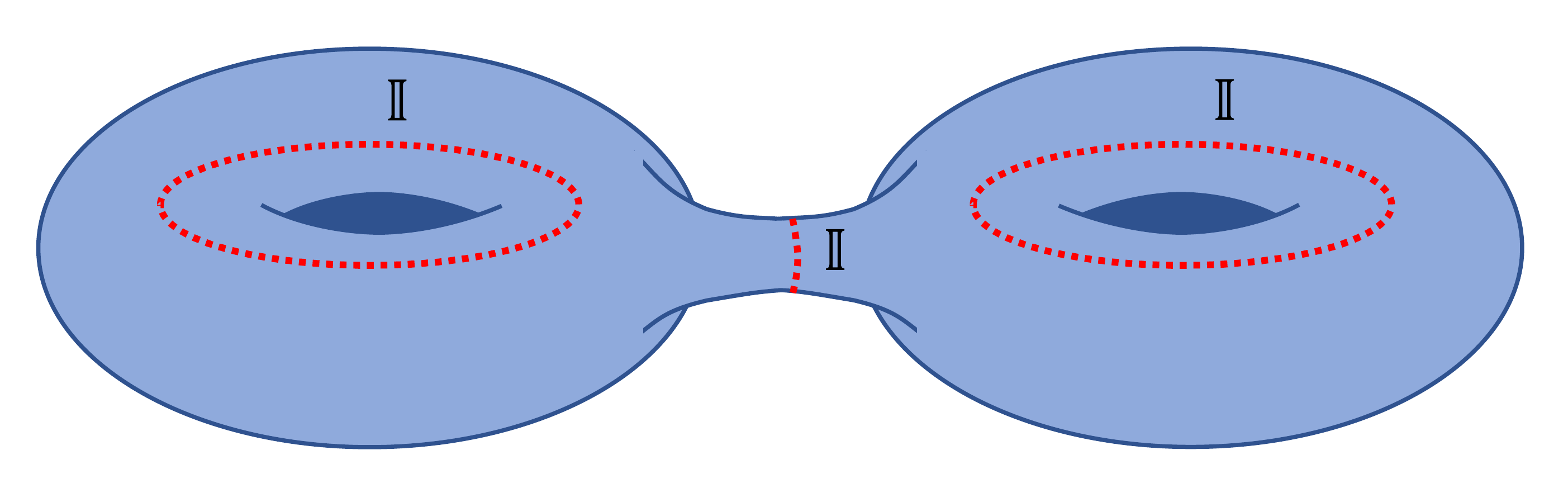}}
\newlength{\boxholecw}
\settowidth{\boxholecw}{\usebox{\boxholec}} 

\newsavebox{\boxholed}
\sbox{\boxholed}{\includegraphics[width=120pt]{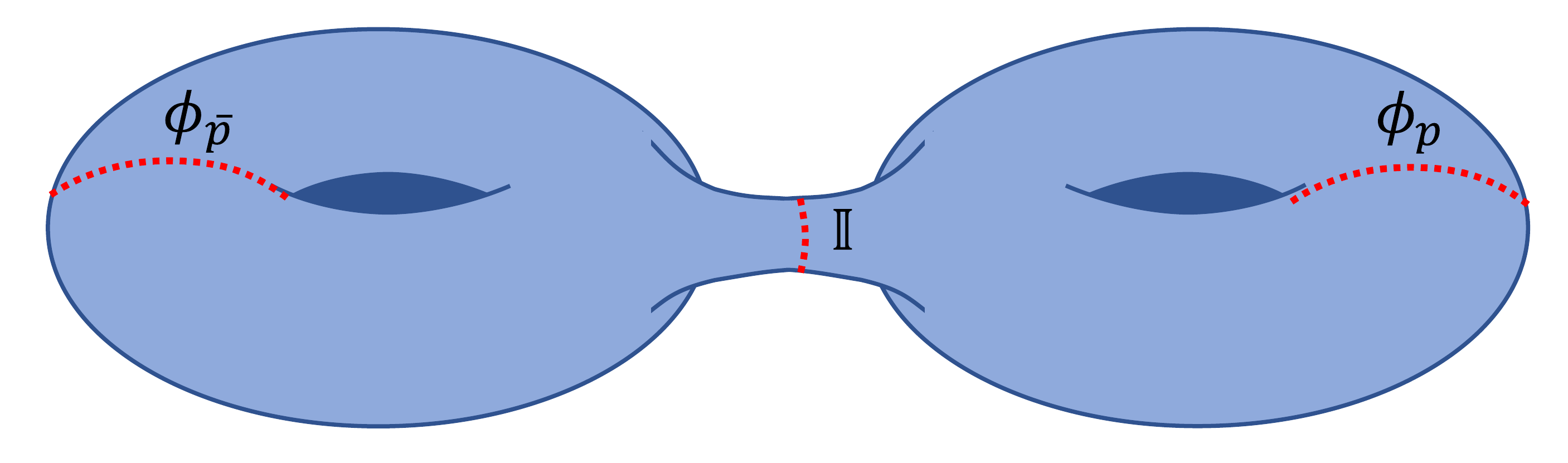}}
\newlength{\boxholedw}
\settowidth{\boxholedw}{\usebox{\boxholed}} 

For a torus partition function with a hole, we can consider the following two choices of how to cut,
\begin{equation}
\parbox{\boxgaw}{\usebox{\boxga}} = \parbox{\boxgbw}{\usebox{\boxgb}}.
\end{equation}
The corresponding bootstrap equation can be expressed by
\begin{equation}
\begin{aligned}
&
\int \dd \a_p \ 
\int \dd \bar{\a}_p \ 
\int \dd \a_Q \ 
\rho (\a_p,\bar{\a}_p)
\rho (\a_Q)
\overline{C_{pQ} C_{pQ} }
\parbox{\boxholebw}{\usebox{\boxholeb}}
\\
&=
\int \dd \a_p \ 
\int \dd \a_q \ 
\int \dd \bar{\a}_q \ \ 
\rho (\a_p)
\rho (\a_q,\bar{\a}_q)
\overline{C_{pqq}C_{p\mathbb{I}}}
\parbox{\boxholeaw}{\usebox{\boxholea}}.
\end{aligned}
\end{equation}
Here, the OPE coefficients $\overline{C_{pQ} C_{pQ} }$ are functions of three parameters $(\a_p,\bar{\a}_p,\a_Q)$,
unlike the examples in Section \ref{sec:bootstrap}. For example, in the zero-point cylinder partition function (\ref{eq:cylinder}),
the OPE coefficients $\overline{C_{p\mathbb{I}} C_{p\mathbb{I}} }$ have only only one parameter because of $\a_p=\bar{a}_p$.
The OPE coefficients $C_{pqq}$ can also have more parameters because the bulk primaries $q$ with the Liouville momentum $\a_q \neq  \bar{a}_q$ can exist in this case.
Let us consider the limit where the hole of the right torus shrinks whereas the left torus remains.
If we assume our theory to be a  $c>1$ compact CFT with no extended current,
the density of the bulk-primaries satisfies
\begin{equation}
\begin{aligned}
\rho (0,\bar{\a}_p)
&=\left\{
    \begin{array}{ll}
      1 ,& \text{if \ }\bar{\a}_p=0   ,\\
      0 ,& \text{otherwise }   .\\
    \end{array}
  \right.\\
\end{aligned}
\end{equation}
Therefore, if the right torus is dominated by the vacuum primary, then the left torus is also dominated by the vacuum.
This is completely the same strategy as the light-cone bootstrap.
Let us take the limit where the hole of the right torus shrinks and the bridge between the two tori has a large distance.
By using the fusion transformation, we obtain
\begin{equation}
\begin{aligned}
&
\int \dd \a_p \ 
\int \dd \bar{\a}_p \ 
\int \dd \a_Q \ 
\rho (\a_p,\bar{\a}_p)
\rho (\a_Q)
\overline{C_{pQ} C_{pQ} }
\parbox{\boxholebw}{\usebox{\boxholeb}}
\\
&\simeq
\parbox{\boxholecw}{\usebox{\boxholec}}\\
&=
\int \dd \a_p \ 
\int \dd \bar{\a}_p \ 
S_{0p}
S_{0\bar{p}}
\parbox{\boxholedw}{\usebox{\boxholed}}\\
&=
\int \dd \a_p \ 
\int \dd \bar{\a}_p \ 
\int \dd \a_Q \ 
S_{0p}
S_{0\bar{p}}
{\bold F}_{0, \a_Q} 
   \left[
    \begin{array}{cc}
    \a_p   & \a_p  \\
     \bar{\a}_p  &   \bar{\a}_p \\
    \end{array}
  \right]
\parbox{\boxholebw}{\usebox{\boxholeb}}.
\end{aligned}
\end{equation}
Thus, the universal formula for the bulk-boundary OPE coefficients is given by

\begin{equation}
\rho (\a_p,\bar{\a}_p)
\rho (\a_Q)
\overline{C_{pQ} C_{pQ} }
\simeq
S_{0p}
S_{0\bar{p}}
{\bold F}_{0, \a_Q} 
   \left[
    \begin{array}{cc}
    \a_p   & \a_p  \\
     \bar{\a}_p  &   \bar{\a}_p \\
    \end{array}
  \right],
  \ \ \ \ \ \ h_p, h_Q \to \infty.
\end{equation}
Here the anti-chiral conformal dimension $\bar{h}_p$ can be free chosen, therefore, this formula tells us the bulk-boundary OPE in the large-spin limit.
By using the Cardy formula for the bulk primary operators and boundary primary operators, we obtain
\begin{equation}
\overline{C_{pQ} C_{pQ} }
\simeq
g^{-2}
S_{0Q}^{-1}
{\bold F}_{0, \a_Q} 
   \left[
    \begin{array}{cc}
    \a_p   & \a_p  \\
     \bar{\a}_p  &   \bar{\a}_p \\
    \end{array}
  \right],
  \ \ \ \ \ \ h_p, h_Q \to \infty.
\end{equation}

\subsection{Boundary two-point function on a cylinder: bulk-boundary OPE coefficients}
\newsavebox{\boxha}
\sbox{\boxha}{\includegraphics[width=70pt]{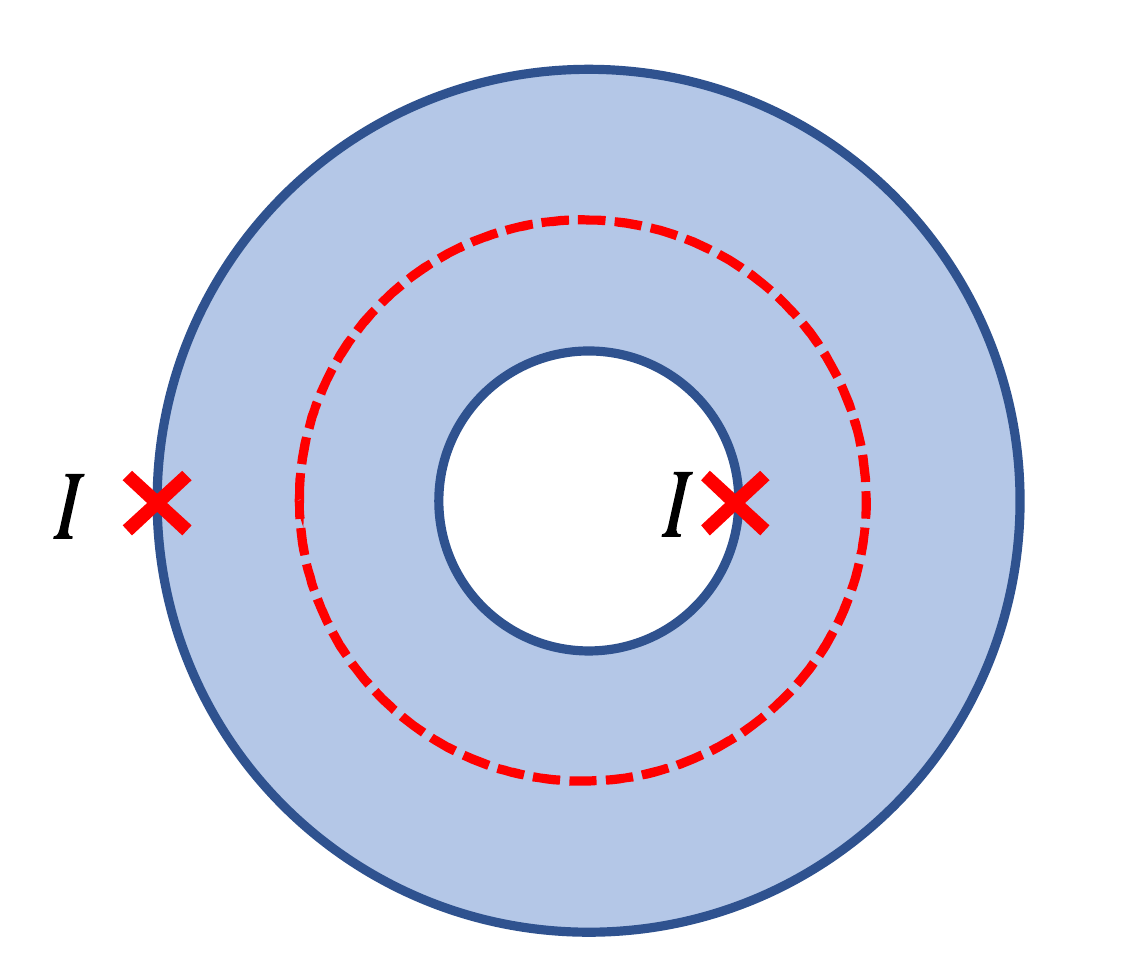}}
\newlength{\boxhaw}
\settowidth{\boxhaw}{\usebox{\boxha}} 

\newsavebox{\boxhb}
\sbox{\boxhb}{\includegraphics[width=70pt]{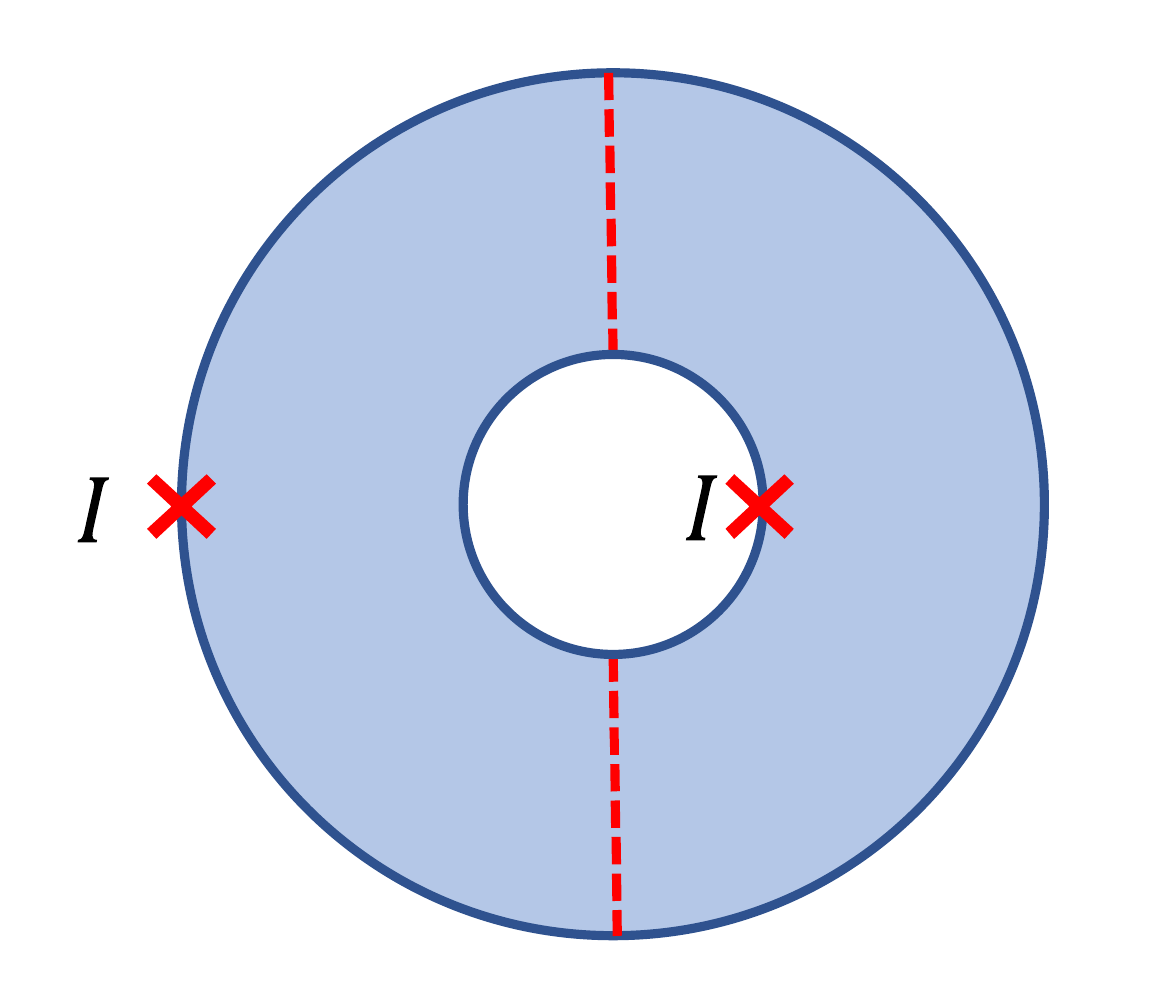}}
\newlength{\boxhbw}
\settowidth{\boxhbw}{\usebox{\boxhb}} 

\newsavebox{\boxtwolighta}
\sbox{\boxtwolighta}{\includegraphics[width=70pt]{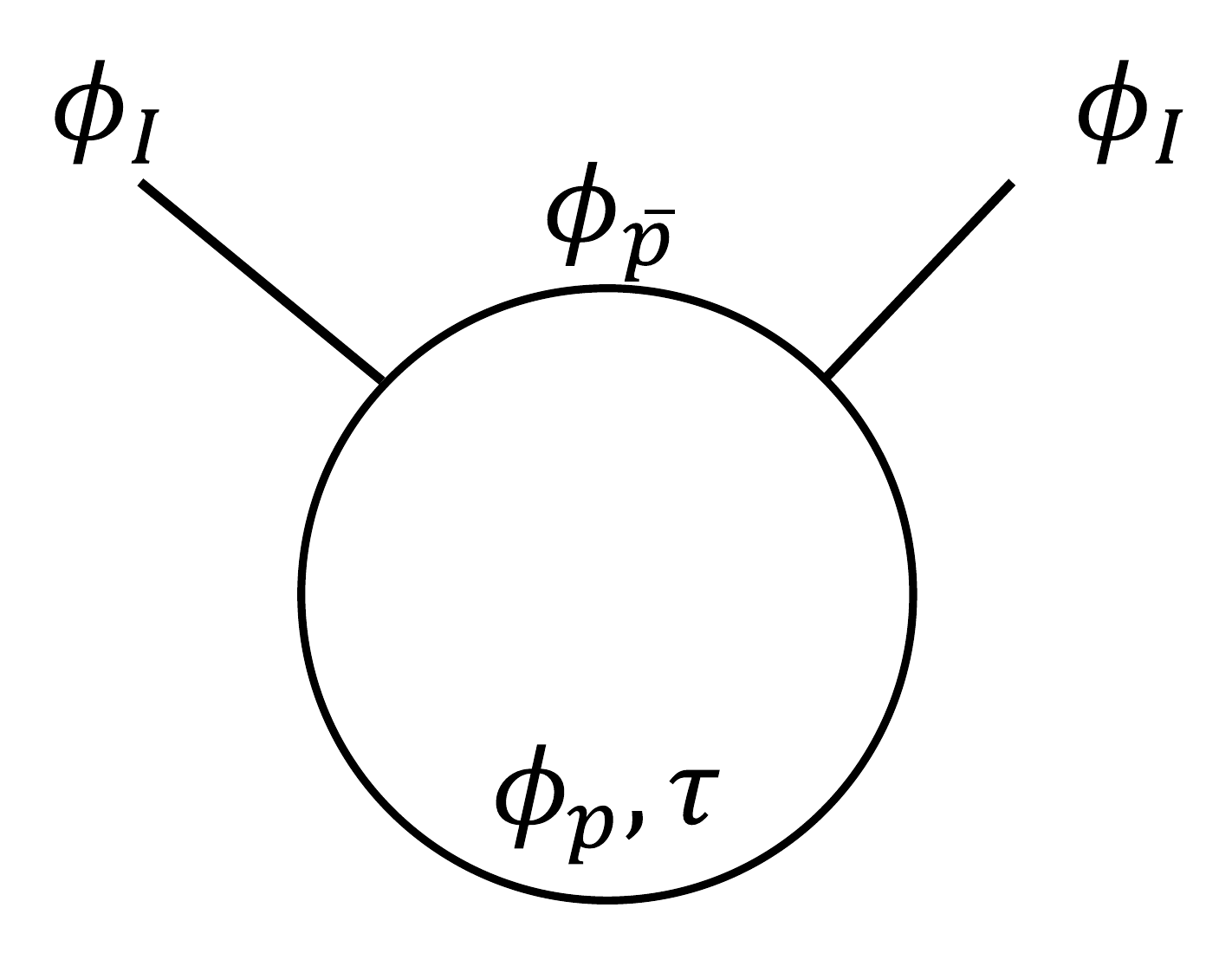}}
\newlength{\boxtwolightaw}
\settowidth{\boxtwolightaw}{\usebox{\boxtwolighta}} 

\newsavebox{\boxtwolightb}
\sbox{\boxtwolightb}{\includegraphics[width=70pt]{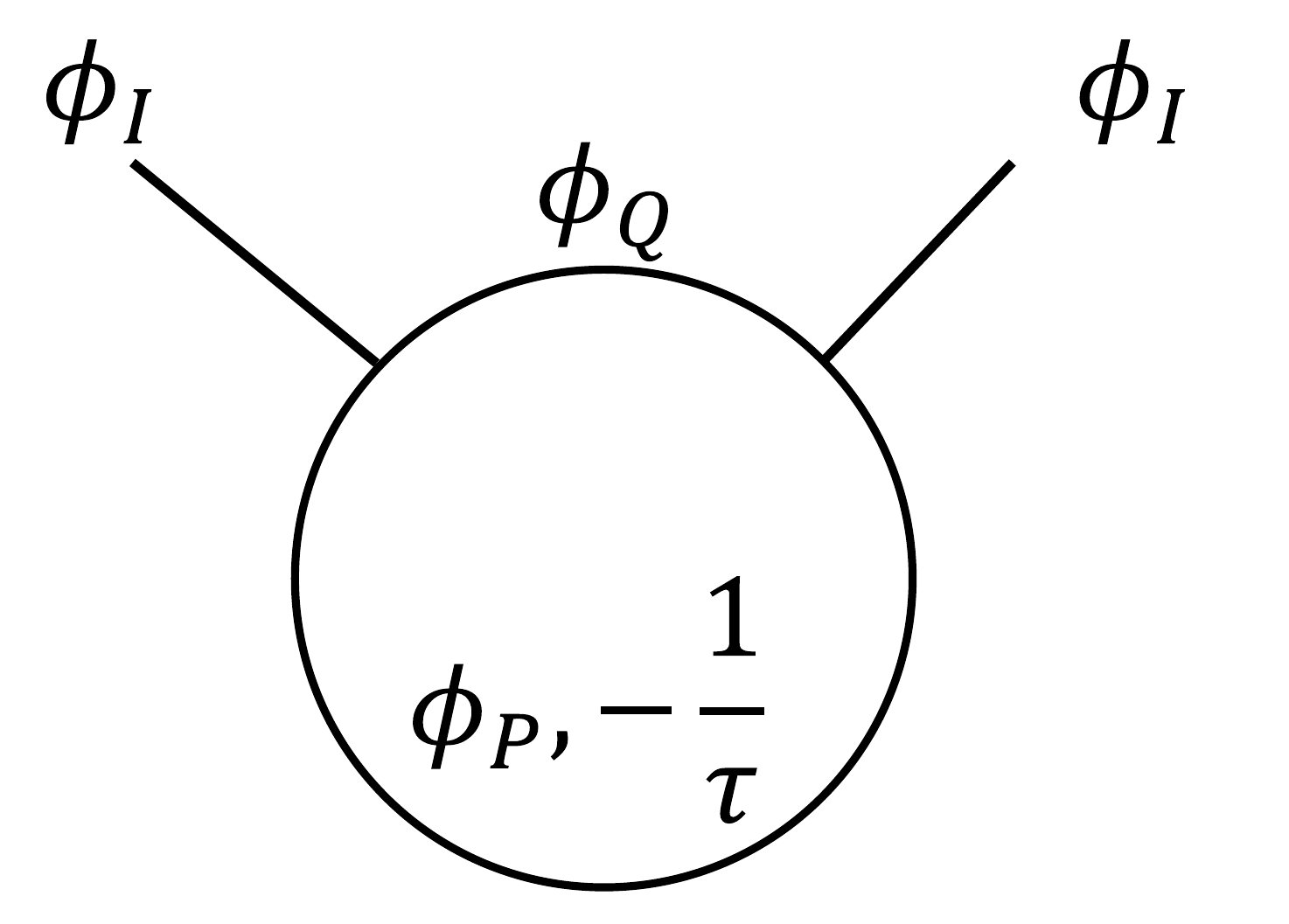}}
\newlength{\boxtwolightbw}
\settowidth{\boxtwolightbw}{\usebox{\boxtwolightb}} 

\newsavebox{\boxtwolightc}
\sbox{\boxtwolightc}{\includegraphics[width=50pt]{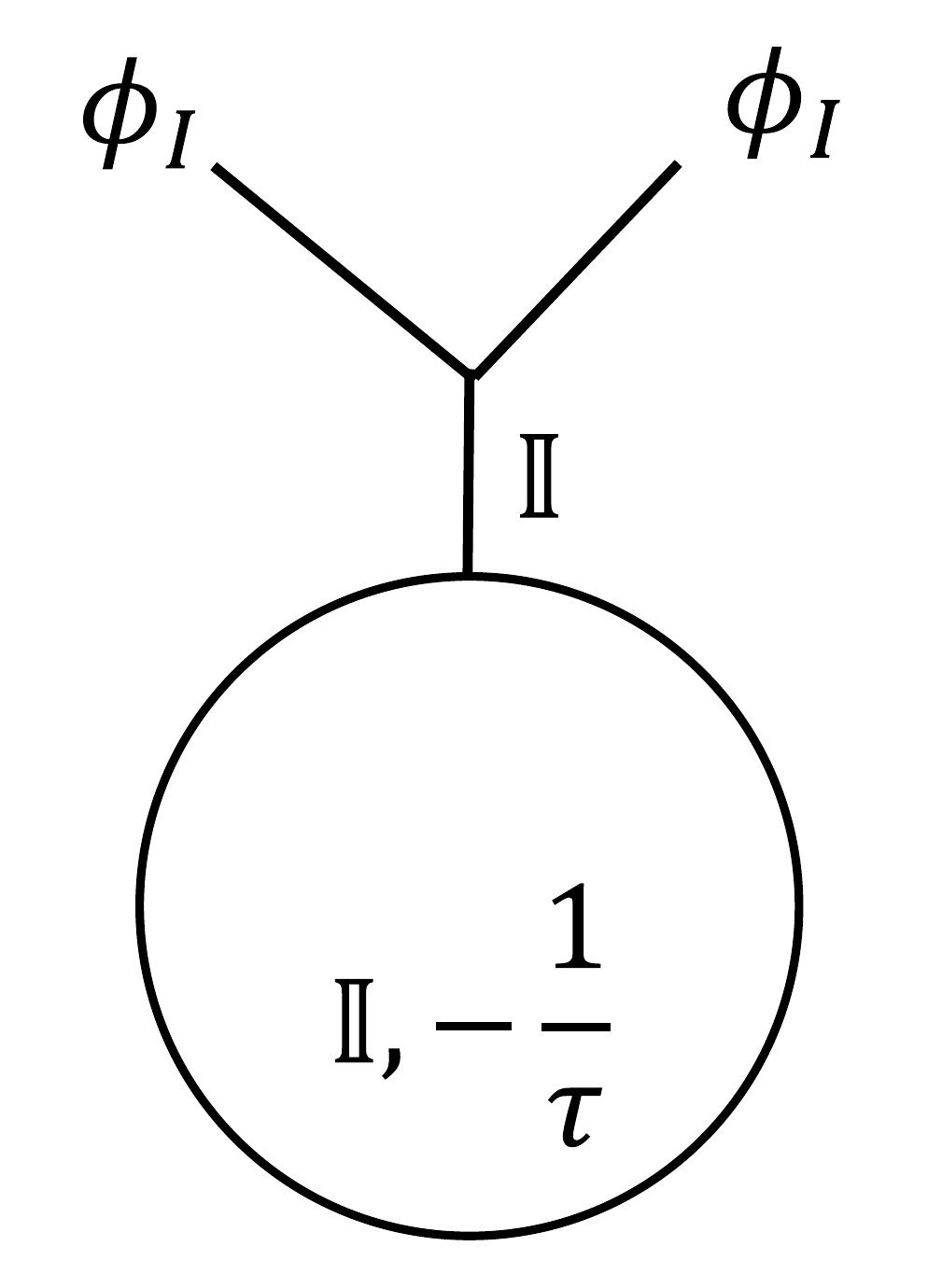}}
\newlength{\boxtwolightcw}
\settowidth{\boxtwolightcw}{\usebox{\boxtwolightc}} 

\newsavebox{\boxtwolightd}
\sbox{\boxtwolightd}{\includegraphics[width=50pt]{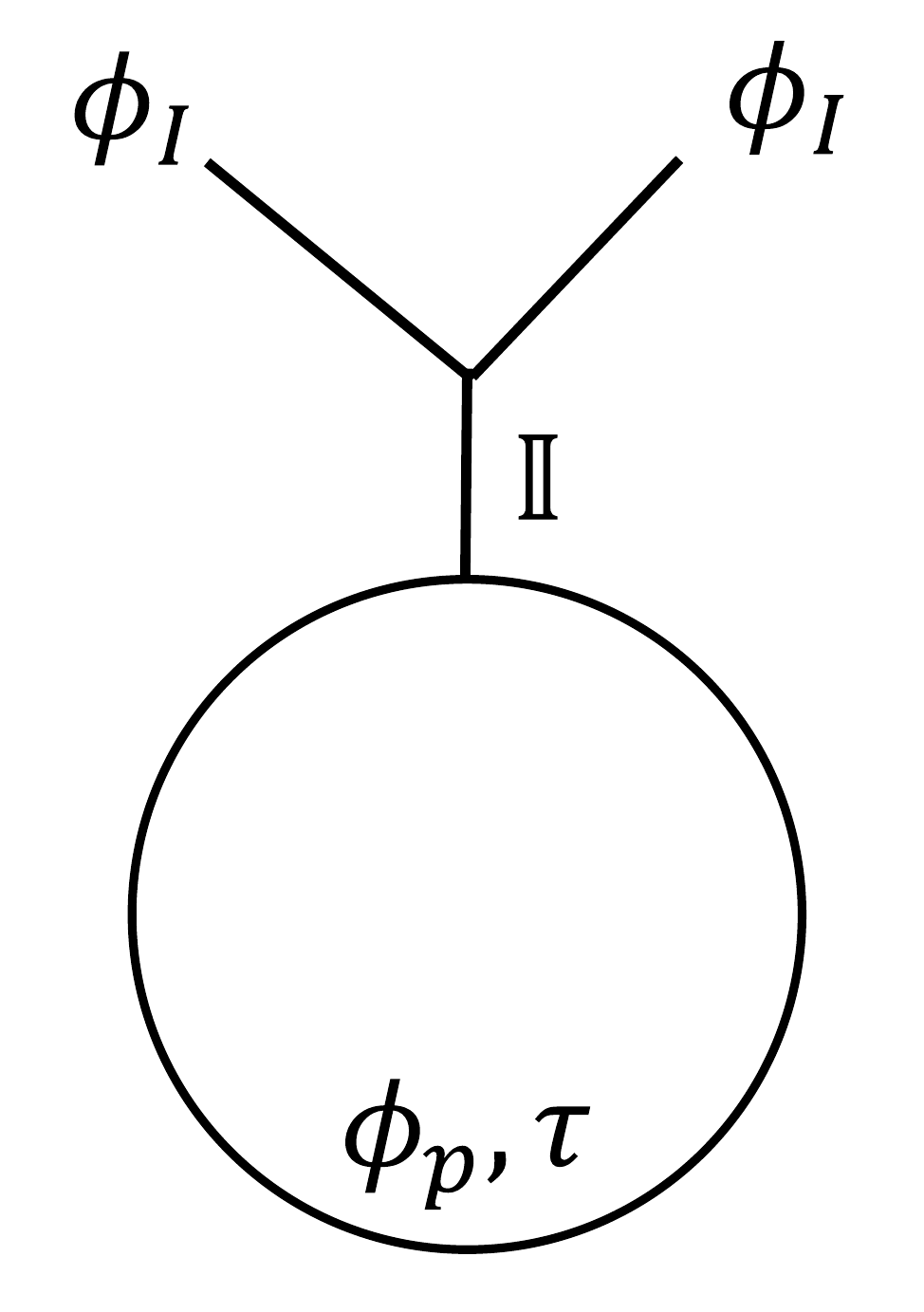}}
\newlength{\boxtwolightdw}
\settowidth{\boxtwolightdw}{\usebox{\boxtwolightd}} 

For a boundary two-point function on a cylinder, we can consider the following two choices of how to cut,
\begin{equation}
\parbox{\boxhaw}{\usebox{\boxha}} = \parbox{\boxhbw}{\usebox{\boxhb}}.
\end{equation}
The corresponding bootstrap equation can be expressed by
\begin{equation}
\begin{aligned}
&
g^2
\int \dd \a_p \ 
\int \dd \bar{\a}_p \ 
\rho (\a_p, \bar{\a}_p)
\overline{C_{pI}C_{pI}}
\parbox{\boxtwolightaw}{\usebox{\boxtwolighta}}\\
&=
\int \dd \a_P \ 
\int \dd \a_Q \ 
\rho (\a_P)
\rho (\a_Q)
\overline{C_{PQI} C_{PQI} }
\parbox{\boxtwolightbw}{\usebox{\boxtwolightb}}.
\end{aligned}
\end{equation}
In the  $\tau \to i0$ limit, the right-hand side can be approximated by the vacuum block.
By using the fusion transformation, we obtain
\begin{equation}\label{eq:twotorus}
\begin{aligned}
&
g^2
\int \dd \a_p \ 
\int \dd \bar{\a}_p \ 
\rho (\a_p, \bar{\a}_p)
\overline{C_{pI}C_{pI}}
\parbox{\boxtwolightaw}{\usebox{\boxtwolighta}}\\
&\simeq
\parbox{\boxtwolightcw}{\usebox{\boxtwolightc}}\\
&=
\int \dd \a_p \ 
S_{0p}
\parbox{\boxtwolightdw}{\usebox{\boxtwolightd}}\\
&=
\int \dd \a_p \ 
\int \dd \bar{\a}_p \ 
S_{0p}
{\bold F}_{0, \bar{\a}_p} 
   \left[
    \begin{array}{cc}
    \a_I   & \a_I  \\
     \a_p  &   \a_p \\
    \end{array}
  \right]
\parbox{\boxtwolightaw}{\usebox{\boxtwolighta}}.
\end{aligned}
\end{equation}
Thus, the universal formula for the bulk-boundary OPE coefficients is given by
\begin{equation}
\rho (\a_p, \bar{\a}_p)
\overline{C_{pI}C_{pI}}
\simeq
g^{-2}
S_{0p}
{\bold F}_{0, \bar{\a}_p} 
   \left[
    \begin{array}{cc}
    \a_I   & \a_I  \\
     \a_p  &   \a_p \\
    \end{array}
  \right],
  \ \ \ \ \ \ h_p \to \infty,
\end{equation}
or equivalently,
\begin{equation}
\overline{C_{pI}C_{pI}}
\simeq
g^{-2}
\pa{\overline{S_{0p}}}^{-1}
{\bold F}_{0, \bar{\a}_p} 
   \left[
    \begin{array}{cc}
    \a_I   & \a_I  \\
     \a_p  &   \a_p \\
    \end{array}
  \right],
  \ \ \ \ \ \ h_p \to \infty.
\end{equation}

Note that if we take the $\tau \to i \infty$ limit, the first line of (\ref{eq:twotorus}) is dominated by the vacuum block with $h_p = \bar{h}_p = 0$.
At first glance, this seems strange because the vacuum torus block with two external operators should vanish unless the external operators are the vacuum.
Indeed, this is consistent with $C_{\mathbb{I} I} = 0$ if $I \neq \mathbb{I}$.

\subsection{Re-derivation of CFT universality from BCFT}
\newsavebox{\boxia}
\sbox{\boxia}{\includegraphics[width=120pt]{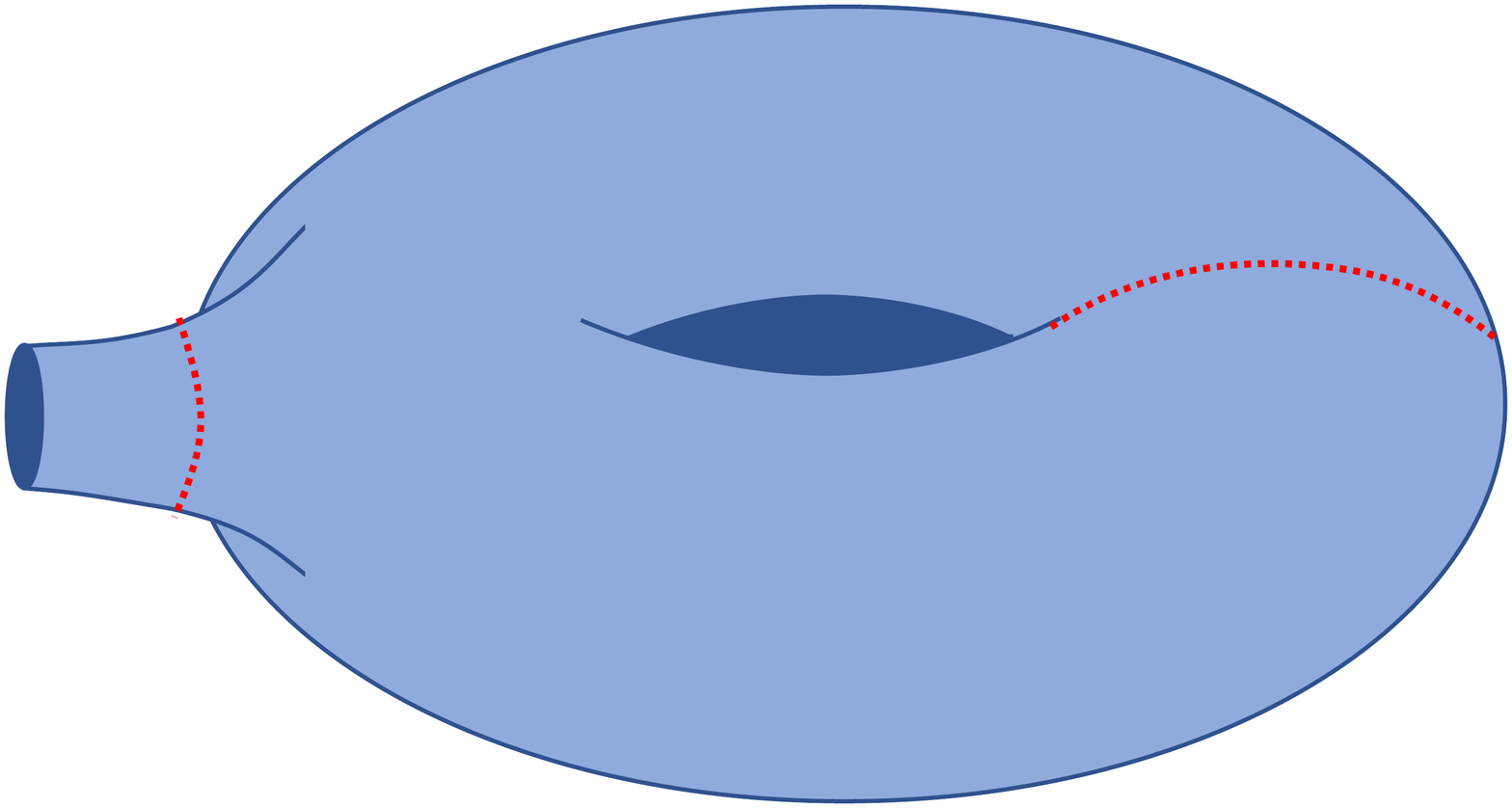}}
\newlength{\boxiaw}
\settowidth{\boxiaw}{\usebox{\boxia}} 

\newsavebox{\boxib}
\sbox{\boxib}{\includegraphics[width=120pt]{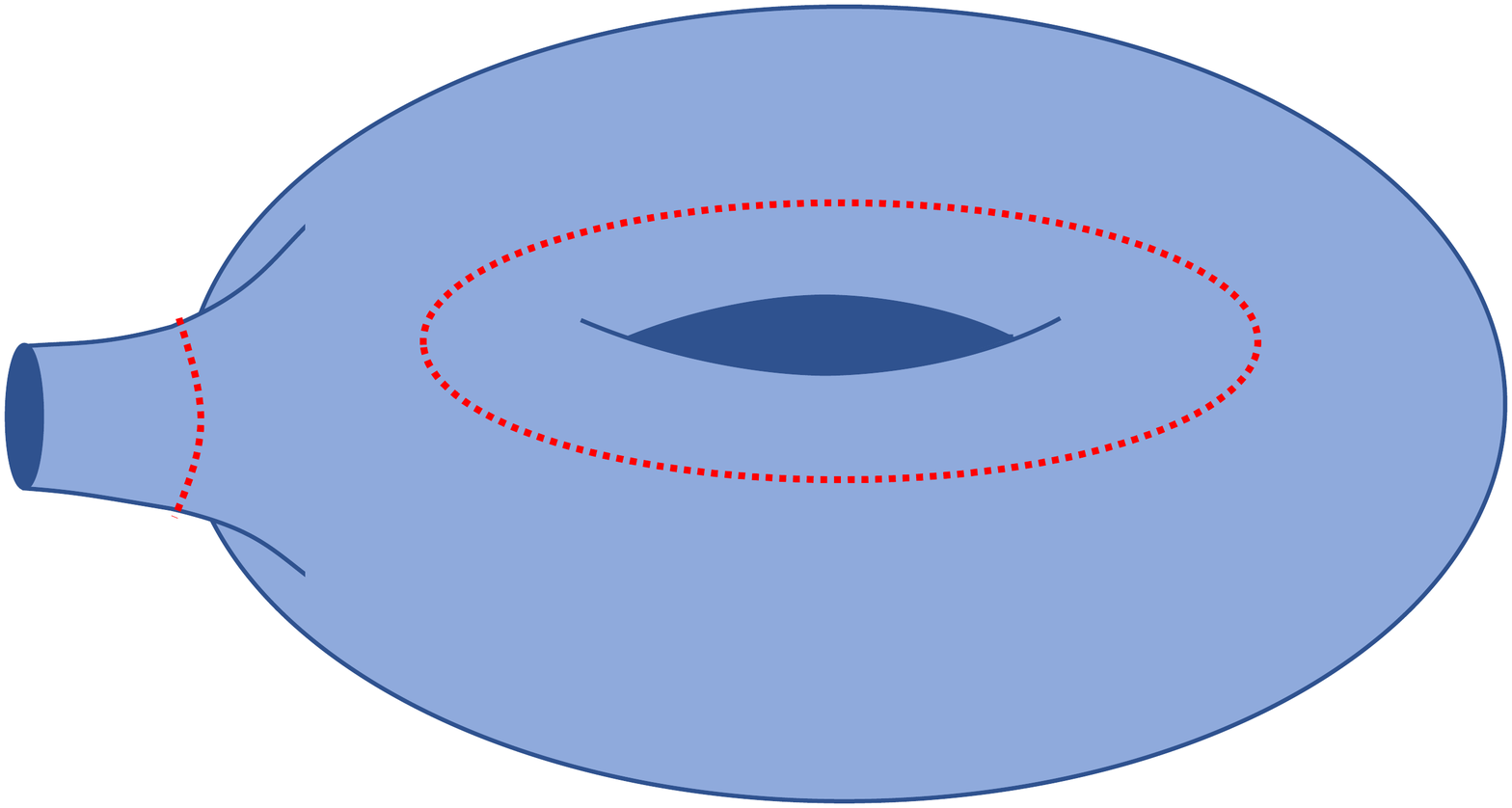}}
\newlength{\boxibw}
\settowidth{\boxibw}{\usebox{\boxib}} 

\newsavebox{\boxmoda}
\sbox{\boxmoda}{\includegraphics[width=120pt]{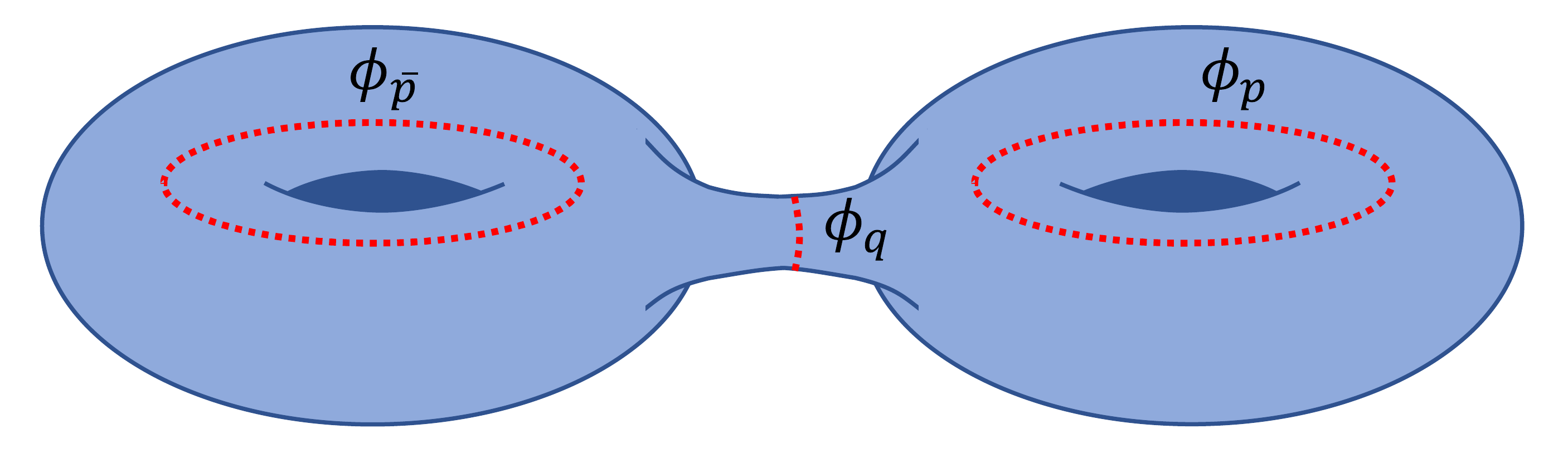}}
\newlength{\boxmodaw}
\settowidth{\boxmodaw}{\usebox{\boxmoda}} 

\newsavebox{\boxmodb}
\sbox{\boxmodb}{\includegraphics[width=120pt]{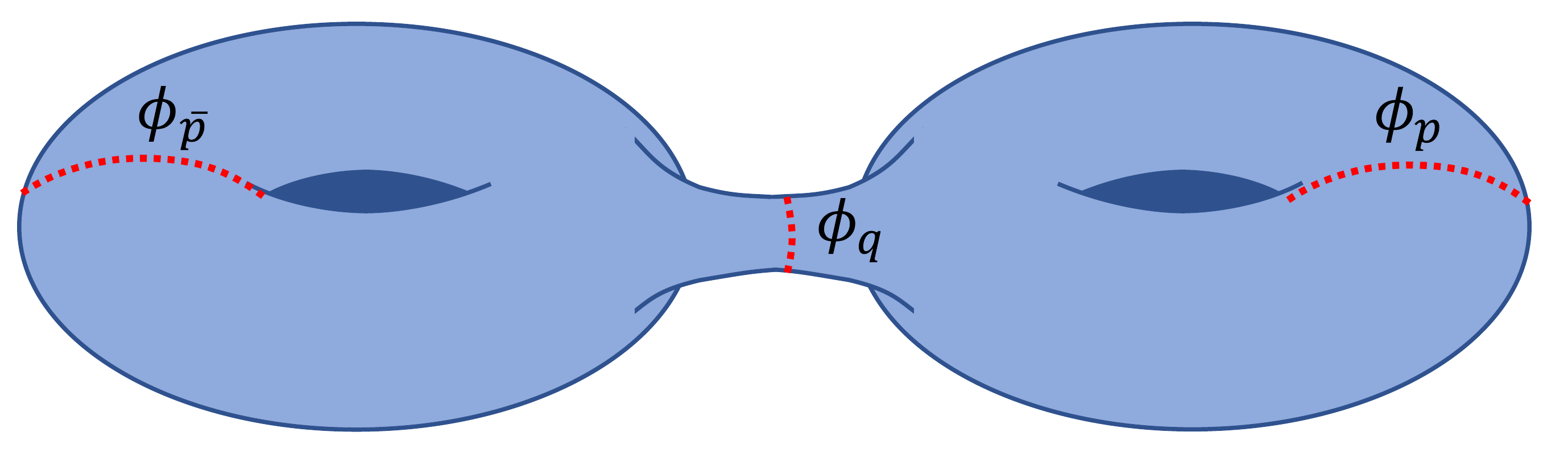}}
\newlength{\boxmodbw}
\settowidth{\boxmodbw}{\usebox{\boxmodb}} 

In the last of this section, we will comment on the relation between the light-cone bootstrap in CFTs and BCFTs.
In fact, we can always obtain a bootstrap equation in CFTs as a special case of a bootstrap equation in BCFTs.
Let us show the simplest case by using a torus partition function with a hole.
For this partition function, we can consider the following two choices of how to cut,
\begin{equation}
\parbox{\boxiaw}{\usebox{\boxia}} = \parbox{\boxibw}{\usebox{\boxib}}.
\end{equation}
The corresponding bootstrap equation can be expressed as
\begin{equation}
\begin{aligned}
&
\int \dd \a_p \ 
\int \dd \bar{\a}_p \ 
\int \dd \a_q \ 
\rho (\a_p,\bar{\a}_p)
\rho (\a_q)
\overline{C_{q \mathbb{I}} C_{q \mathbb{I}} }
\parbox{\boxmodbw}{\usebox{\boxmodb}}
\\
&=
\int \dd \a_p \ 
\int \dd \bar{\a}_p \ 
\int \dd \a_q \ 
\rho (\a_p,\bar{\a}_p)
\rho (\a_q)
\overline{C_{q \mathbb{I}} C_{q \mathbb{I}} }
\parbox{\boxmodaw}{\usebox{\boxmoda}}.
\end{aligned}
\end{equation}
In the large distance limit between two tori, this bootstrap equation reduces to
\begin{equation}
\begin{aligned}
&
\int \dd \a_p \ 
\int \dd \bar{\a}_p \ 
\rho (\a_p,\bar{\a}_p)
\chi_{h_p}(\tau)
\chi_{\bar{h}_p} (\bar{\tau})
\\
&=
\int \dd \a_p \ 
\int \dd \bar{\a}_p \ 
\rho (\a_p,\bar{\a}_p)
\chi_{h_p} \pa{-\fr{1}{\tau}}
\chi_{\bar{h}_p} \pa{-\fr{1}{\bar{\tau}}}.
\end{aligned}
\end{equation}
This is just the modular bootstrap equation without boundary.
In this sense, the limit considered in Section \ref{sec:hole} is definitely the light-cone limit,
and the result is a slight generalization of the light-cone modular bootstrap.

\section{Implication for Braneworld Holography}\label{sec:braneworld}
The CFT${}_2$ with boundaries has new ingredients, the boundary primary operators, which live only on the boundary.
We may think of the boundary primaries as primaries in a CFT${}_1$.
If the duality between the boundary and the braneworld is true,
the boundary primaries should satisfy some properties expected from the gravity side as explained in the introduction.
In this section, we will present some thoughts on them.

\subsection{Cardy formula}
We have shown that the density of the boundary primary states at high energy is given by
\begin{equation}
\rho(h_P) \simeq \ex{2\pi\sqrt{\fr{c}{6}\pa{h_P-\fr{c}{24}}}},  \ \ \ \ \ \ \ h_P \to \infty.
\end{equation}
In other words, the entropy is given by
\begin{equation}
S(E) \simeq 2\pi\sqrt{\fr{c}{6}E},
\end{equation}
where the energy for the boundary primaries is defined as $E=h-\fr{c}{24}$.
This is just a chiral part of the Cardy formula in 2D CFTs.
We should mention that this is not the spectrum of the CFT${}_1$ (i.e, the braneworld) itself but the CFT${}_1$ coupled to the heat bath CFT${}_2$, like the island model.
Note that this result matches with the entropy of a 2D gravity \cite{Cadoni2002}.

So far, the Cardy formula for the open strings is universal in any CFT.
Let us consider a particular class of CFTs, which have sparse spectrum and large central charge.
These properties are expected if a given CFT is dual to classical gravity.
If we restrict ourselves to such a CFT, we may extend the validity regime of our universal formula to $E=O(c)$.
The reason is because the vacuum dominance is crucial in our derivation and naively the sparseness and the large $c$ may support this vacuum dominance beyond the limit $E \to \infty$.
In fact, this extension of the validity regime has been shown for the Cardy formula in CFTs without boundaries in \cite{Hartman2014},
\begin{equation}
S(E) \simeq 2\pi\sqrt{\fr{c}{3}E},   \ \ \ \ \ \text{if} \ \ \ E>\fr{c}{12},
\end{equation}
where the energy $E$ is defined as $E=h+\bar{h}-\fr{c}{12}$.
One can expect that something similar happens in BCFTs.
In that, if we assume that the boundary primary spectrum satisfies an analog of the sparseness condition (see (2.14) in \cite{Hartman2014}),
then the Cardy formula for the boundary primaries can hold in the regime $h_P=O(c)$.
In fact, this is true in some sense.
The main difference between the modular bootstrap equations for CFTs with no boundaries and BCFTs is that
the former is self-dual under the modular-S transformation but the latter is not.
Nevertheless, one can consider a symmetrized version of the cylinder partition function,
\footnote{We would like to thank Dalimil Mazac for pointing this out.}
\begin{equation}\label{eq:sym}
\begin{aligned}
Z(\tau)
\equiv
\int \dd \a_P \ 
\pa{
\rho^{bdy} (\a_P)
+
g^2
\rho^{bulk} (\a_P)
\overline{C_{P\mathbb{I}}  C_{P\mathbb{I}} }
}
\parbox{\boxtorussw}{\usebox{\boxtoruss}},
\end{aligned}
\end{equation}
where we abuse the capital letters, which describe the boundary primaries, to label the bulk primaries in the second term.
That is, $\rho^{bulk} (\a_P) (\equiv \rho^{bulk} (\a_P, \a_P)) $ is the spectrum for the ``bulk'' primary operators and
$\overline{C_{P\mathbb{I}}  C_{P\mathbb{I}} }$ are the ``bulk''-boundary OPE coefficients.
This partition function is self-dual under the modular-S transformation, therefore, we can show the validity regime extension for the holographic BCFT in the same way as \cite{Hartman2014}.
We have to mention that $\rho^{bulk} (\a_P)$ is the spectrum for the scalar bulk primaries, which is different from the spectrum as found in the original paper, $\rho^{bulk}(\Delta_P)$ with $\Delta_P = h_P + \bar{h}_P$.
Consequently, we obtain the following statement,
\begin{quote}
If we assume our theory to have large central charge and satisfy the following sparseness condition,
\begin{equation}
\rho^{bdy} (E), \rho^{bulk} (E) \leq \ex{2\pi \pa{E+\fr{c}{24}}},\ \ \ \ \ \text{if} \ \ \ E\leq 0, 
\end{equation}
then the spectrum satisfies the Cardy formula in the extended regime,
\begin{equation}
S(E) \simeq 2 \pi \sqrt{\fr{c}{6}E}, \ \ \ \ \ \text{if} \ \ \ E\geq\fr{c}{24}.
\end{equation}
Here we define $E=h-\fr{c}{24}$, following the notation in \cite{Hartman2014}.
\end{quote}
Note that the sparseness condition (the large gap condition) on the boundary primary spectrum is conjectured in \cite{Reeves2021} from a different viewpoint.

\subsection{Eigenstate Thermalization Hypothesis}\label{sec:ETH}

The eigenstate thermalization hypothesis (ETH) is one criterion of quantum chaos \cite{Srednicki1998, DAlessio2016, Mondaini2017}.
The statement is as follows:
\begin{quote}
If a given system exhibits thermalization, the matrix elements of (few-body) observables $O$ in the energy eigenstate basis should follow the equation,
\begin{equation}\label{eq:ETH}
\braket{n|O|m} = f_O(E)\delta_{nm} + \ex{-\fr{S(R)}{2}}g_O(E_n, E_m) R_{nm},
\end{equation}
where S(E) is the entropy at the every $E=\fr{E_n+E_m}{2}$ and the functions $f_O(E)$ and $g_O(E_n,E_m)$ are smooth and $\ca{O}(1)$ functions.
The matrix $R_{nm}$ is a random matrix variable following a Gaussian distribution with zero mean and unit variance.
\end{quote}
One can easily check a system satisfying the ETH thermalizes
\begin{equation}
\overline{\braket{\phi(t)|O|\phi(t)}} = \braket{O}_{\text{micro}}
\end{equation}
and
\begin{equation}
\overline{\pa{\braket{\phi(t)|O|\phi(t)}  -  \overline{\braket{\phi(t)|O|\phi(t)}} }^2 } \simeq \ex{-S(E)},
\end{equation}
where the overline denotes time averaging.

In fact, our results are consistent with this ETH in the following way.
We regard low-body operators as light operators and take the states $\ket{n}$ to be at high energy, which are responsible for thermalization.
\footnote{The ETH in 2D CFTs without boundaries was investigated by the bootstrap in \cite{Hikida2018, RomeroBermudez2018, Brehm2018, Collier2020}.}
Then, what we are interested in are the heavy-heavy-light OPE coefficients.
For convenience, we again exhibit our results for the boundary-boundary-boundary OPE coefficients in terms of the boundary primary entropy:
\begin{itemize}
\item boundary-boundary-boundary OPE coefficients for heavy-heavy-heavy weights,
\begin{equation}
\sqrt{\overline{C_{PQR}^2}} \simeq \pa{\fr{27}{16}}^{\fr{3}{2}h_P} \ex{-\fr{3}{4}S(h_P)},  \ \ \ \ \ \ \ h_P, h_Q, h_R \gg c, \abs{h_P-h_Q}, \abs{h_Q-h_R},
\end{equation}
\item boundary-boundary-boundary OPE coefficients for heavy-heavy-light weights,
\begin{equation}\label{eq:P!=Q}
\sqrt{\overline{C_{IPQ}^2}} \simeq \ex{-\fr{1}{2}S(h_P)},  \ \ \ \ \ \ \ h_P, h_Q \gg c, \abs{h_P-h_Q},
\end{equation}
\begin{equation}\label{eq:P=Q}
\overline{C_{IPP}} \simeq \ex{-\pa{1-\fr{12}{c}\sqrt{\fr{c}{6}\pa{\fr{c}{24}-h_{p_{min}}}}}S(h_P)},  \ \ \ \ \ \ \ h_P \gg c,
\end{equation}
\item boundary-boundary-boundary OPE coefficients for heavy-light-light weights,
\begin{equation}
\sqrt{\overline{C_{IJP}^2}} \simeq \pa{\fr{1}{16}}^{\fr{1}{2}h_P} \ex{-\fr{1}{4}S(h_P)},  \ \ \ \ \ \ \ h_P \gg c.
\end{equation}
\end{itemize}
One can find that the result (\ref{eq:P!=Q}) is consistent with the off-diagonal part of the ETH (\ref{eq:ETH}).
\footnote{
Note that this is not a proof of the ETH.
If our theory is really chaotic,
the exponential suppression of the OPE coefficients should be found in the average over an exponentially small window, not over an order one window.
The dependence on the size of the window can be investigated by using the Tauberian theorem as in \cite{Mukhametzhanov2020, Das2021}.
It would be interesting to consider an analog of this issue in BCFTs.
}
In general, the dimension $h_{p_{min}}$ is of order $O(1)$, therefore, in the large $c$ limit, we have
\begin{equation}
\overline{C_{IPP}} \simeq O(1).
\end{equation}
That is, although the ETH is not satisfied in general CFTs, it is satisfied in holographic CFTs.
The point is that even though the off-diagonal part (\ref{eq:P!=Q}) is universal, the diagonal part (\ref{eq:P=Q}) is theory-dependent.

It is hard to directly check the randomness in the off-diagonal part.
Nevertheless, one can indirectly check it by considering a $n$-point function on a cylinder.
\begin{figure}[H]
 \begin{center}
  \includegraphics[width=4.0cm,clip]{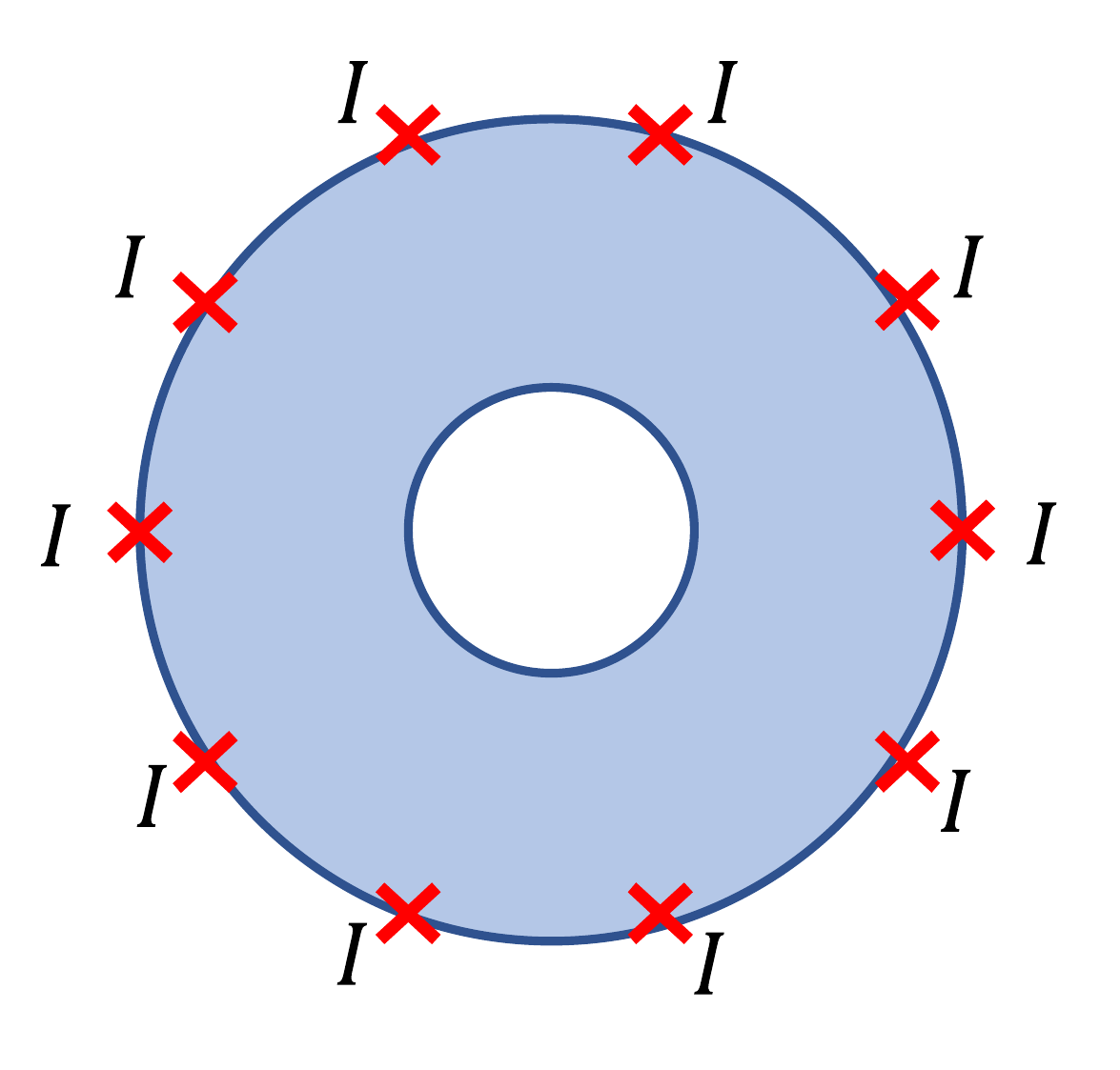}
 \end{center}
\end{figure}
\noindent
If the OPE coefficients reflect the randomness in the ETH, the following property should be satisfied
\begin{equation}
\overline{C_{IP_1 P_2} C_{IP_2 P_3} C_{IP_3 P_4} \cdots C_{IP_{n-1} P_n}} \simeq \ex{\pa{n-1}S(h_P)},
\end{equation}
where we just substitute the ETH and use the Wick theorem for the random matrices.
Indeed, the conformal bootstrap for the $n$-point function on a cylinder gives the same result.
This is one consistency check for the randomness.

Let us move on to the holographic CFT case.
Like the Cardy formula, the key point in the derivation of the OPE asymptotics is the vacuum block approximation.
Therefore, one can naturally expect that sparseness and large central charge support the extension of the vacuum block dominance
, and then the OPE asymptotics can hold beyond the regime $h_P,h_Q,h_R\gg c$.
We do not prove it here in an explicit way, but it would be interesting to give a rigorous proof of this statement.

\section{Discussion}

We propose some remaining questions and interesting future works at the end of this paper:

\begin{itemize}

\item Hellerman bound

One interesting topic is to investigate the largest gap between the vacuum and the first primary state.
The value of the largest gap is important because if there exists a CFT dual to pure gravity on AdS${}_3$,
it should have the gap $\Delta_{gap} = \fr{c}{12}+O(1)$, which corresponds to the BTZ threshold.
This problem was investigated in \cite{Hellerman2011},
which showed that the largest gap obeys $\Delta_{gap}\leq\fr{c}{6}+O(1)$ by utilizing an approximated modular bootstrap equation (improved versions can be found in  \cite{Friedan2013, Collier2018, Hartman:2019pcd, AfkhamiJeddi2019}).
It would be interesting to consider an analog of this result in BCFTs.
We would like to mention that the weakest version of the gap can be straightforwardly obtained by applying the tricks in \cite{Hellerman2011, Collier2018, Hartman:2019pcd, AfkhamiJeddi2019} to the symmetrized partition function (\ref{eq:sym}),
\begin{equation}
\min \pa{h_{gap}^{bdy},h_{gap}^{bulk}} \leq \fr{c}{12},
\end{equation}
where $h_{gap}$ is the chiral spectral gap between the vacuum and the first primary state, and $\fr{c}{12}$ comes from the Hellerman bound (we can take $\fr{c}{18.2}$ instead, following \cite{AfkhamiJeddi2019}).
It would also be interesting to consider the gap in the OPE \cite{Besken:2021eps}, which is the generalization of the Hellerman bound into four-point functions.

Another progress in this direction can be found in \cite{Benjamin2019},
which shows that we need primary fields with twist less than $\fr{c}{12}$ to be consistent with positivity of the density of states.
The key tool to show it is the fusion matrix bootstrap in the light-cone limit.
As shown in Section \ref{sec:light-cone},
we can take a generalization of the light-cone limit in BCFTs.
Therefore, it may be possible to show a similar no-go theorem from the partition function on a manifold with boundaries.

We may be able to consider an analog of the Maloney-Witten partition function \cite{Maloney2010} in BCFTs (see \cite{Suzuki2021}),
in which one can find a negative density of states.
In a similar way as \cite{Benjamin2019}, we may find a resolution of this problem by looking at the fusion matrix approach as shown in this paper.

\item Boundary condition

An important remaining task is to fix the boundary entropy in some way.
In this paper, we do not focus on the details of the boundary condition.
However, it is worth investigating the boundary condition for reproducing the braneworld.
We can find some progress in this direction in \cite{Belin2021}

In particular, one can find interesting results in \cite{Collier2021}, one of which provides the upper and lower bound on the boundary entropy by making use of  semindefinite programming using the program SDPB \cite{SimmonsDuffin2015}.
This is complementary to our results.
\footnote{
The information at low energy (including the boundary entropy) can be studied by the bootstrap equation at the fixed point as in \cite{Collier2021}. On the other hand, the information at high energy can be studied by the bootstrap equation in the high-low temperature limit as in our results.
}
It would be interesting to consider an analytic (Virasoro) bootstrap for the boundary entropy as a next step and for this purpose,
the fusion matrix could be useful.
\footnote{
As shown in \cite{Hartman:2019pcd}, one can show that the analytic bootstrap in the $sl(2,\mathbb{R})$ block basis can be utilized for giving the maximal gap for the ``Virasoro'' primary spectrum. The reason is because the Virasoro block can be expanded in terms of the $sl(2,\mathbb{R})$ blocks with positive coefficients.
However, there is a possibility that while the $sl(2,\mathbb{R})$ block expansion has negative coefficients, the Virasoro block expansion has no negative coefficients.
It implies that the analytic Virasoro bootstrap has the capacity to give more strict constraints. Indeed, the numerical Virasoro bootstrap provides more strict constraints on the maximal gap \cite{Collier2018, AfkhamiJeddi2019}.
We would like to thank David Simmons-Duffin for explaining this to us.
}

\begin{figure}[t]
 \begin{center}
  \includegraphics[width=13.0cm,clip]{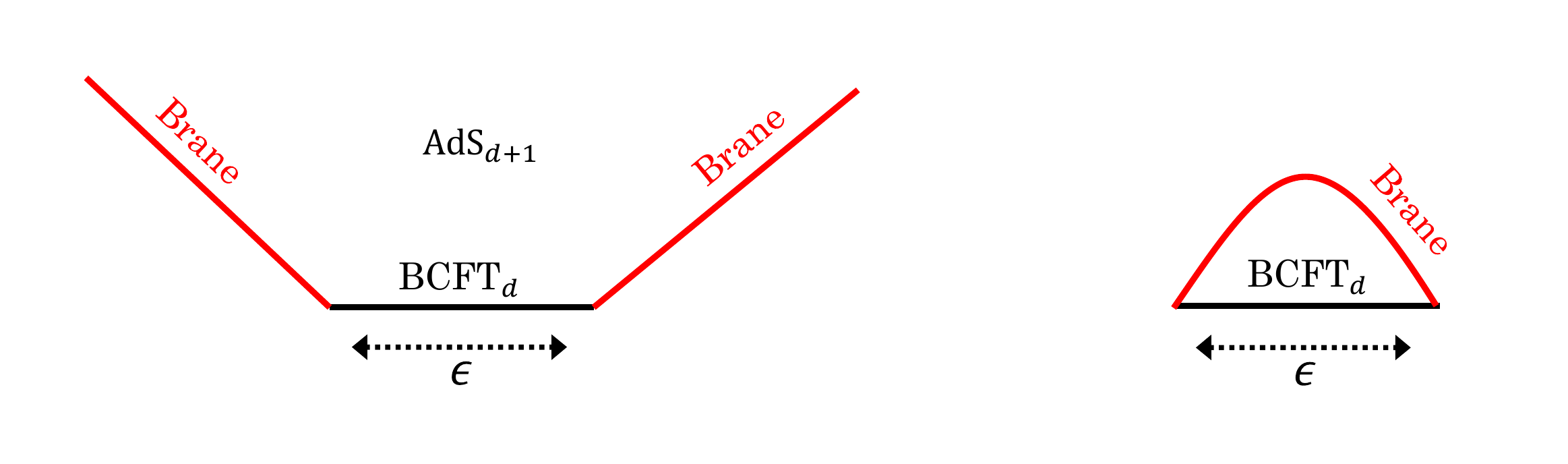}
 \end{center}
 \caption{BCFT with two boundaries. In the $\epsilon \to 0$ limit, the left setup leads to the duality between AdS${}_{d+1}$ gravity and CFT${}_{d-1}$, so-called the wedge holography. On the other hand, the right setup vanishes in this limit.}
 \label{fig:wedge}
\end{figure}

\item Wedge Holography

Recently, a new holographic duality has been proposed in \cite{Akal2020}, called wedge holography or codimension-two holography (see also \cite{Miao2021,Miao2021a}).
This is a duality between AdS${}_{d+1}$ gravity and CFT${}_{d-1}$.
The basic idea comes from the AdS/BCFT correspondence.
Let us consider a BCFT${}_d$ with two boundaries whose charges are different from each other (see the left of Figure \ref{fig:wedge}). If we take the limit as the distance between two boundaries (i.e., the parameter $\epsilon$ in Figure \ref{fig:wedge}) approaches zero, we can obtain a CFT${}_{d-1}$, which is dual to the AdS${}_{d+1}$ gravity.
Since this new holography is in an early phase of investigation, many parts remain unexplained yet.
For example, it is very interesting but still unclear what difference can be found in the Virasoro spectrum for the AdS${}_2$/CFT${}_1$ and AdS${}_3$/CFT${}_1$ correspondences.
Our fusion matrix bootstrap in BCFTs could be useful in such a situation where we have little knowledge about the spectrum.
We might be able to obtain characteristic properties of the CFT dual to the wedge,
which can be compared to the AdS${}_3$ side calculations.
This type of challenge would provide new support on the wedge holography.

\item Impossible brane configuration

A brane configuration where a brane overlaps itself should be prohibited.
We can translate this constraint into the CFT language by the AdS/BCFT correspondence.
With this idea, some constraints have been given (for example, \cite{Cooper2019, Geng:2021iyq }, which investigate properties of the braneworld by the AdS/BCFT).
It would be interesting to reproduce these types of constrains from the conformal bootstrap,
because the conformal bootstrap is particularly useful to exclude impossible spectra.
We would like to comment that the value $\fr{c}{32}$ discovered in \cite{Geng:2021iyq} sometimes appear in the light-cone conformal bootstrap \cite{Kusuki2018b, Kusuki2018,Kusuki2018a,Kusuki2019a} and the light-cone modular bootstrap \cite{Benjamin2019}.
Now that we have formulated the light-cone bootstrap in BCFTs,
it would be interesting to investigate the relation between them.

\item Some assumptions from braneworld

In order for gravity to be localized on the brane,
we need a large brane tension. Moreover, the more we decrease the brane tension, the larger the mass the graviton gets.
Therefore, we have to consider the brane with the large tension if we would like to use the braneworld holography \cite{Karch2001}.
We also have some constraints from the existence of the bulk causality \cite{Omiya2021, Reeves2021}.
This type of assumption appears as some constraints on the BCFT data on the BCFT side (for example, see \cite{Aharony2003, Reeves2021}).
It would be interesting to make use of these assumptions to solve the bootstrap equation.
In particular, it would be interesting to study whether such a theory can exist in a consistent way with the bootstrap equation.

We can also consider a similar approach in the AdS/BCFT setup.
\footnote{
Note that although the AdS/BCFT is similar to the braneworld holography, they are different from each other in a precise sense.
For example, while in the braneworld holography, all degrees of freedom of the conformal boundary on the BCFT side are localized on the brane on the bulk side,
this is not true in the AdS/BCFT.
}
It is naturally expected that not all boundary states can be dual to classical gravity with an EOW, only some do.
Thus, we have a question, which boundary condition can have a good classical gravity description.
To answer this question, it is natural to use the bootstrap equation with few known constraints from the consistency between BCFT calculations and gravity calculations (e.g., \cite{Kastikainen2021}).  It would be interesting to further explore this direction.

\item Generalization to BCFTs with other symmetries

It would be interesting to consider a generalization of our results into a BCFT with a rich structure.
For example, recently, the asymptotic behavior of the density of states transforming under a given irreducible representation of a global symmetry has been studied in \cite{Harlow2021,Magan2021,Cao2021,Pal2020a}.
From its generalization into BCFTs, we may be able to give statements about black holes with finite group hair in the doubly holographic scenario.

\end{itemize}

\section*{Acknowledgments}
We are grateful to Nathan Benjamin,  Dalimil Mazac, Hirosi Ooguri, Sridip Pal, David Simmons-Duffin, and Yifan Wang for useful discussions.
We would like to thank Nathan Benjamin, Sridip Pal, and Zixia Wei very much for valuable comments on this paper.
YK is supported by Burke Fellowship (Brinson Postdoctoral Fellowship).

\clearpage

\appendix

\section{Explicit Form of Fusion Matrix} \label{app:FM}

In the following, we introduce the notation usually found in Liouville CFTs.
\begin{equation}
c=1+6Q^2, \ \ \ \ \ Q=b+\fr{1}{b}, \ \ \ \ \ h_i=\a_i(Q-\a_i).
\end{equation}
Note that we can relate the parameter $\eta_i$ appearing in \cite{Kusuki2019a} to $\a_i$ as $\a_i=Q\eta_i$.

The fusion matrix is defined by the  invertible fusion transformations between $s$ and $t$- channel conformal blocks \cite{Teschner2001} as follows:
\begin{equation}\label{eq:fusiontrans}
\begin{aligned}
\ca{F}^{21}_{34}(h_{\a_s}|z)=\int_{\bb{S}} \dd \a_t {\bold F}_{\a_s, \a_t} 
   \left[
    \begin{array}{cc}
    \a_2   & \a_1  \\
     \a_3  &   \a_4\\
    \end{array}
  \right]
  \ca{F}^{23}_{14}(h_{\a_t}|1-z),
\end{aligned}
\end{equation}
where the contour $\bb{S}$ runs from $\fr{Q}{2}$ to $\fr{Q}{2}+ i\infty$, and also runs anti-clockwise around $\a_t=\a_1+\a_4+mb+nb^{-1}<\fr{Q}{2}$ and $\a_t=\a_2+\a_3+mb+nb^{-1}<\fr{Q}{2}$ for $m, n\in \mathbb{Z}_{\geq 0}$. The kernel $ {\bold F}_{\a_s, \a_t} $ is called the {\it crossing matrix} or {\it fusion matrix}. The explicit form of the fusion matrix is given in \cite{Ponsot1999,Teschner2001}as follows:
\begin{equation}\label{eq:crossing}
\begin{aligned}
{\bold F}_{\a_s, \a_t} 
   \left[
    \begin{array}{cc}
    \a_2   & \a_1  \\
     \a_3  &   \a_4\\
    \end{array}
  \right]
=\fr{N(\a_4,\a_3,\a_s)N(\a_s,\a_2,\a_1)}{N(\a_4,\a_t,\a_1)N(\a_t,\a_3,\a_2)}
   \left\{
    \begin{array}{cc|c}
    \a_1   & \a_2    &  \a_s  \\
     \a_3  & \a_4    &  \a_t   \\
    \end{array}
  \right\}_b,
\end{aligned}
\end{equation}
where the function $N(\a_3,\a_2,\a_1)$ is
\begin{equation}
N(\a_3,\a_2,\a_1)=\fr{\G_b(2\a_1)\G_b(2\a_2)\G_b(2Q-2\a_3)}{\G_b(2Q-\a_1-\a_2-\a_3)\G_b(Q-\a_1-\a_2+\a_3)\G_b(\a_1+\a_3-\a_2)\G_b(\a_2+\a_3-\a_1)},
\end{equation}
and 
$ \left\{
    \begin{array}{cc|c}
    \a_1   & \a_2    &  \a_s  \\
     \a_3  & \a_4    &  \a_t   \\
    \end{array}
  \right\}_b$
is the Racah--Wigner coefficient for the quantum group $U_q(sl(2,\bb{R}))$, which is given by
\footnote{Ponsot--Teschner have derived a more symmetric form of the Racah--Wigner coefficient \cite{Teschner2014} than the traditional expression found in \cite{Ponsot1999,Teschner2001}. In this study, we used the new expression derived in \cite{Teschner2014}.}
\begin{equation}\label{eq:6j}
\begin{aligned}
&\left\{
    \begin{array}{cc|c}
    \a_1   & \a_2    &  \a_s  \\
     \a_3  & \bar{\a_4}    &  \a_t   \\
    \end{array}
  \right\}_b\\
&= \fr{S_b(\a_1+\a_4+\a_t-Q)S_b(\a_2+\a_3+\a_t-Q)S_b(\a_3-\a_2-\a_t+Q)S_b(\a_2-\a_3-\a_t+Q)}{S_b(\a_1+\a_2-\a_s)S_b(\a_3+\a_s-\a_4)S_b(\a_3+\a_4-\a_s)}\\
&\times \abs{S_b(2\a_t)}^2 \int^{2Q+i \infty}_{2Q-i \infty} \dd u 
\fr{S_b(u-\a_{12s})S_b(u-\a_{s34})S_b(u-\a_{23t})S_b(u-\a_{1t4})}{S_b(u-\a_{1234}+Q)S_b(u-\a_{st13}+Q)S_b(u-\a_{st24}+Q)S_b(u+Q)},
\end{aligned}
\end{equation}
where we have used the notations $\bar{\a}=Q-\a$, $\a_{ijk}=\a_i+\a_j+\a_k$ and $\a_{ijkl}=\a_i+\a_j+\a_k+\a_l$.
The functions $\G_b(x)$ and $S_b(x)$ are defined as
\begin{equation}
\G_b(x)= \fr{\G_2(x|b,b^{-1})}{\G_2\pa{\fr{Q}{2}|b,b^{-1}}}, \ \ \ \ \ S_b(x)=\fr{\G_b(x)}{\G_b(Q-x)},
\end{equation}
$\G_2(x|\w_1,\w_2)$ is the double gamma function,
\begin{equation}
\log \G_2(x|\w_1,\w_2)=\pa{\pd{t}\sum^{\infty}_{n_1,n_2=0} \pa{x+n_1 \w_1+n_2\w_2}^{-t}}_{t=0}.
\end{equation}
Note that the function $\G_b(x)$ is introduced such that $\G_b(x)=\G_{b^{-1}}(x)$ and satisfies the following relationship:
\begin{equation}
\G_b(x+b)=\fr{\s{2 \pi}b^{bx-\fr{1}{2}}}{\G(bx)}\G_b(x).
\end{equation}
By substituting the explicit form of the Racah--Wigner coefficients (\ref{eq:6j}) into (\ref{eq:crossing}), we can simplify the expression for the fusion matrix into
\begin{equation}\label{eq:crossing2}
\begin{aligned}
&{\bold F}_{\a_s, \a_t} 
   \left[
    \begin{array}{cc}
    \a_2   & \a_1  \\
     \a_3  &   \a_4\\
    \end{array}
  \right]\\
&=\fr{\G_b(Q+\a_2-\a_3-\a_t)\G_b(Q-\a_2+\a_3-\a_t)\G_b(2Q-\a_1-\a_4-\a_t)\G_b(\a_1+\a_4-\a_t)}{\G_b(2Q-\a_1-\a_2-\a_s)\G_b(\a_1+\a_2-\a_s)\G_b(Q+\a_3-\a_4-\a_s)\G_b(Q-\a_3+\a_4-\a_s)}\\
&\times \fr{\G_b(Q-\a_2-\a_3+\a_t)\G(-Q+\a_2+\a_3+\a_t)\G_b(\a_1-\a_4+\a_t)\G_b(-\a_1+\a_4+\a_t)}{\G_b(\a_1-\a_2+\a_s)\G_b(-\a_1+\a_2+\a_s)\G_b(Q-\a_3-\a_4+\a_s)\G_b(-Q+\a_3+\a_4+\a_s)}\\
&\times \abs{S_b(2\a_t)}^2 \fr{\G_b(2Q-2\a_s)\G_b(2\a_s)}{\G_b(2Q-2\a_t)\G_b(2\a_t)} \\
&\times \int^{2Q+i \infty}_{2Q-i \infty} \dd u 
\fr{S_b(u-\a_{12s})S_b(u-\a_{s34})S_b(u-\a_{23t})S_b(u-\a_{1t4})}{S_b(u-\a_{1234}+Q)S_b(u-\a_{st13}+Q)S_b(u-\a_{st24}+Q)S_b(u+Q)}.
\end{aligned}
\end{equation}

\clearpage
\bibliographystyle{JHEP}
\bibliography{bcft}

\end{document}